\documentclass[twocolappendix]{emulateapj}
\usepackage{graphicx}
\usepackage{amsmath}
\pdfoutput=1

\def\solmass{$M_{\sun}$}
\def\solmassyr{$M_{\sun}$ yr$^{-1}$}

\def\reff{$R_{\rm eff}$}
\def\ClusterFrac{$f_{\rm cluster}$}
\def\ClusterFracW{$f_{\rm cluster, W}$}
\def\GalFrac{$f_{\rm galaxy}$}
\def\ViewFrac{$f_{\rm clst+gal}$}
\def\AvgClstFrac{$\bar{f}_{\rm cluster}$}
\def\GoodFrac{$f_{\rm consensus}$}

\begin{document}

\title{PHAT Stellar Cluster Survey. II. Andromeda Project Cluster Catalog}

\author{L. Clifton Johnson$^{1}$, Anil C. Seth$^{2}$, Julianne J. Dalcanton$^{1}$, Matthew L. Wallace$^{2}$, Robert J. Simpson$^{3}$, Chris J. Lintott$^{3}$, Amit Kapadia$^{4}$, Evan D. Skillman$^{5}$, Nelson Caldwell$^{6}$, Morgan Fouesneau$^{1,7}$, Daniel R. Weisz$^{1,10}$, Benjamin F. Williams$^{1}$, Lori C. Beerman$^{1}$, Dimitrios A. Gouliermis$^{7,8}$, Ata Sarajedini$^{9}$}

\email{lcjohnso@astro.washington.edu}
\affil{$^{1}$Department of Astronomy, University of Washington, Box 351580, Seattle, WA 98195, USA}
\affil{$^{2}$Department of Physics and Astronomy, University of Utah, Salt Lake City, UT 84112, USA}
\affil{$^{3}$Oxford Astrophysics, Denys Wilkinson Building, Keble Road OX1 3RH Oxford, UK}
\affil{$^{4}$Astronomy Department, Adler Planetarium, 1300 S. Lake Shore Drive Chicago, IL 60605 USA}
\affil{$^{5}$Department of Astronomy, University of Minnesota, 116 Church Street SE, Minneapolis, MN 55455, USA}
\affil{$^{6}$Harvard-Smithsonian Center for Astrophysics, 60 Garden Street, Cambridge, MA 02138, USA}
\affil{$^{7}$Max-Planck-Institut f\"ur Astronomie, K\"onigstuhl 17, 69117 Heidelberg, Germany}
\affil{$^{8}$Institut f\"ur Theoretische Astrophysik, Zentrum f\"ur Astronomie der Universit\"at Heidelberg, Albert-Ueberle-Stra{\ss}e~2, 69120 Heidelberg, Germany}
\affil{$^{9}$Department of Astronomy, University of Florida, 211 Bryant Space Science Center, Gainesville, FL 32611-2055, USA}
\altaffiltext{10}{Hubble Fellow}


\begin{abstract}

We construct a stellar cluster catalog for the Panchromatic Hubble Andromeda Treasury (PHAT) survey using image classifications collected from the Andromeda Project citizen science website.  We identify 2,753 clusters and 2,270 background galaxies within $\sim$0.5 deg$^2$ of PHAT imaging searched, or $\sim$400 kpc$^2$ in deprojected area at the distance of the Andromeda galaxy (M31).  These identifications result from 1.82 million classifications of $\sim$20,000 individual images (totaling $\sim$7 gigapixels) by tens of thousands of volunteers.  We show that our crowd-sourced approach, which collects $>$80 classifications per image, provides a robust, repeatable method of cluster identification.
The high spatial resolution Hubble Space Telescope images resolve individual stars in each cluster and are instrumental in the factor of $\sim$6 increase in the number of clusters known within the survey footprint.
We measure integrated photometry in six filter passbands, ranging from the near-UV to the near-IR.  PHAT clusters span a range of $\sim$8 magnitudes in F475W ($g$-band) luminosity, equivalent to $\sim$4 decades in cluster mass.
We perform catalog completeness analysis using $>$3000 synthetic cluster simulations to determine robust detection limits and demonstrate that the catalog is 50\% complete down to $\sim$500 \solmass\ for ages $<$100 Myr.
We include catalogs of clusters, background galaxies, remaining unselected candidates, and synthetic cluster simulations, making all information publicly available to the community.
The catalog published here serves as the definitive base data product for PHAT cluster science, providing a census of star clusters in an $L^\star$ spiral galaxy with unmatched sensitivity and quality.

\end{abstract}


\newpage

\section{Introduction} \label{intro}

Observations of our Local Group neighbor, M31, present the best opportunity for a detailed yet comprehensive study of a large spiral galaxy, providing a local analog to the disk-dominated systems that populate wide-field galaxy surveys.  While the Milky Way allows analysis at the highest level of detail, studying our host galaxy on the whole proves difficult due to distance ambiguities and large amounts of dust attenuation within the Galactic plane.  Conversely, studying galaxies beyond the local group necessitates a substantial decrease in data quality and content due to reduced spatial resolution and rising photometric completeness limits.

Similarly, Andromeda is an excellent target for obtaining a big picture view of a galaxy's stellar cluster population.  While many extragalactic cluster samples exist, each offering galaxy-wide coverage unattainable in the Milky Way, M31's proximity provides a number of sensitivity-based advantages.  Using the power of the Hubble Space Telescope (HST), we can obtain a census of Andromeda's star cluster population that extends deep into the low-mass regime while simultaneously resolving individual stars within each cluster. The ability to resolve individual stars also allows for thorough analysis of M31's field star populations, leading to detailed comparisons of field and cluster populations, enabling studies of cluster formation and dissolution in the context of the galaxy's overall star formation activity.

The Panchromatic Hubble Andromeda Treasury survey \citep[PHAT; ][]{Dalcanton12} provides contiguous, high spatial resolution imaging of approximately one-third of the M31 disk using the HST, observed in six broadband passbands that span from the near-UV to the near-IR.  The Year 1 cluster catalog \citep[][hereafter, Paper I]{Johnson12} presented cluster results from the first 20\% of the survey data.  In this paper, we present a final, survey-wide cluster catalog created through a crowd-sourced, visual search of the data.  The contribution of citizen scientists to astronomical research is not novel: projects such as Galaxy Zoo \citep{Lintott08, Willett13}, the Milky Way Project \citep{Simpson12}, and Planet Hunters \citep{Schwamb12} have previously made use of crowd-sourcing.  In this work we analyze image classifications collected from the Andromeda Project, a website established explicitly for the identification of star clusters in the PHAT dataset.

We utilize these data to assemble a cluster catalog that reaches cluster masses below $10^{3}$ \solmass.  This level of catalog completeness represents a significant extension to previous ground-based studies of M31, which mainly focused on old massive globular clusters, presented in the compilations of \citet[][and updates via its corresponding website\footnote{{\scriptsize \url{http://www.cfa.harvard.edu/oir/eg/m31clusters/M31\_Hectospec.html}}}]{Caldwell09}, the Revised Bologna Catalog\footnote{{\scriptsize \url{http://www.bo.astro.it/M31/}}} \citep[RBC;][last updated 2012 August to v5]{Galleti04}, \citet{Huxor14}, and references therein.  The new catalog also builds upon previous space-based efforts in M31 by \citet{Williams01} and the series of Hodge-Krienke catalogs \citep[][hereafter collectively referred to as the HKC]{Krienke07,Krienke08,Krienke13,Hodge09,Hodge10}.  The HST's high spatial resolution imaging allows for the identification of less massive clusters through its ability to differentiate between single stars and compact clusters, but previous HST-based studies were limited to isolated targeted observations.  In contrast, PHAT's contiguous wide-area coverage allows us to study cluster populations across the entire northeast quadrant of M31.

The catalog presented here serves as the basis for future work that will further characterize the sample: basic cluster parameter determinations \citep[age, mass, $A_V$;][Beerman et al., in prep]{Beerman12,Fouesneau14}, spatial profiles (Fouesneau et al., in prep), and comparison to spectroscopically-derived properties of the globular cluster population (Caldwell et al., in prep).  Once characterized, the star clusters presented here will be used as input for a variety of explorations by the PHAT collaboration and others.  As part of PHAT, we will place constraints on the high-mass stellar initial mass function (D. Weisz et al., in prep), and measure cluster formation efficiency throughout the galactic disk (L.C. Johnson et al., in prep) to test theoretical model predictions \citep{Kruijssen12}.  Further, we will constrain cluster dissolution time scales (M. Fouesneau et al., in prep) in an effort to differentiate between competing models \citep[mass-dependent versus mass-independent dissolution;][]{Fall09, Boutloukos03, Bastian12, Chandar10-M83}.

We begin with a description of the citizen science website and data in Section \ref{ap}.  Section \ref{catintro} discusses the process of converting contributions from citizen scientists into a catalog of objects, while Section \ref{comp} characterizes the make-up and completeness of the final catalogs.  We present our cluster catalog and accompanying integrated photometry in Section \ref{results}.  Section \ref{discuss} includes a comparison of the current catalog with our previous Year 1 work and a discussion of how this cluster sample fits within the context of other well-known cluster catalogs.  We conclude with a summary of our work in Section \ref{summary}.  Throughout this work, we assume a distance modulus for M31 of 24.47 \citep[785 kpc;][]{McConnachie05}, where 1~arcsec corresponds to a physical size of 3.81~pc.

\subsection{Cluster Definition} \label{clusterdef}

A star cluster can be defined in the most general sense as a grouping of stars that are spatially and temporally correlated.  Beyond this broad definition, the notion of a star cluster can vary significantly, depending mostly on whether the system is still embedded in its natal gas or exposed \citep{Lada03}.  Older ($>$10--30 Myr) gas-free systems are relatively straightforward to classify using a criterion based on the gravitational boundedness of individual members to a larger group.  In contrast, young groupings of stars that are still embedded within the ISM make classification a difficult, uncertain task.  These embedded clusters are still forming through hierarchical merging of sub-clumps \citep{Allison10}, and the application of various stellar density thresholds to identify distinct features of a continuous (scale-free) distribution leads to interpretative challenges \citep{Bressert10, Gieles12}.  Embedded environments are dynamically evolving and membership within a particular gravitational grouping is neither well-defined nor unique.  

For the PHAT cluster catalog, we work mostly in the exposed, gas-free regime because our identification is based on optical imaging.  Once the gas has been expelled from a star cluster and its stars have evolved through multiple dynamical times, it becomes possible to infer whether a grouping of stars is either gravitationally bound or expanding and dissolving \citep{Gieles11}.  Therefore, uncertainties pertaining to boundedness are minimal for our sample because a majority of PHAT clusters are already many dynamical times old, as inferred from the age and mass distributions of the Year 1 catalog \citep{Fouesneau14}.

At young ages ($<$10 Myr), the use of boundedness as a selection criterion for clusters becomes difficult.  Due to the similar appearance (i.e., radial spatial profile) of bound clusters and unbound stellar associations at young ages, determining boundedness becomes an uncertain and contentious enterprise \citep[e.g., see][]{Chandar10, Bastian12, Whitmore14}.  In the work that follows, we include all objects identified as part of our search.  As a result, our catalog may include a heterogeneous mix of bound and unbound objects at ages $<$10--30 Myr.  We choose this approach in an effort to maximize the return for science cases that do not depend on the differentiation between bound clusters and unbound associations, while allowing open discussion of differing cluster definitions where they affect the resulting scientific interpretation.  Overall, we seek a catalog of objects that are spatially and temporally correlated and can be reasonably approximated as simple stellar populations.  While this goal is easily achieved for a majority of the sample, we will make a point to identify regions of parameter space that contain debatable objects, allowing the reader to make informed decisions with regards to boundedness.  A full exploration of the question of boundedness requires detailed age and spatial structure information \citep{Gieles11}, which is beyond the scope of this work.

The inclusive philosophy that we adopt in this work represents a shift from the approach we took in Paper I, where we discarded objects that were classified as likely associations.  This paper's inclusive methodology leads to a modest $\sim$15\% increase in clusters when compared to the Year 1 catalog within their shared imaging footprint.  We discuss these catalog differences in detail in Section \ref{yr1diff}, but find good overall agreement between the two samples.

\begin{figure*}
\centering
\includegraphics[scale=0.5]{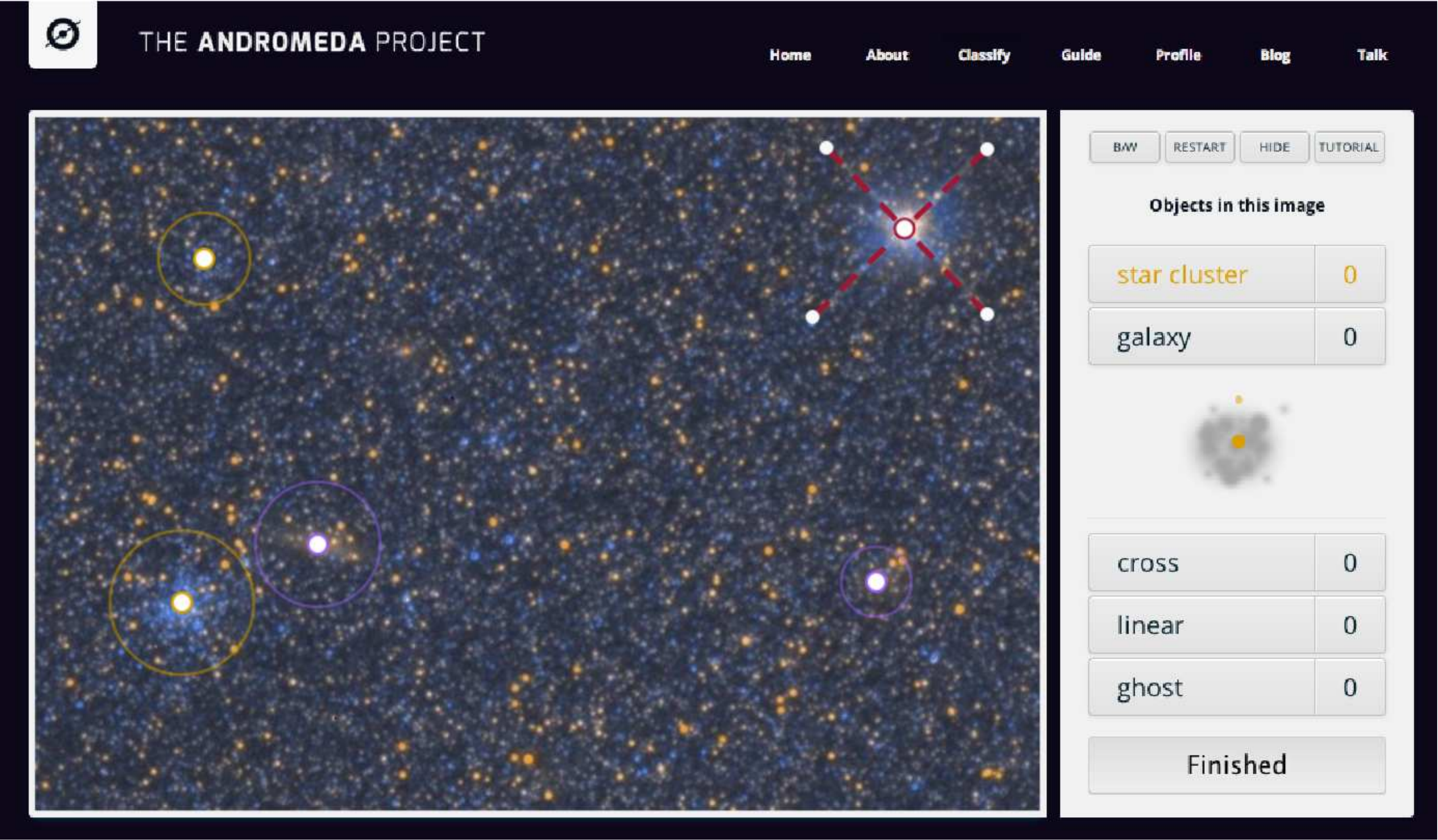}
\caption{The web-based classification interface for the Andromeda Project.  The tutorial image used to train participants is shown here, which includes all three object types: clusters (yellow), background galaxies (purple), and artifacts (i.e., a saturated star with diffraction spikes; red).}
\label{figclassify}
\end{figure*}

\section{The Andromeda Project} \label{ap}

In Paper I, we presented a sample of 601 clusters identified in a visual search carried out by eight professional astronomers, which examined the first 20\% of imaging acquired by the PHAT survey.  This task was time consuming; the initial identification of cluster candidates and subsequent quality ranking of the candidates required more than a month of effort from each scientist involved.  This cost limited our cluster search in two significant ways.  First, only 3--4 people looked at each image to make initial identifications of cluster candidates.  Of the 601 clusters, 23 were originally identified by just a single person, suggesting that a small number of additional good cluster candidates were probably missed in our initial search.  Second, characterization of the cluster completeness was done with a sample of 550 synthetic (artificial) clusters.  This relatively small sample of synthetic clusters limited our ability to track the completeness as a function of age, mass, cluster size, and galactocentric radius.    

Our original plan for extending the cluster search to the full PHAT footprint was to devise an automated algorithm to identify clusters using the Year 1 sample as a training set.  This approach proved challenging because all of the automated techniques we tested produced samples with at least as many contaminants as true clusters.  Expert by-eye verification would have been necessary to reduce the number of contaminants to an acceptable level.  This verification would have been time consuming and the resulting catalog would still suffer from subjectivity issues.
In addition, the goal of robustly characterizing the catalog selection function becomes difficult, requiring an understanding of human and machine behavior and their joint interaction.

The failure to devise a fully automated cluster identification technique, combined with the difficulty of scaling our original by-eye techniques to the full dataset, led us to create the Andromeda Project.  This crowd-sourced solution allows us to scale a by-eye search to the volume of data available from PHAT, improve the robustness and repeatability of cluster identifications, and accurately characterize the catalog completeness function.

\subsection{Interface} \label{interface}

The Andromeda Project\footnote{\url{http://www.andromedaproject.org}} (AP) is a website built and hosted by the Zooniverse\footnote{\url{http://www.zooniverse.org}} citizen science platform.  The AP interface is based on previous tools and code used for the Seafloor Explorer project, another Zooniverse project that aims to survey scallops, seastars, and other aquatic life using underwater imaging.

Upon entering the AP website, visitors are presented with the primary option to start classifying data, as well as links to find out more about the project.  Individuals who start classifying for the first time are directed to a tutorial image, where the basic functionality of the classification screen is explained.  The classification screen is shown in Figure \ref{figclassify}.  By default, the site displays a color image constructed from F475W and F814W imaging.  By clicking on the ``B/W'' button, participants can change the image to an inverted F475W gray scale image in which it is often easier to distinguish individual stars and faint image features.  The site's marking tool is set for cluster identification by default; modes for identifying background galaxies and three types of image artifacts are also available.  Markers for clusters, galaxies, or ghost artifacts are circular, positioned by clicking the center of an image feature and dragging outward to select the desired radius.  Only the cluster and galaxy markings are utilized in this paper.  

After clicking on the ``Finished'' button, volunteers are shown the location of the field they were classifying within M31 and given the option to discuss the images in the AP Talk\footnote{\url{http://talk.andromedaproject.org}} forum.  This feature enables new volunteers to get help identifying clusters, and allows participants to highlight interesting or confusing objects and discuss them with other volunteers and the science team.  After choosing whether or not to enter the Talk forum, volunteers are presented a new search image; the AP image database ensures that no user sees the same image twice.

Volunteers are urged to log-in or create a Zooniverse account, but participants are allowed to classify an unlimited number of images as an unregistered user.  Unregistered users do, however, receive periodic messages suggesting that they log-in or create an account.  Registration allows analysis of volunteers' classification behavior using consistent (anonymous) identifiers.  Input from unregistered users can still be aggregated from within a single classification session, however the (anonymous) identifiers tend not to carry over from session to session and could be shared by multiple unregistered users, limiting the depth of analysis we can perform.

\subsection{Input Data \& Synthetic Clusters} \label{datainput}

\begin{figure}
\includegraphics[scale=0.45]{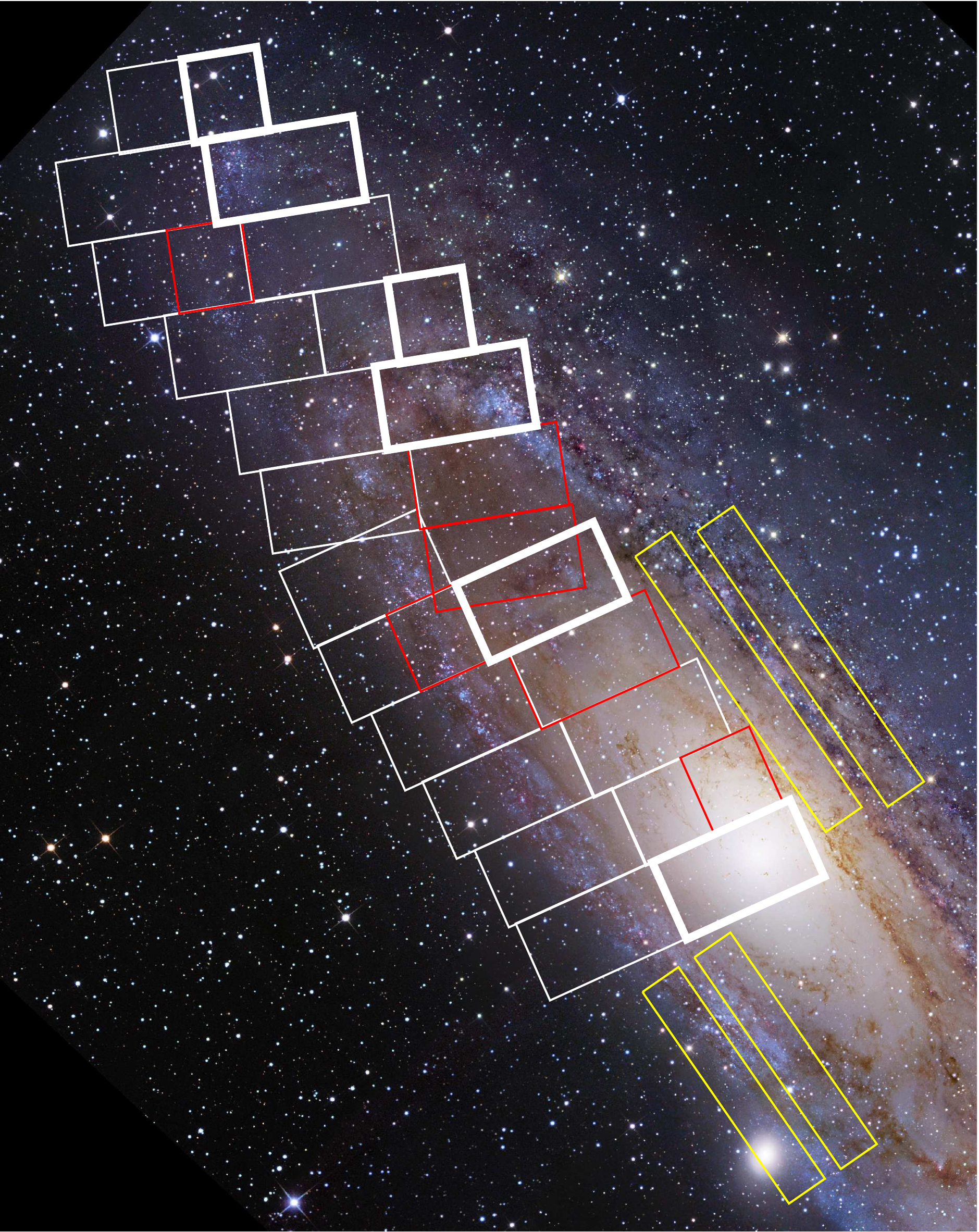}
\caption{Map of M31 showing HST imaging footprints, oriented such that north is up and east is left.  Color-coded regions denote various subsets of data: PHAT images searched during 2012 AP campaign (white); PHAT images searched during 2013 AP campaign (red); HST archival images searched during 2013 AP campaign (yellow).  Bold white regions show areas searched during Year 1 effort (Paper I).  Image Credit: Robert Gendler.}
\label{fig_footprint}
\end{figure}

Each search image shown on the AP site was extracted from high-resolution (0.05 arcsec pixel$^{-1}$) HST/ACS images of M31.  A vast majority of these images came from the PHAT dataset; we show the survey's imaging footprint in Figure \ref{fig_footprint}.  The prominent rectangular regions in Figure \ref{fig_footprint} that divide the survey into 23 parts are referred to as ``bricks''; their numbering increases from SW to NE along the major axis, starting with the brick enclosing the galaxy nucleus, B01 \citep[see Fig.~1 in][]{Dalcanton12}.  In addition to the optical (F475W, F814W; equivalent to $g$, $I$) ACS images, PHAT also obtained near-UV (F275W, F336W; the latter is equivalent to $U$) and near-IR (F110W, F160W; similar to $J$, $H$) imaging using the HST/WFC3 instrument.  Additional information about PHAT imaging data and survey design is available in \citet{Dalcanton12} and Paper I.

In addition to the PHAT data, we also processed and prepared ACS data from the HST archive (PID: 10273; PI: Crotts) that covered portions of M31 not imaged by PHAT.  The imaging footprint for these data are also shown in Figure \ref{fig_footprint}.  This program obtained two-filter optical imaging using filters (F555W, F814W) similar to those used by PHAT, allowing easy incorporation into the AP search.  Due to significant differences in data richness for objects identified in the archival dataset compared to the PHAT imaging, we choose not to include these objects in any further analysis, but we present object catalogs in Appendix \ref{archivecat}.

We created AP search images using 725$\times$500 pixel (36.25$\times$25 arcsec; $\sim$6.9$\times$4.8 pc in projected size) extractions from single-field ACS images.  This subimage size efficiently divides the image area, includes 100 pixels of overlap between neighboring subimages to reduce incompleteness and biases caused by image edges, and allows participants to search images at full resolution.  The parent ACS images have missing data due to the camera's chip gap that we filled using overlapping data from neighboring ACS images.  Gaps, edges, and other artifacts are still present in some images, but our efforts mitigated most issues concerning missing data.  We created a total of 13,017 subimages (4.7 gigapixels) from imaging that spans the entire PHAT survey region, as well as 1,728 addition subimages from archival imaging.

In addition to the normal imaging, we also produced additional search images that included synthetic clusters.  The primary reason for inserting these synthetic test objects is to measure the cluster catalog completeness as a function of age, mass, size, and environment.  In addition, the synthetics provided feedback to our volunteers: when a participant identified a synthetic cluster, they were notified that the object was synthetic and congratulated on their find.  Participants on the site's Talk forum confirm that these notifications acted as positive reinforcement that they were performing the task they set out to accomplish.

We used the Year 1 cluster catalog results and its small number of accompanying synthetic cluster tests to create a realistic variety of clusters for insertion into AP search images.  To begin, we choose age, mass, metallicity, attenuation, and effective radius values for the synthetic clusters drawn from distributions in each parameter:
\begin{itemize}
\item Ages were drawn from a flat distribution of discrete log(Age/yr) values between 6.6 and 10.1, spaced at an increment of 0.05 dex to match the grid of stellar isochrones from the Padova stellar evolution models \citep{Marigo08,Girardi10}.
\item Masses were drawn randomly from a continuous flat distribution of log(Mass/\solmass) between 2.0 and 5.0, yielding usable sample sizes across the full range of masses.
\item Solar metallicity ($Z=0.019$) was assumed for ages less than 5 Gyr.  For ages greater than 5 Gyr, the metallicity was selected from a grid of $Z$ to simulate the presence of metal-poor globular clusters: 0.0001 (0.005 $Z_{\sun}$), 0.001 (0.05 $Z_{\sun}$), 0.004 (0.2 $Z_{\sun}$), 0.008 (0.4 $Z_{\sun}$), 0.019 ($Z_{\sun}$).
\item Extinctions were drawn from an exponential $A_V$ distribution ranging from 0.17 mag \citep[foreground Galactic extinction;][]{Schlafly11} to 3.0 mag following the expression 
\begin{equation}
P(A_V) \propto e^{-A_V/1.34}.
\end{equation}
This distribution was chosen to match the extinction distribution derived by \citet{Fouesneau14} from their integrated light fitting of the Year 1 cluster catalog.
\item Spatial profiles are defined using a \citet{King62} profile with a fixed concentration ($R_{\rm tidal}/R_{\rm core}=30$) and an effective radius (\reff; equivalent to half-light radius) drawn from a distribution of measured half-light radii presented in Paper I, but with a linear bias towards larger radii.  We include this bias to boost the number of extended objects and ensure our ability to characterize the completeness of diffuse clusters.  The resulting \reff\ distribution peaks at 1.5 pc (0.39 arcsec) and extends from 0.5--9.0 pc (0.13--2.4 arcsec).
\end{itemize}

After drawing cluster parameter combinations, we populated individual cluster star lists using the Padova models, assuming a \citet{Kroupa01} IMF. We computed total luminosities for each cluster and selected a subset for insertion into search images that straddle the detection limit, as computed for the Year 1 catalog.  In Paper I, we found that the sample was 100\% complete for clusters brighter than $m_{F475W} = 18.5$ and 0\% complete for clusters fainter than $m_{F475W} = 23.5$.  Furthermore, when we take cluster age into account we can narrow the range of acceptable $m_{F475W}$ values even more: for 6.6 $< $ log(Age/yr) $<$ 8.0 we adopted $18.5 < m_{F475W} < 22.0$; for 8.0 $< $ log(Age/yr) $<$ 9.0 we adopted $19.5 < m_{F475W} < 22.5$;1 and for 9.0 $< $ log(Age/yr) $<$ 10.0 we adopted $20.0 < m_{F475W} < 23.0$.  These ranges allow us to efficiently map the functional form of completeness as a function of F475W magnitude at all ages.

Once we were satisfied with the sample, we inserted synthetic clusters into F475W and F814W images using the DOLPHOT photometry package, an updated version of HSTphot \citep{Dolphin00} that is used by the PHAT collaboration for point-spread function photometry.  These synthetic clusters were added into search images, one cluster per subimage, positioned pseudo-randomly within the image but always $>$120 pixels from the image edge.
We spatially distributed the synthetic clusters across the PHAT survey footprint, covering a wide range of galactic environments to ensure our ability to evaluate completeness throughout M31.  We selected fields that sample the survey's image variety, as defined by per-image red giant branch (RBG) star counts\footnote{RGB stars are defined as sources with F110W$-$F160W $>$ 0.5 and F160W $<$ 21.0; see Section \ref{comp}.}.  We inserted synthetic clusters into fields with $10^2 < N{\rm(RGB)} < 10^3$, and inserted proportionally less synthetics into fields with $N{\rm(RGB)} < 400$ to achieve a uniform number of synthetic clusters per $N$(RGB) bin within this range.  This selection results in the placement of synthetic clusters into regions where a majority of cluster identifications are made.

\subsection{Data Collection \& Classification Statistics} \label{userstats}

We obtained AP data during two rounds of collection; the first ran from 5--21 December 2012 and included 72\% of the PHAT images.  The remaining PHAT images and archival images were searched between 22--30 October 2013.  Defining a classification as a volunteer's submitted response to a single image (containing zero to many individual markings), AP volunteers performed a total of 1.82 million image classifications.  This corresponds to an average rate of about 70,000 classifications per day; our peak classification rate was over 80,000 classifications per hour.

\begin{figure}
\includegraphics[scale=0.44]{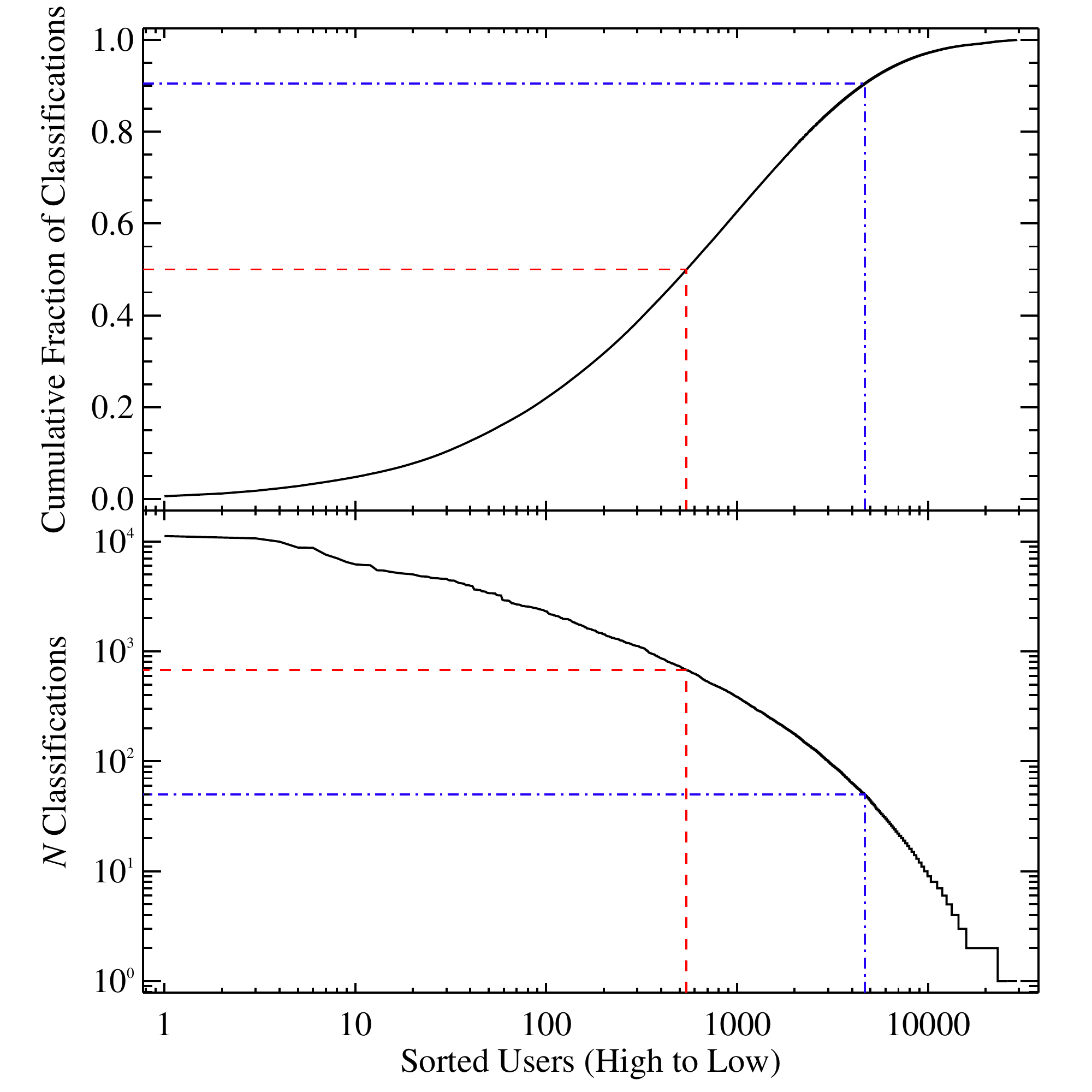}
\caption{Classification statistics for AP.  Participants are sorted as a function of decreasing contribution to the project and plotted logarithmically on the x-axis.  Top: Cumulative fraction of 1.82 million classifications submitted by the top $N$ volunteers.  Bottom: The number of classifications submitted individually by the $N$th volunteer.  The red dashed lines highlight that half of the classification work (cumulative fraction = 0.5) was performed by the top 543 participants, who each classified $\ge$678 images.  The blue dash-dotted lines highlight that 90.5\% of the total classifications were submitted by the 4,671 participants who each classified $\ge$50 images.}
\label{figusers}
\end{figure}

A total of 29,262 unique users participated in the AP; 9,663 of these participants logged in using a Zooniverse account.  While the median number of classified images among all users was only 3 images (27 when only considering registered participants), 90.5\% of our image classifications were performed by volunteers who examined at least 50 images.  The distribution of work completed by the AP volunteer community is shown in Figure \ref{figusers}.  The combined effort of Andromeda Project volunteers totals approximately 24 months of constant human attention.

Each image was classified a minimum of 80 times, but the distribution of classifications per image extends up to 108 with a median of 88.  The classification counts vary slightly between the two rounds of data collection: the median for the 2012 campaign is 86, while the median for the 2013 campaign is 101.  In all, participants made $>$2 million individual cluster and galaxy identifications.  

\section{Catalog Construction} \label{catintro}

The primary goal of this work is to construct a catalog of clusters from the identifications provided by the project's participants.  In this section we describe the process of converting clicks to scientifically-valuable data products.  We evaluate the reliability of the crowd-sourced results and choose appropriate catalog thresholds by comparing to the PHAT Year 1 catalog (Paper I), an expert-derived ``gold standard'' reference.

The first step of catalog construction is to synthesize a merged list of identifications.  We describe the details of our catalog creation algorithm and show examples of its application in Appendix \ref{buildcat}.  To briefly summarize: we spatially merge object identifications on an image-by-image basis, then merge these intermediate results into a survey-wide catalog. The resulting raw catalog includes $\sim$54,000 candidate clusters and galaxies, although a vast majority of these are low significance detections as we discuss below.  Synthetic cluster tests are analyzed using outputs from the per-image catalogs.  Also, artifact identifications are processed separately from the cluster and galaxy identifications and will not be discussed as part of this work.

After assembling a set of candidate objects, we use three metrics to identify cluster candidates and separate them from galaxies:
\begin{itemize}
\item \ClusterFrac\ -- the fraction of volunteers who viewed the search image and identified the object as a cluster.
\item \GalFrac\ -- the fraction of classifications for an object that identified it as a galaxy.
\item \ViewFrac\ -- the fraction of volunteers who viewed the search image and identified the object as either a cluster or a galaxy.
\end{itemize}
These quantities are related by:
\begin{equation}
f_{\rm cluster}=f_{\rm clst+gal} \times (1 - f_{\rm galaxy})
\end{equation}

\begin{figure}
\centering
\includegraphics[scale=0.8]{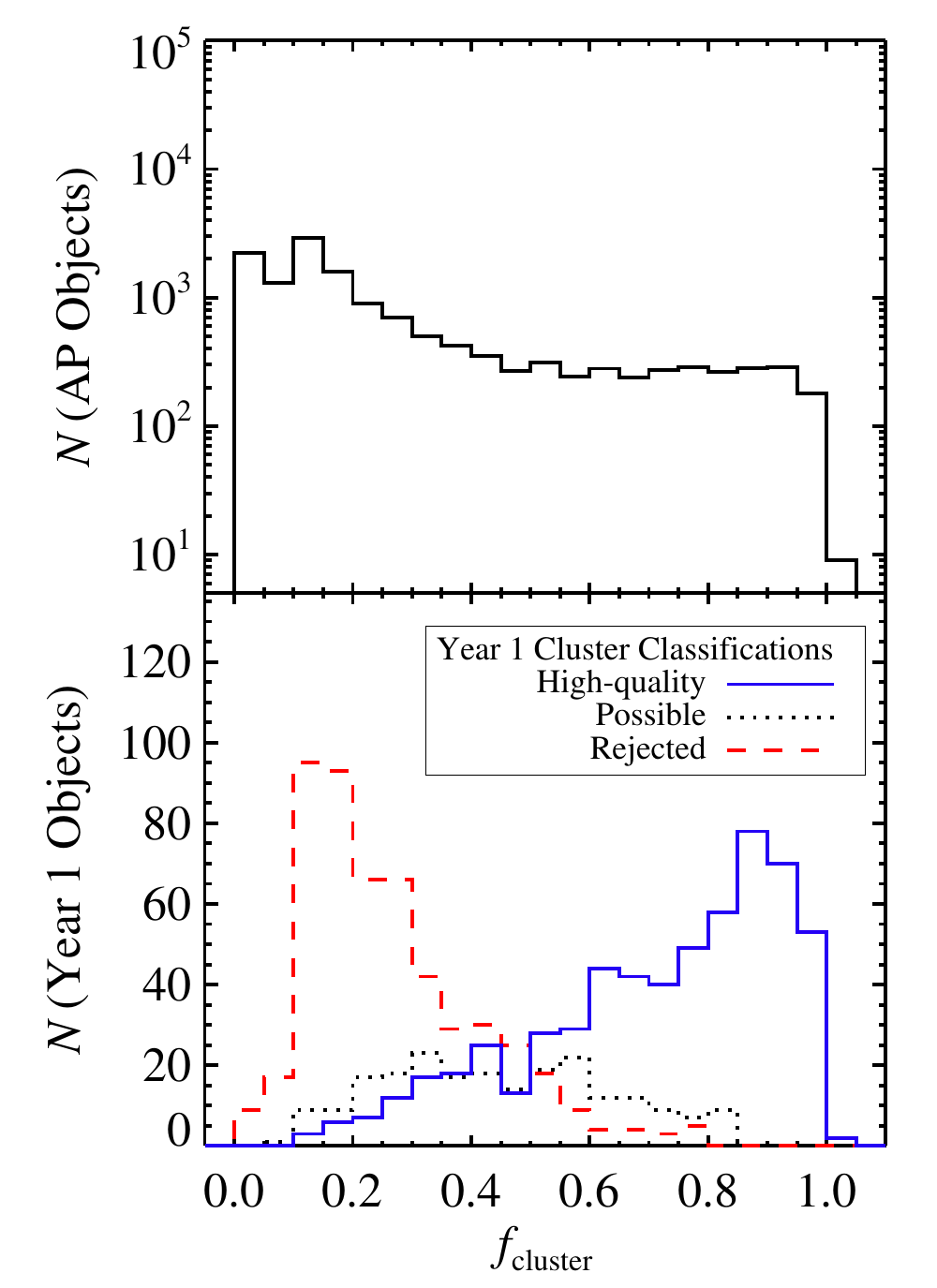}
\caption{Top: Histogram of \ClusterFrac\ values for the full catalog of AP identifications.  Bottom: Histograms of \ClusterFrac\ values for cross-matched high-quality (blue solid), possible (black dotted), and rejected (red dashed) Year 1 cluster candidates.}
\label{figclstfrac}
\end{figure}

The \ClusterFrac\ scores provide relative rankings for AP cluster candidates.  The top panel of Figure \ref{figclstfrac} shows the overall distribution of \ClusterFrac\ scores for all AP identifications.  This plot shows a large number of low significance detections with respect to higher significance detections.  The distribution of \ViewFrac\ values is nearly identical to the \ClusterFrac\ distribution.

We begin our comparison between AP and Year 1 results by cross-matching the two catalogs.  The bottom panel of Figure \ref{figclstfrac} compares the distribution of AP \ClusterFrac\ scores for three categories of Year 1 cluster cluster classifications.  We confirm the expectation that increasing AP \ClusterFrac\ scores correlate with a greater likelihood that candidates are clusters.

\begin{figure}
\centering
\includegraphics[scale=0.8]{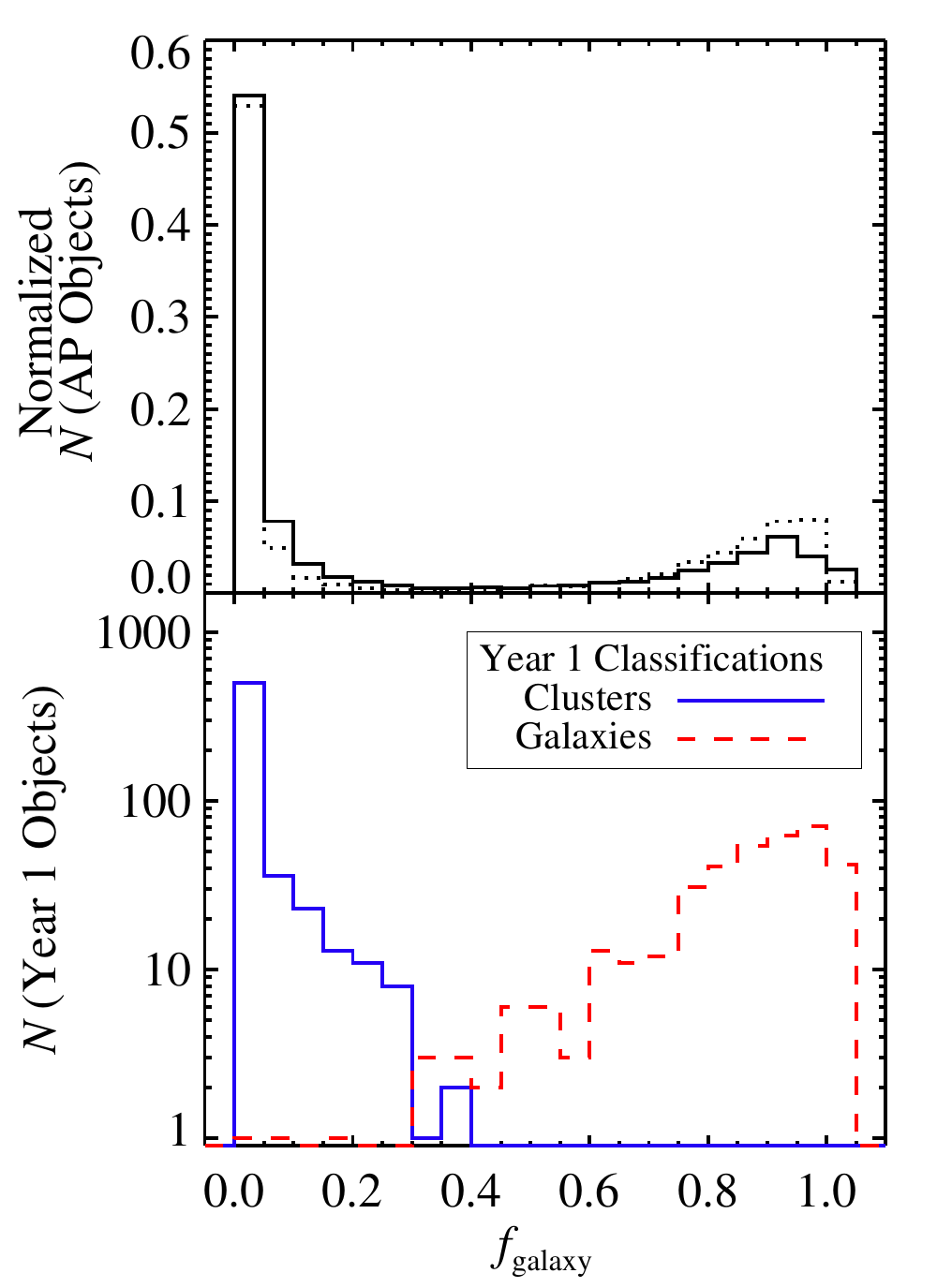}
\caption{Top: Histogram of \GalFrac\ values, where the solid (dotted) lines represent the 13,801 (4,449) AP identifications with \ViewFrac\ $\ge$ 0.1 (0.5). The bimodality of this distribution shows the tendency for AP classifiers to strongly differentiate between clusters and galaxies. Bottom: Histogram of \GalFrac\ values for AP identifications cross-matched to the Year 1 galaxy (red dashed line) and high-quality cluster (blue solid line) catalogs.  An \GalFrac\ threshold of 0.3 divides clusters and galaxies with minimal classification errors.}
\label{figgalfrac}
\end{figure}

The distribution of \GalFrac\ values is presented in Figure \ref{figgalfrac}.  The top panel shows a clear bimodality in \GalFrac\ values, signaling a clear cluster-versus-galaxy classification preference for a majority of candidate objects.  The bottom panel confirms the accuracy of these classification preferences; the expert-derived cluster and galaxy classifications from the Year 1 catalog map to low and high \GalFrac\ scores, respectively.  We also observe that \GalFrac\ = 0.3 defines a division between clusters and galaxies that leads to a minimal number of misclassifications.

It is interesting to note that there is an apparent bias at intermediate \GalFrac\ values (0.3$<$ \GalFrac\ $<$ 0.5), such that a majority vote of AP participants would not classify these objects accurately, according to expert-derived labels.
We hypothesize that this bias may be caused by the default cluster setting for the site's marking tool, leading to the tendency to mark candidates, particularly questionable ones, as clusters.  Whatever the cause may be, only a small number of objects in this range of \GalFrac\ could plausibly be considered for inclusion in the AP catalog as a cluster instead of as a galaxy: there are 125 (13) objects with \ClusterFrac\ $>$ 0.2 (0.5) in the full catalog of AP identifications that fall within 0.3$<$ \GalFrac\ $<$ 0.5.
Nevertheless, we adopt an \GalFrac -based selection criterion to account for this bias and incorporate as much information as possible during classification. We use the observed \GalFrac\ = 0.3 threshold throughout the remainder of the paper to differentiate cluster and galaxy candidates.

To select a catalog of likely clusters from the set of AP identifications, we use selection criteria based on the cluster candidate's \ClusterFrac\ and \GalFrac\ values.  While we've clearly defined an \GalFrac -based selection criterion, we now need to define an \ClusterFrac\ threshold that maximizes the number of clusters identified while minimizing the number of non-cluster contaminants.  As the bottom panel of Figure \ref{figclstfrac} shows, these are directly competing goals; decreasing the \ClusterFrac\ threshold to include a greater number of high-quality clusters necessarily introduces additional contaminants as well.

To evaluate how our choice of \ClusterFrac\ threshold affects the resulting cluster catalog, we calculate completeness and contamination fractions based on a comparison between the AP and Year 1 catalogs within their shared search footprint in the disk of M31.  We exclude the bulge region (Brick 1) from our comparison as its classification results differ sufficiently from the rest of the survey (see Section \ref{catdata} for further discussion).  We define completeness as the fraction of high-quality Year 1 clusters accepted by the AP selection criteria.  Contamination is quantified as the fraction of accepted AP clusters that were previously classified as non-clusters or galaxies by the Year 1 catalog, or are new AP-only objects not identified or classified during the Year 1 search.

We note that these definitions of completeness and contamination make an imperfect assumption that the Year 1 search is flawless, in which no worthy clusters escaped identification and every high-quality cluster tabulated deserves that distinction.  While this expert-derived catalog serves as a useful standard against which we can compare, it is inevitable that the completeness and contamination fractions we calculate with respect to the Year 1 catalog are approximate: 100\% completeness will not be attained, and we expect a modest, non-zero contamination fraction.  
To evaluate previously unidentified objects, we could perform an expert review to individually assess these possible contaminants, however this strategy cannot remove the element of researcher subjectivity.  Instead, we adopt an explicitly conservative stance that affects the absolute values of the contamination fractions we derive, but which do not impact the analysis choices we make due to the relative nature of most of these decisions.

\begin{figure}
\centering
\includegraphics[scale=0.55]{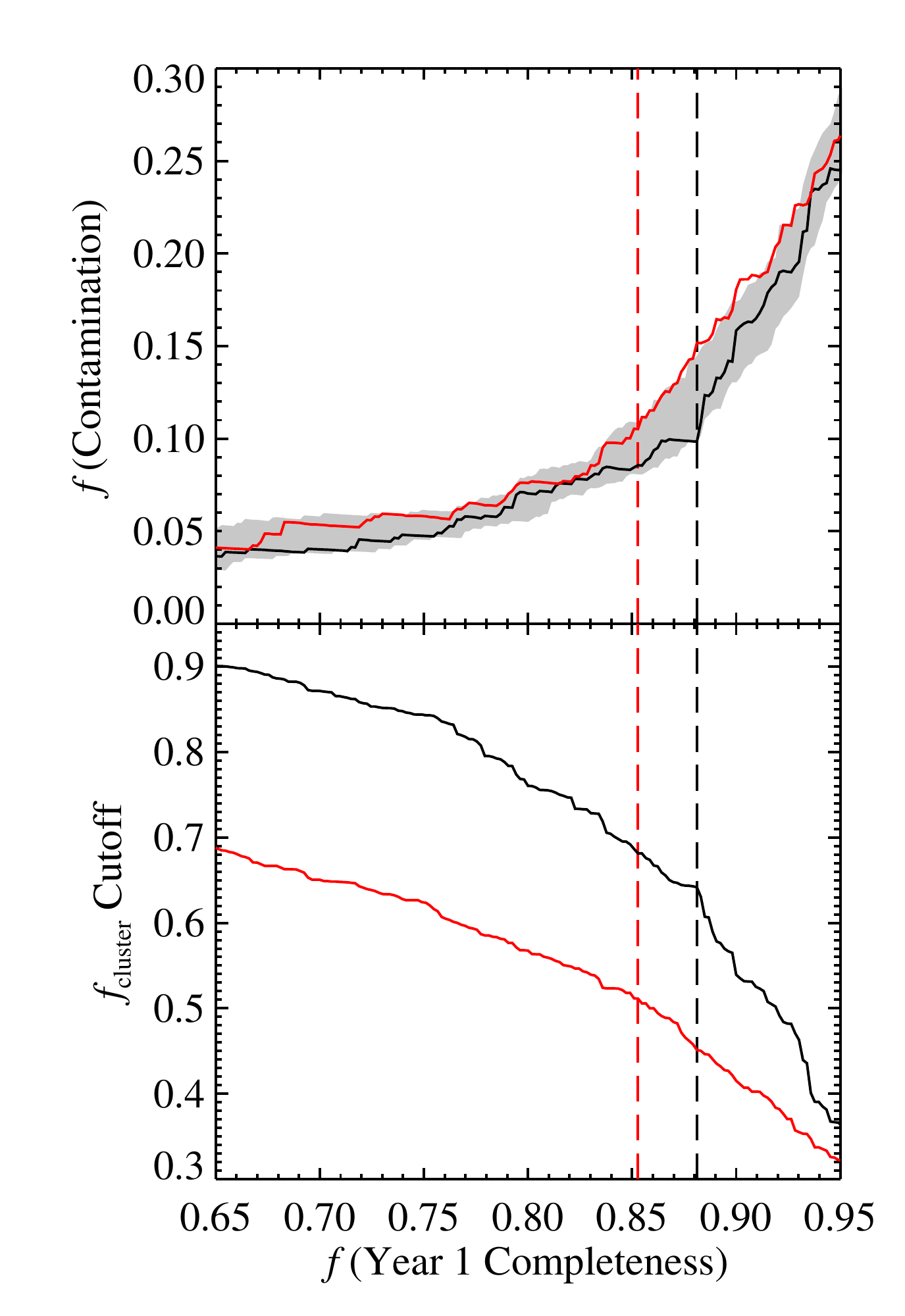}
\caption{Top: Completeness versus contamination curves that result from uniform user weighting (red) and the optimal user weighting system (black).  The gray shaded region denotes the parameter space covered by the sum of all curves derived for the grid of weighting systems we tested (see Section \ref{userweights}).  Bottom: The \ClusterFrac\ thresholds used by the uniform and adopted weighting systems as a function of Year 1 completeness.  The vertical dashed lines in both panels denote the catalog limits adopted for each system based on the $d_{\rm optimal}$ metric.}
\label{fig_compcont}
\end{figure}

We calculate a completeness versus contamination curve with respect to the expert-derived Year 1 catalog, akin to a receiver operator characteristic (ROC) curve.  By continuously lowering the \ClusterFrac\ threshold for the definition of AP clusters, we increase the completeness of Year 1 objects identified (bottom panel of Fig.~\ref{fig_compcont}).  However, the decreasing \ClusterFrac\ threshold also increases the contamination, defined as the fraction of the cluster catalog objects that are either Year 1 non-clusters or new AP-only clusters (top panel of Fig.~\ref{fig_compcont}).  In addition to the initial uniformly-weighted set of object identifications, we construct completeness versus contamination curves assuming different user weighting schemes, as discussed in Section \ref{userweights}.  We compare the result from uniformly-weighted inputs (red) to the range of results obtained from a grid of weighting systems (gray), including the curve derived for our optimal weighting scheme (black).

To choose a catalog cutoff, we seek a metric that identifies the \ClusterFrac\ cutoff value for which the resulting catalog achieves a balance between completeness and contamination.  We choose to work directly with the completeness versus contamination curve and define $d_{\rm optimal}$, the distance from each point along the curve to the optimal corner of the plot (completeness and contamination fractions are 1.0 and 0.0, respectively).  We note that our choice of metric, which values the minimization of false positives and false negatives equally, is somewhat arbitrary; given a specific use-case, one might prefer a metric that optimizes for a greater number of classifications at the expense of additional contamination.  Our choice to weight completeness and contamination equally is grounded in the goal of creating a general-purpose catalog.  Also, when we considering the specific shape of the completeness versus contamination curves we are working with, we find that this metric also tends to select the approximate point of diminishing return, the limit beyond which relaxing the catalog threshold tends to add more contaminants than additional good objects.  On the completeness versus contamination plot this limit corresponds to the point at which the curve is tangent to a line with a slope of unity.  In addition, it is also comforting that our choice of metric also tends to approximately conserve the number of clusters within the Year 1 footprint, yielding a similar number of clusters as found in Paper I.  Together, the similarity of these limits gives us confidence that our specific choice of cutoff is appropriate.

We use the $d_{\rm optimal}$ metric to identify an optimal completeness and contamination combination of 85.3\% and 10.5\%, respectively, for the case of uniform user weighting; the corresponding \ClusterFrac\ cutoff is plotted in Figure \ref{fig_compcont}, which is tabulated along with other corresponding information in Table \ref{compconttable}.  We improve sample completeness and contamination fractions using a user weighting system, as we discuss in Section \ref{userweights}.

We select a catalog of likely background galaxies using a combination of \ViewFrac\ and \GalFrac\ selection criteria in a process similar to the one described here for the clusters.  We document that analysis and its accompanying details in Appendix \ref{bckgal}.

\subsection{\ClusterFrac\ Uncertainties \& Robustness}

\begin{figure}
\centering
\includegraphics[scale=0.37]{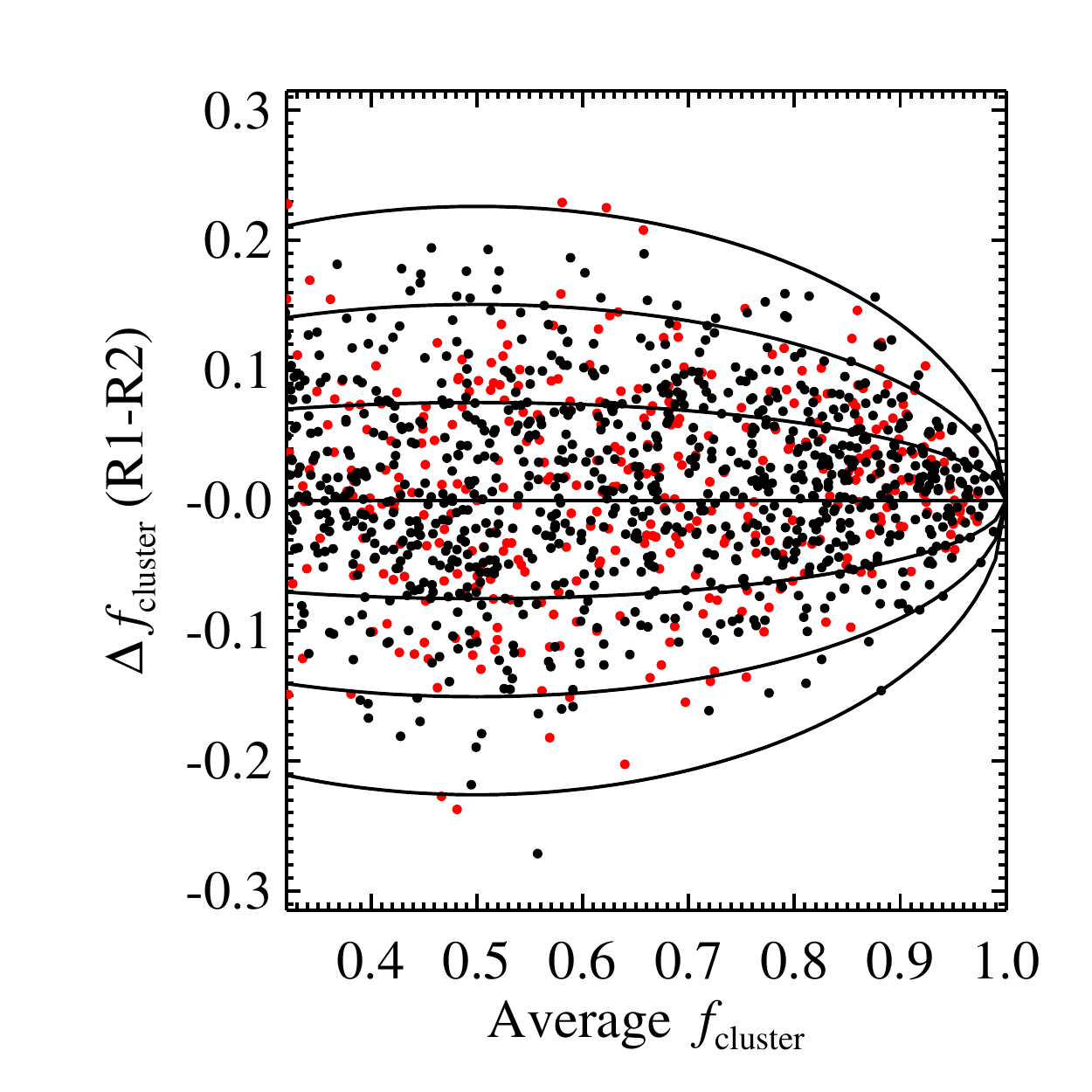}
\includegraphics[scale=0.37]{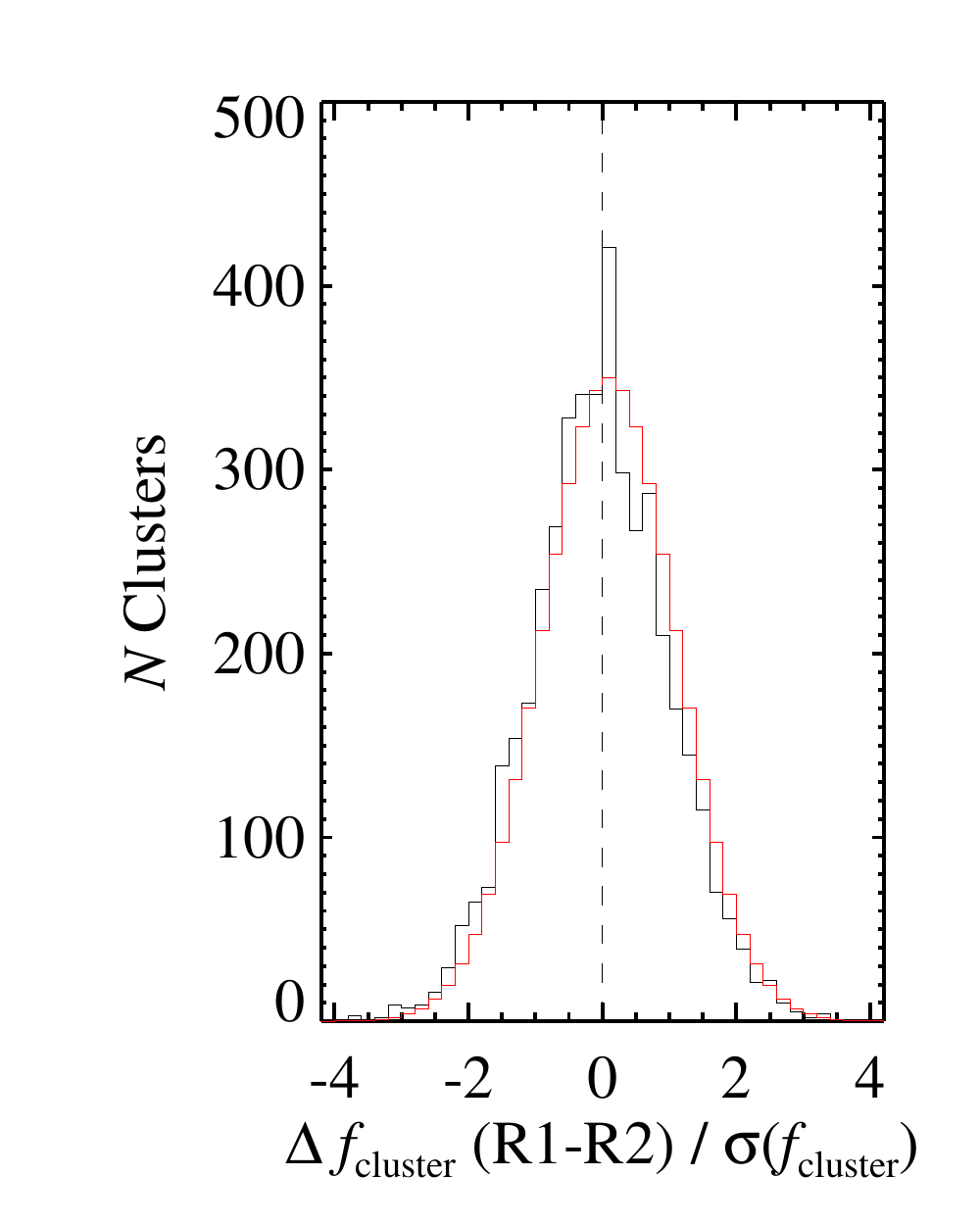}
\caption{Left: Comparison of \ClusterFrac\ for clusters derived from the 2012 campaign (R1) data versus those from the 2013 campaign (R2).  Black points reflect measurements made in normal images, red points reflect measurements made in synthetic images.  We plot 1, 2, and 3$\sigma$ contours showing the scatter predicted by our noise model.  Right: Histogram of \ClusterFrac\ differences scaled by the expected dispersion.  A Gaussian function with $\sigma$=1 and a peak value of 350 is overlaid for reference.  The dispersion of the \ClusterFrac\ differences between the two rounds matches the statistical expectation of the noise model.}
\label{figrepeat}
\end{figure}

To demonstrate the robustness of our \ClusterFrac\ metric, quantify its associated uncertainties, and establish its consistency across two separate rounds of data collection, we carried out a repeatability experiment during the 2013 campaign.  We selected 741 images (397 normal, 344 synthetic) searched during the 2012 campaign (Round 1; R1) that included highly-ranked cluster candidates and repeated data collection for these images during the 2013 campaign (Round 2; R2).  We match the catalogs that emerge from each run and compare \ClusterFrac\ scores for 1,241 objects whose R1 and R2 scores average to $>$0.35 (891 from normal data, 350 from synthetic data) to test the repeatability of \ClusterFrac\ scores for likely clusters.  We present the distribution of \ClusterFrac\ differences between the two rounds in Figure \ref{figrepeat}.  We model the $\Delta$\ClusterFrac$({\rm R1-R2})$ scatter using an expression for the combined variance of two drawing experiments governed by the binomial distribution:
\begin{equation} \sigma(p) =  \sqrt{\frac{2p(1-p)}{N}}, \end{equation}
where $N=88$, representing the median number of image views, and $p$ is the averaged R1 and R2 \ClusterFrac\ score for each object, representing our best estimate of an object's ``true'' \ClusterFrac\ value.  We plot 1, 2, and 3$\sigma$ contours as predicted by our noise model, which accurately captures the scatter shown in the data.

These results demonstrate that image classifications collected during the 2012 and 2013 campaigns are functionally equivalent, allowing us to easily combine data from the two rounds.  This experiment also shows that our procedure of combining $>$80 image classifications from the pool of AP participants provides consistent \ClusterFrac\ results with minimal systematic biases.

\subsection{User Weighting} \label{userweights}

Up to this point, we have assumed that the abilities of all classifiers are equal on average.  In this section we investigate whether weighting individual volunteers based on the quality of their classifications can improve the cluster sample.  User weighting has been applied in several other citizen science projects \citep[][]{Lintott08, Schwamb12} and seems naturally applicable to our AP data.  In line with these previous implementations, we calculate weightings based on the level of agreement between a participant's classifications and the consensus opinion of all the volunteers.  Individuals who agree with the consensus opinion are up-weighted, while those who disagree with the consensus opinion are down-weighted.  Expanding beyond previous implementations, we vary the strength and form of weighting, evaluating the success of each iteration by comparing completeness versus contamination curves (derived through comparison to the Year 1 sample) to the unweighted case presented earlier in Section \ref{catintro}.

We could have chosen another way to assign weights, such as assessing a volunteer's performance with respect to expert-derived Year 1 results, or basing weights on a participant's recovery rate of synthetic clusters.  One downside suffered by both of these alternative methods: resulting weights are based on only a fraction of the available classification data.  Decreasing the volume of classifications considered by the weighting system leads to an increasing number of participants with little or no assessment information, and noisier ability estimates for every volunteer.  Additionally, weighting systems tend to produce catalogs that resemble the data used for training and calibration.  We were concerned that defining weights based on data that did not sample the variety and parameter ranges included in the full cluster sample might result in unwanted biases.  Particularly in the case of the synthetic cluster data, which was specifically designed to characterize cluster recovery near the detection limit, these biases could be significant.  To exploit the unique benefits provided by our crowd-sourced methodology, we utilize the unfiltered opinion of AP volunteers.

\begin{figure}
\centering
\includegraphics[scale=0.6]{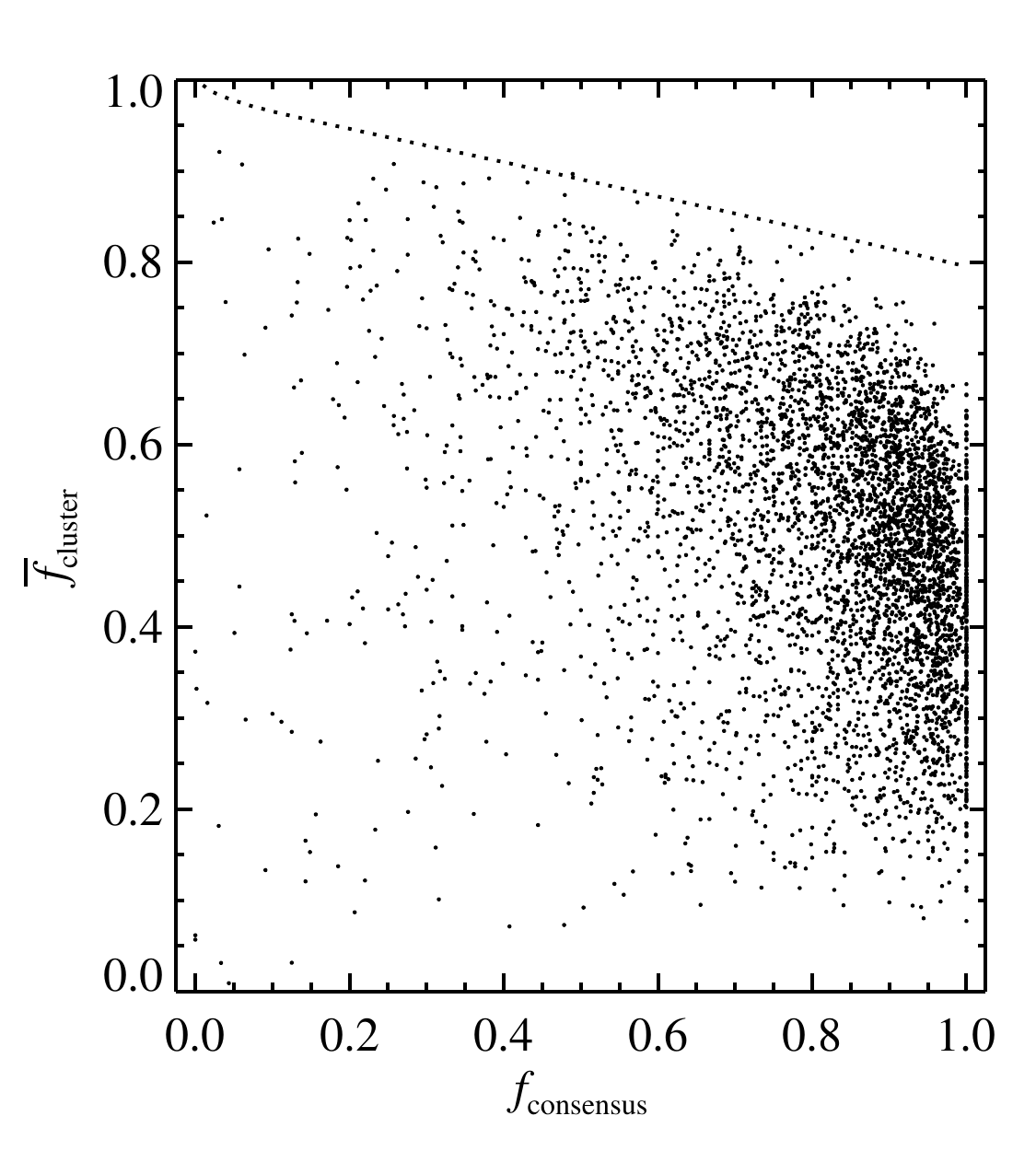}
\caption{Performance metrics for 4,671 volunteers who classified $\ge$50 AP search images.  The x-axis represents \GoodFrac, the fraction of consensus clusters (\ClusterFrac\ $\ge$ 0.6 and \GalFrac\ $<$0.2) identified by each participant out of the total number of consensus clusters they saw.  The y-axis represents \AvgClstFrac, the average \ClusterFrac\ value of all clusters identified by that volunteer.  The dotted line represents an approximate ceiling to \AvgClstFrac\ values as a function of \GoodFrac, calculated by considering the intrinsic \ClusterFrac\ distribution of the consensus cluster sample.   Conservative classifiers, those that identify only the best cluster candidates, lie in the upper left portion of the plot.  Liberal classifiers, those that identify most good clusters but also identify many low-ranked candidates, lie in the bottom right portion of the plot.}
\label{figuserweights}
\end{figure}

Figure~\ref{figuserweights} shows two quantities that we use to characterize the performance of our volunteers: the fraction of consensus clusters a volunteer identified, \GoodFrac, and the mean \ClusterFrac\ of all cluster identifications made by a volunteer, \AvgClstFrac.  We define consensus clusters as objects that show a high degree of agreement among AP participants, where \ClusterFrac\ $>0.6$ and \GalFrac\ $<0.2$; these limits provide a sample with a sufficient number of clusters to enable weighting of individual participants while ensuring that weights are not based on questionable candidates (see Figure \ref{figclstfrac}).

Examination of Figure \ref{figuserweights} reveals that there is wide variation of classification behavior among AP volunteers.  Individuals that lie in the upper left part of the plot are conservative classifiers; everything they clicked was an obvious cluster, leaving many consensus clusters unmarked.  On the other hand, participants in the lower right are liberal classifiers; they identified a large fraction of consensus cluster sample, but also identified many other low-ranked objects that are not likely clusters.  Volunteers with scores that lie in the upper right portion of Figure \ref{figuserweights} are desirable classifiers, obtaining high completeness but with little sacrifice to the overall quality of their identifications.  We note that because of the intrinsic \ClusterFrac\ distribution of the good clusters, volunteers who identify a large fraction of the good clusters cannot have an average \ClusterFrac\ of 1.0; we compute the upper limit to average \ClusterFrac\ based on the \ClusterFrac\ distribution of good clusters and plot this envelope as a dashed line in Figure \ref{figuserweights}.

To make best use of classifications from both conservative and liberal cluster identifiers, we apply separate weightings to volunteer's detections and non-detections.  Specifically, we weight a participant's detections based on the average \ClusterFrac\ of clusters they identified, while non-detections are weighted based on \GoodFrac, the fraction of consensus clusters the volunteer identified.  As an example: in the case where a liberal classifier in the lower right corner of Fig.~\ref{figuserweights} did not click on a cluster, their non-detections are up-weighted because they are known to identify most good clusters.  The detections from the same classifier, however, are down-weighted because this individual identifies many low-quality cluster candidates in addition to the high-quality ones.

We adopt a threshold number of subimage classifications above which we can assume we have adequately characterized a participant's classification behavior.  Volunteers with fewer than 50 subimage classifications are distributed with greater randomness across the \AvgClstFrac\ versus \GoodFrac\ plane, suggesting large uncertainties in the values of their performance metrics; we adopt 50 classifications as the threshold.  Individuals who fall below this classification threshold are assigned mean detection and non-detection weights.  Even when this limit is imposed, $\sim$90\% of all image classifications are weighted using individually determined user weights.  We note that anonymous accounts from unregistered users are treated in the same way as those from registered users for weighting purposes.  Most of these users are assigned mean detection and non-detection weights due to the fact that they submit a small number of classifications (median number of classifications is 2); $\sim$5\% of unregistered users surpass the minimum subimage threshold for individual weight assignment.

Next we determine how to translate performance metric scores into relative user weights.  We adopt a general form for the transformation based on the generalized logistic function.  Favorable aspects of this functional form include its tunable scaling and that it allows for the saturation of weights at high and low input metric scores. Our ``constrained'' logistic function is defined as:
\begin{equation}
W(x) = B \times \left( A + \frac{1}{1+e^{-m_{\rm{logistic}}(x-b_{\rm{logistic}})}} \right),
\end{equation}
where $x$ represents the input performance metric (either \AvgClstFrac\ or \GoodFrac) while $m_{\rm{logistic}}$ and $b_{\rm{logistic}}$ are the slope (growth rate) and the offset (position of maximum growth) of the logistic curve, respectively.  The variables $A$ and $B$ are normalizations set such that $W$ varies from 0 to 1 over the interval $x=[0,1]$, providing the constrained aspect of this function.  Once a set of logistic function parameters have been chosen for the detection and non-detection weighting functions, we apply user weightings to individual cluster votes on an image-by-image basis and recalculate weighted \ClusterFrac\ values, \ClusterFracW.

\begin{figure}
\centering
\includegraphics[scale=0.5]{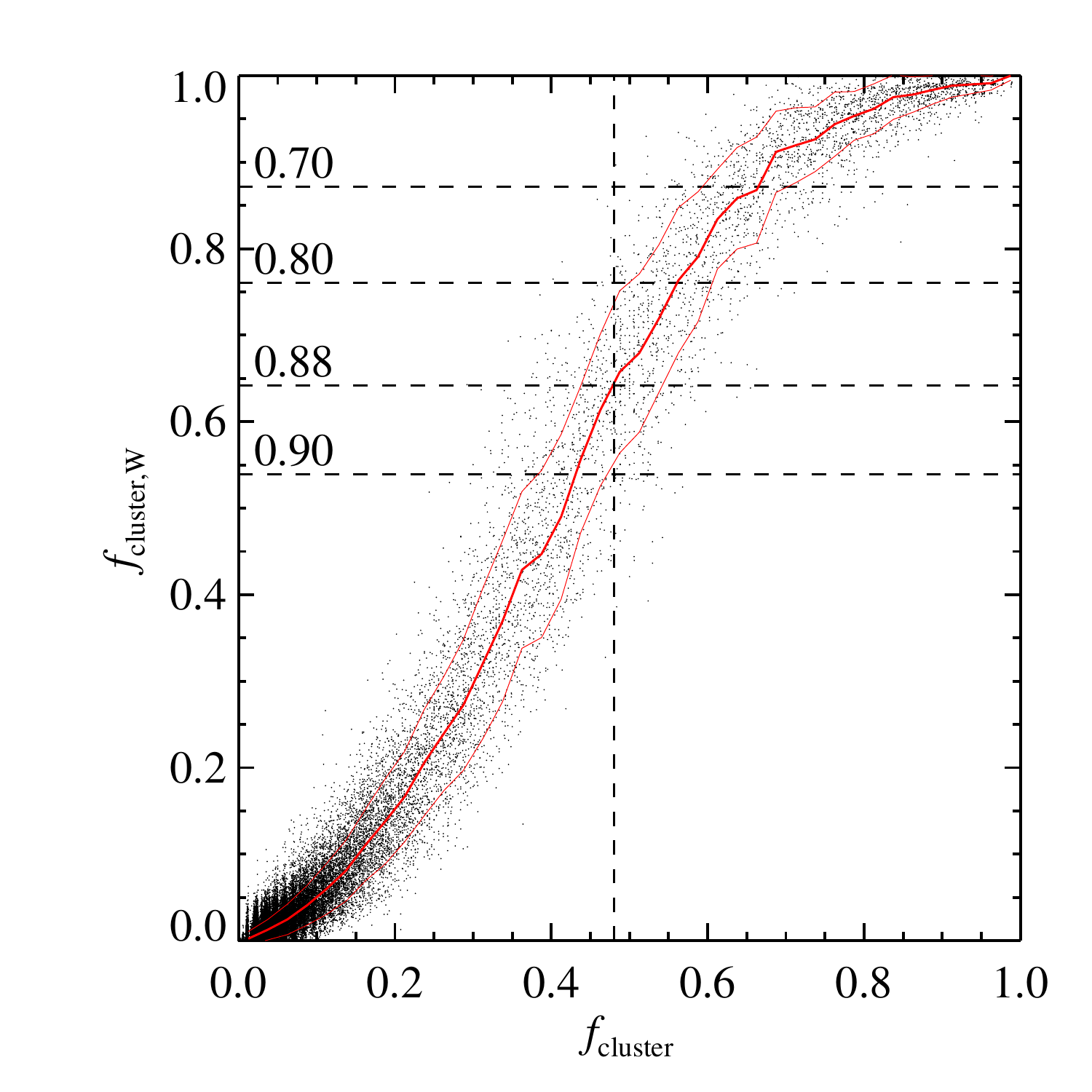}
\caption{A comparison between optimally-weighted \ClusterFracW\ scores and uniformly-weighted \ClusterFrac\ values, showing the impact of user weighting on individual object scores.  The red lines show the median trend and one standard deviation around the median.  Horizontal dashed lines denote the \ClusterFracW\ cutoffs corresponding to each of the printed Year 1 completeness fractions, while the vertical dashed line denotes the approximate \ClusterFrac\ value that corresponds to the optimal \ClusterFracW\ cutoff.}
\label{fig_orig-v-weight}
\end{figure}

We vary the input logistic function parameters to search for a set of values that produce the best possible weighted catalog.  We construct a grid of weighting systems by varying the values of the four free parameters: the slope and offset values for both the detection and non-detection weights.  For each set of parameters, we calculate a completeness versus contamination curve and its corresponding minimum distance to the corner of optimal completeness and contamination, $d_{\rm optimal}$.  We gradually extended the weighting grid to include an increasing range of logistic parameter values until we identified a minimum $d_{\rm optimal}$ value that was unsurpassed.  We defined the set of parameters that yielded this minimum $d_{\rm optimal}$ value as our optimal AP weighting system.

The range of completeness versus contamination curves is represented by the gray region in the top panel of Figure \ref{fig_compcont}.  We also plot the individual curve derived for the optimal weighting system and list its logistic function parameters in Table \ref{compconttable}.  The optimal weighting system provides a contamination fraction of 9.8\% at a completeness of 88.1\%.  When compared to the uniform weighting results, applying user weighting decreases the number of contaminants by 36\% (from $f_{\rm contamination}$ of 0.152 to 0.098 at completeness of 88.1\%), or alternately increases completeness from 84.6\% to 88.1\% (at $f_{\rm contamination}$ of 0.098).  While user weighting does not dramatically change the total number of cataloged clusters or the Year 1 completeness percentage, we are able to reduce the number of possible contaminants by a significant amount.

We compare original versus weighted \ClusterFrac\ values to illustrate the impact of user weighting on individual clusters.  Figure \ref{fig_orig-v-weight} shows that user weighting tends to increase the separation between high and low \ClusterFrac\ objects, providing better differentiation at moderate \ClusterFrac\ values that lie near the catalog cutoff.  To visualize how the choice of \ClusterFracW\ cutoff affects the output cluster catalog, we represent four different threshold values as horizontal lines in the figure, each labeled according to its corresponding Year 1 completeness fraction.  We also plot a vertical line in Figure \ref{fig_orig-v-weight} representing the approximate \ClusterFrac\ cutoff that best approximates the optimal \ClusterFracW\ threshold.

The user weighting applied here enhanced final AP catalog results by achieving small but quantifiable improvements through a combination of decreased contamination and increased completeness.  We note that we were fortunate to obtain a large number of classifications per image ($>$80) allowing us to account for variations in participant performance by averaging over a large number of classifications.  Many citizen science projects cannot afford to collect a similar number of per-image classifications because they need to distribute effort over a larger volume of data, or because the project is working on time-sensitive tasks that cannot wait for additional input to be collected.  In these cases, we expect that user weighting would play an essential role in obtaining high-quality results.

We utilize the \ClusterFracW\ values as defined by the optimal user weighting system throughout the rest of the paper.

\section{Catalog Completeness} \label{comp}

\begin{figure*}
\centering
\includegraphics[width=.3\textwidth]{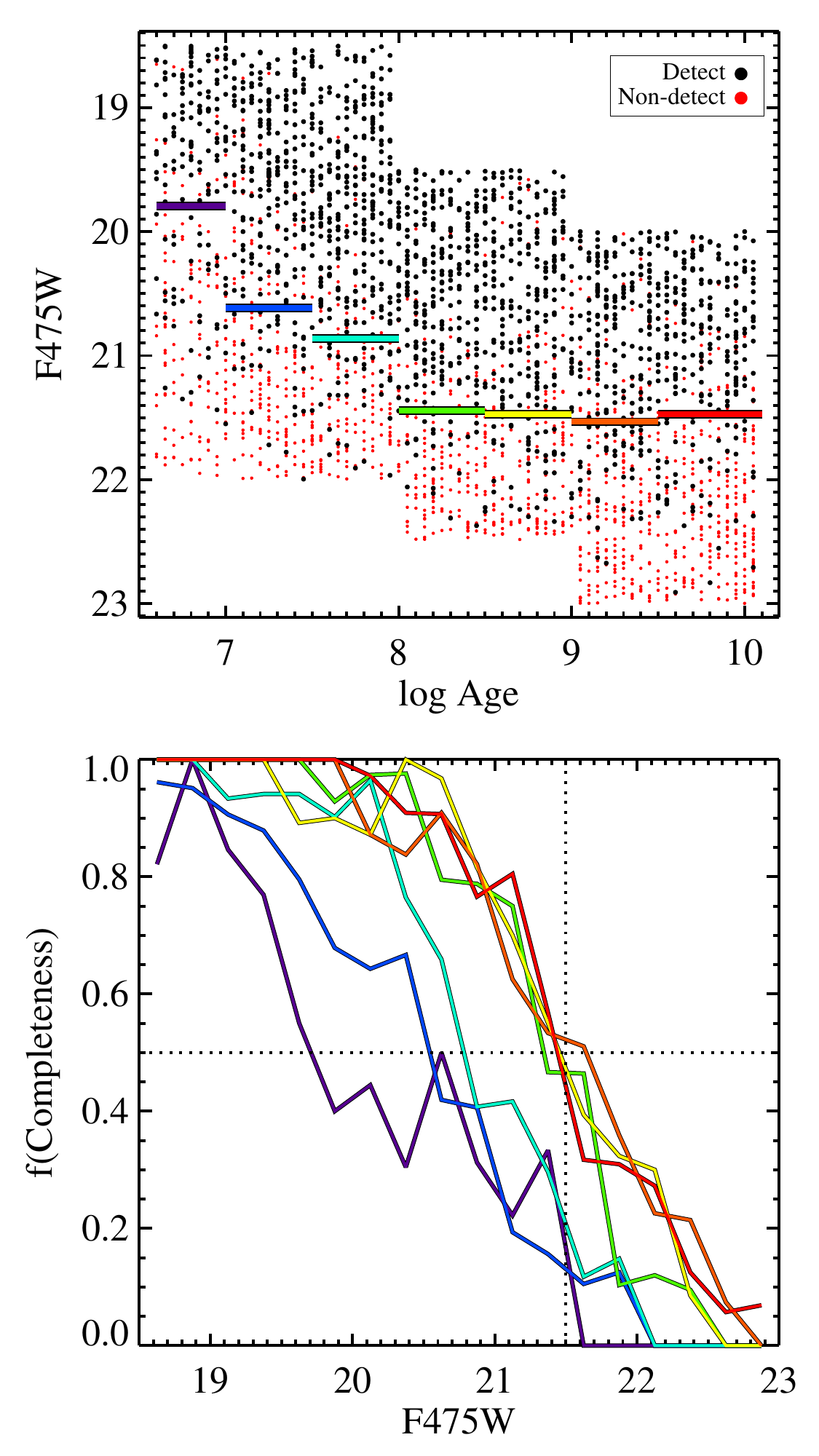}\hfill
\includegraphics[width=.3\textwidth]{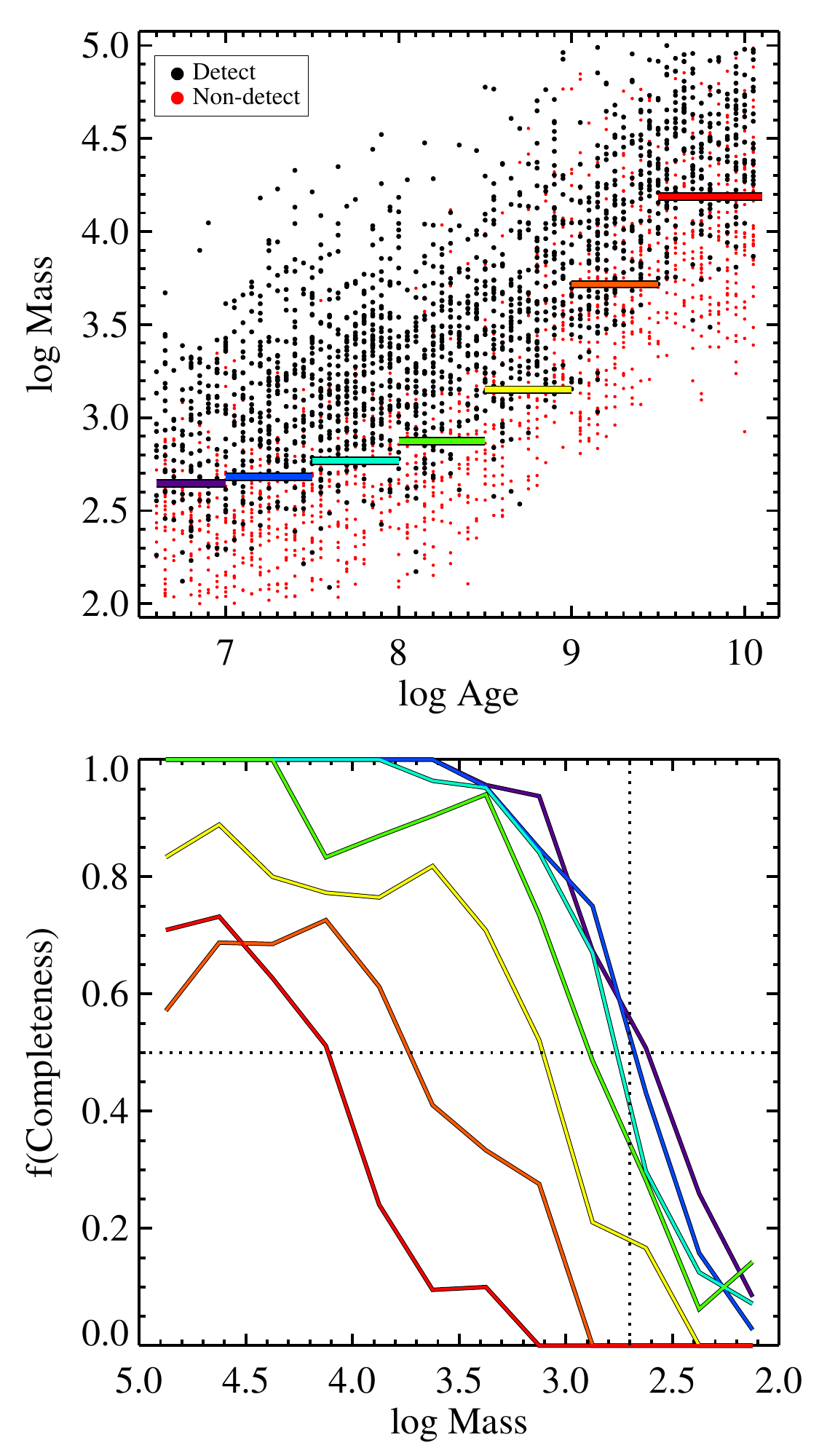}\hfill
\includegraphics[width=.3\textwidth]{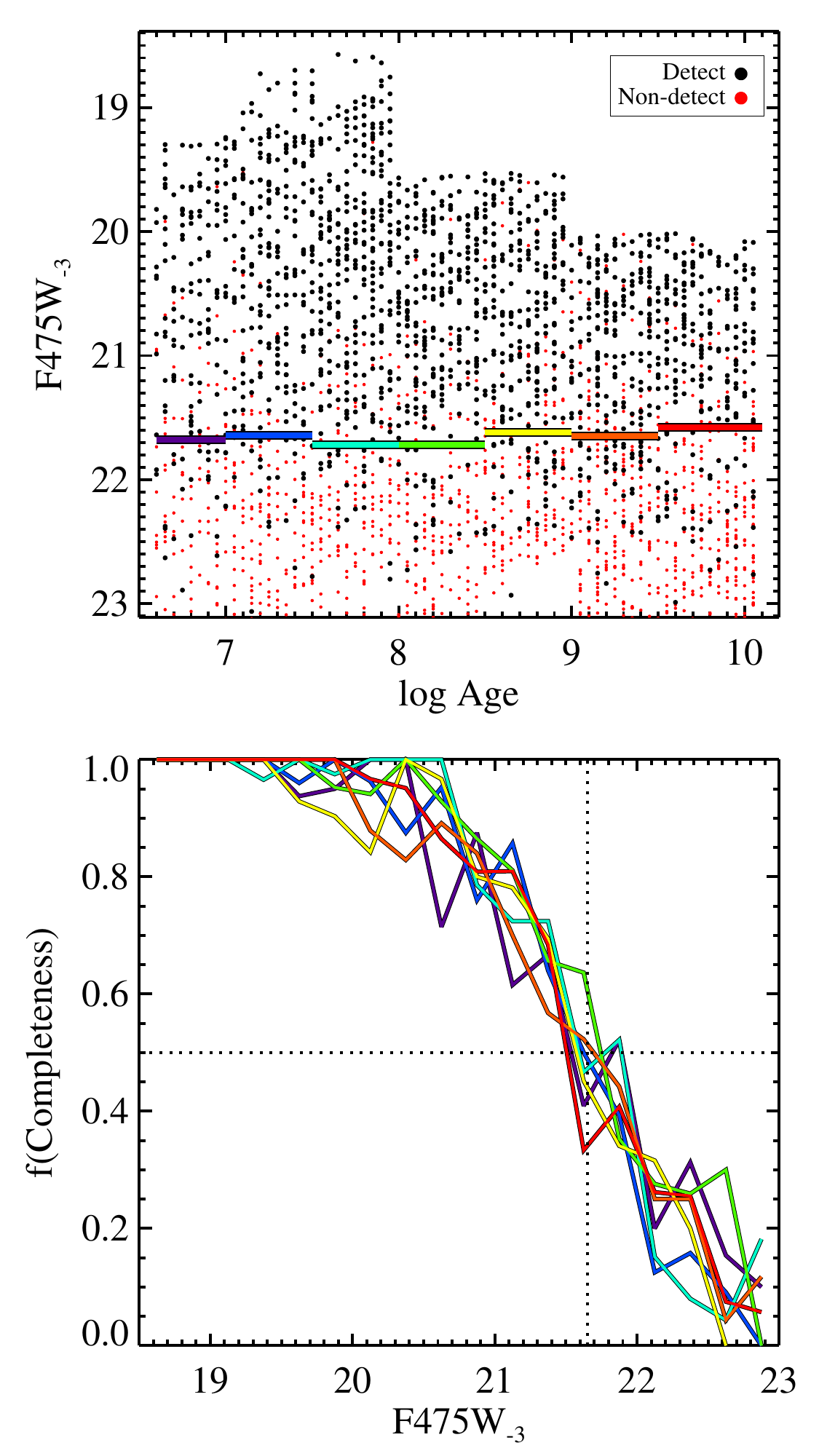}
\caption{Completeness results from synthetic cluster analysis.  Top Panels: Detection results for individual synthetic clusters (black = detected, red = not detected), as well as 50\% completeness limits calculated for each age bin.  Bottom Panels: Completeness functions for each age bin, color-coded to match their corresponding bin in the top plot.  Results as a function of F475W magnitude, mass, and F475W$_{-3}$ magnitude are presented in the left, center, and right columns, respectively.  F475W$_{-3}$ magnitudes represent the cluster flux that remains after subtracting the contribution of the cluster's three most luminous members.}
\label{figcomp}
\end{figure*}

We introduced our set of synthetic cluster tests in Section \ref{datainput}; here we present catalog completeness results derived from those tests, including how catalog completeness correlates with properties of the clusters and their surrounding fields.

The traditional method of characterizing the completeness of a cluster catalog is to identify the 50\% completeness limit as a function of cluster luminosity.  The two plots in the left column of Figure \ref{figcomp} show the behavior of the 50\% completeness limit in F475W as a function of cluster age for the full sample of synthetic clusters.  These plots show that while the synthetic results at log(Age/yr) $>$ 8.0 agree with a single, age-independent magnitude limit at F475W $\sim$ 21.5, there is an apparent age dependence at younger ages.  This result conflicts with the standard assumption that luminosity-based completeness limits for cluster catalogs are independent of age. 

To understand why we find brighter, non-constant completeness limits at ages $<$100 Myr, we examine our completeness results as a function of cluster mass, presented in the middle column of Figure \ref{figcomp}.  Under the assumption of an age-independent, constant luminosity completeness limit, we would expect a continuous increase in the 50\% mass completeness with increasing age due to stellar evolution driven fading of the cluster's stars.  In contrast to these expectations, we find a near-constant 50\% completeness limit for log(Age/yr) $<$ 8.0 of $\sim$500 \solmass.  It appears that catalog completeness correlates with cluster mass rather than luminosity at ages $<$100 Myr.

To explain the observed completeness behavior, it is important to note that nearly every synthetic cluster we tested with an age $<$100 Myr has a mass $<$3$\times$10$^3$\solmass.  The integrated light of young low mass clusters is dominated by a small number of bright stars.  This fact leads to large stochastic variations in the total integrated light for a sample of clusters with identical masses \citep[see][]{Fouesneau10, Beerman12, Popescu12}.  In addition, cluster identification in HST imaging of M31 relies greatly on the presence of an over-density of individually resolved stars, such that the number of observable stars might correlate better with a cluster's detection probability than its total luminosity in this low-mass regime.  In this case, the correlation between completeness and mass is explained by a strong correlation between mass and the number of bright, observable cluster members.

We conclude that there are two regimes for AP cluster catalog completeness: for ages $<$100 Myr, cluster detection is limited by the number of observable member stars; for ages $>$100 Myr, cluster detection is governed by the total cluster luminosity.  To bridge these regimes, we devise a single cluster metric that correlates strongly with the 50\% catalog completeness limit, independent of cluster age: F475W$_{-3}$, the F475W magnitude remaining after subtracting the flux contribution from the cluster's three brightest stars.  By excluding the contribution of the three brightest cluster stars, we significantly reduce the stochastic variation in cluster luminosity that imprinted an age-dependence into the completeness results.  We experimented with the number of stars to exclude and found that three provided the best correction.  The plots in the right column in Figure \ref{figcomp} show that our data are consistent with a single, age-independent 50\% completeness limit at a F475W$_{-3}$ magnitude of 21.65, where the new metric successfully unifies the two completeness regimes.

Using the results derived from the full set of synthetic cluster tests as a baseline, we can test whether completeness depends on two other important factors: the spatial profile of the cluster and the characteristics of the field surrounding the cluster.  At a fixed luminosity, we expect the completeness to worsen for larger, more extended clusters because the same total luminosity is spread over a larger area, causing the contrast between cluster and underlying background to decrease.  Likewise, the cluster to background contrast also decreases as the background surface brightness and stellar density increase, which also leads to a prediction of brighter cluster luminosity completeness limits.

\begin{figure}
\centering
\includegraphics[scale=0.7]{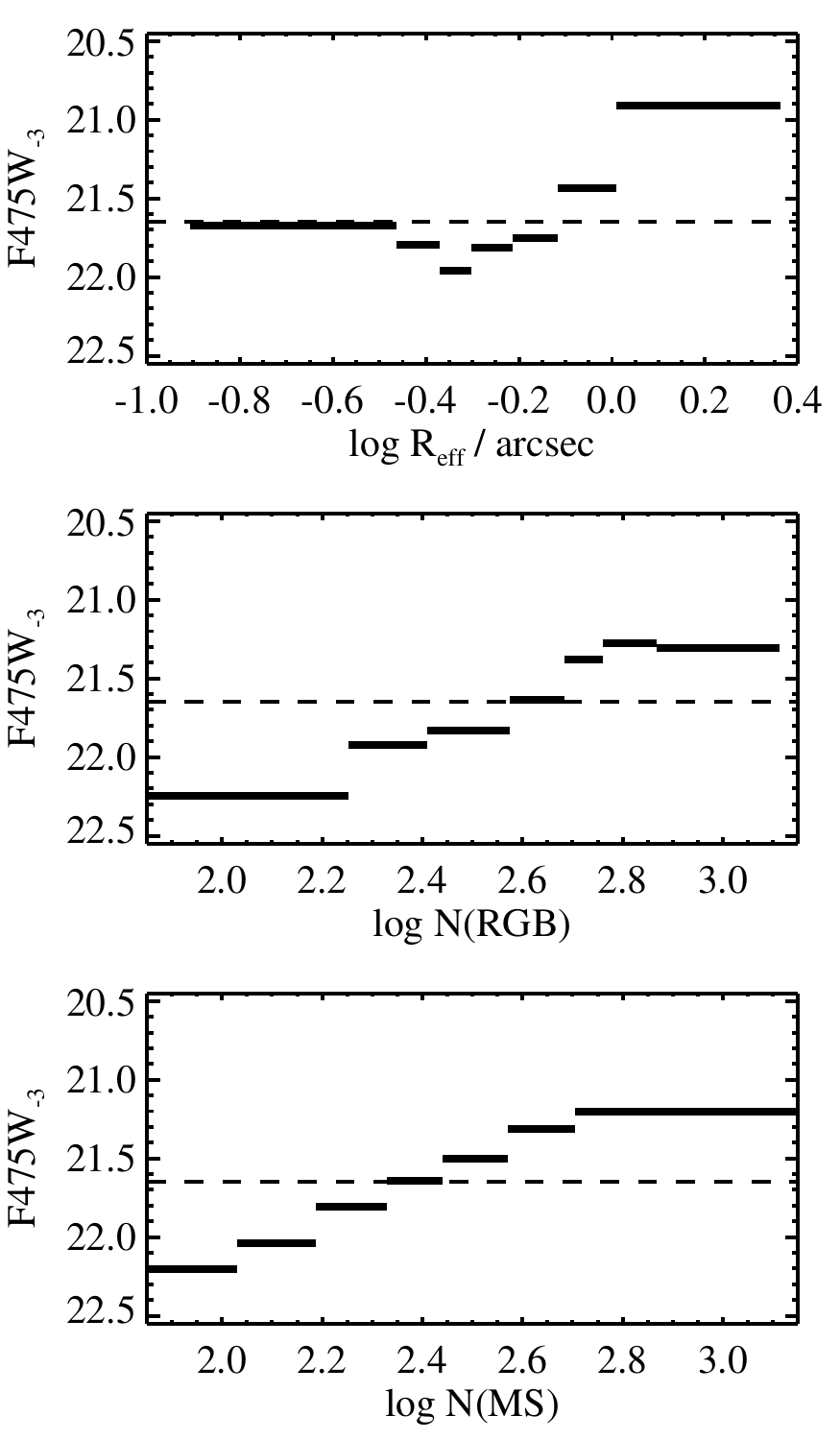}
\caption{Deviations from average completeness in F475W$_{-3}$ magnitude as a function of \reff, $N$(RGB), and $N$(MS) in the top, middle, and bottom panels, respectively.  The dashed line represents the baseline 50\% completeness level of F475W$_{-3}$ of 21.65.  Seven bins divide the synthetic cluster sample into equal parts ($N \sim 440$) as a function of each cluster variable.}
\label{figcompother}
\end{figure}

Contrary to the simple expectation, we observe non-monotonic behavior in the 50\% completeness limit as a function of a cluster's effective radius ($R_{\rm eff}$; equivalent to the half-light radius), as shown in the top panel of Figure \ref{figcompother}.  While the 50\% completeness limit reaches its faintest value at log($R_{\rm eff}$/arcsec) $\sim -0.35$, detection limits worsen as clusters become more extended, the limits also worsen as clusters become more compact.  Detection becomes more difficult at small \reff\ due to the inability for an image classifier to distinguish between single sources and a compact collection of individual stars.  This behavior was also seen by \citet{Silva-Villa11} in their study of extragalactic clusters.  The variation in \reff\ can cause F475W$_{-3}$ 50\% completeness limits to deviate by $>$0.5 mag from the baseline level, translating to a mass completeness difference of up to 0.15-0.2 dex.

Background stellar density, on the other hand, shows the expected behavior that higher stellar density makes cluster detection more difficult.  We quantify local stellar density by counting the number of red giant branch (RGB) and main sequence (MS) stars that lie within the search images ($36.25\times25$ arcsec) that host each synthetic cluster.  These counts are based on the survey-wide 6-band GST photometric catalogs \citep{Williams14}, where we define RGB stars as sources with F110W-F160W $>$ 0.5 and F160W $<$ 21.0, and MS stars as sources with F475W-F814W $<$ 1.0 and F814W $<$ 25.0.  The middle and bottom panels of Figure \ref{figcompother} show 50\% catalog completeness limits as a function of $N$(RGB) and $N$(MS).  As a function of  $N$(RGB) and $N$(MS), the F475W$_{-3}$ 50\% completeness limits vary by $\sim$0.5 mag, translating to a mass completeness difference of up to 0.15-0.2 dex.  This dependency affects the detection of PHAT clusters in the inner disk and bulge, as well as those within dense star forming regions -- especially those located within the $\sim$10 kpc ring.

To supplement the above description of overall, sample-wide completeness behavior, we present a table of object-by-object completeness test results in Appendix \ref{altcat}.  These results allow catalog users to calculate completeness functions for specific spatial regions or over a custom range in parameter space.

\section{Results} \label{results}

\subsection{AP Cluster Catalog} \label{catdata}

\begin{figure*}
\centering
\includegraphics[scale=1.0]{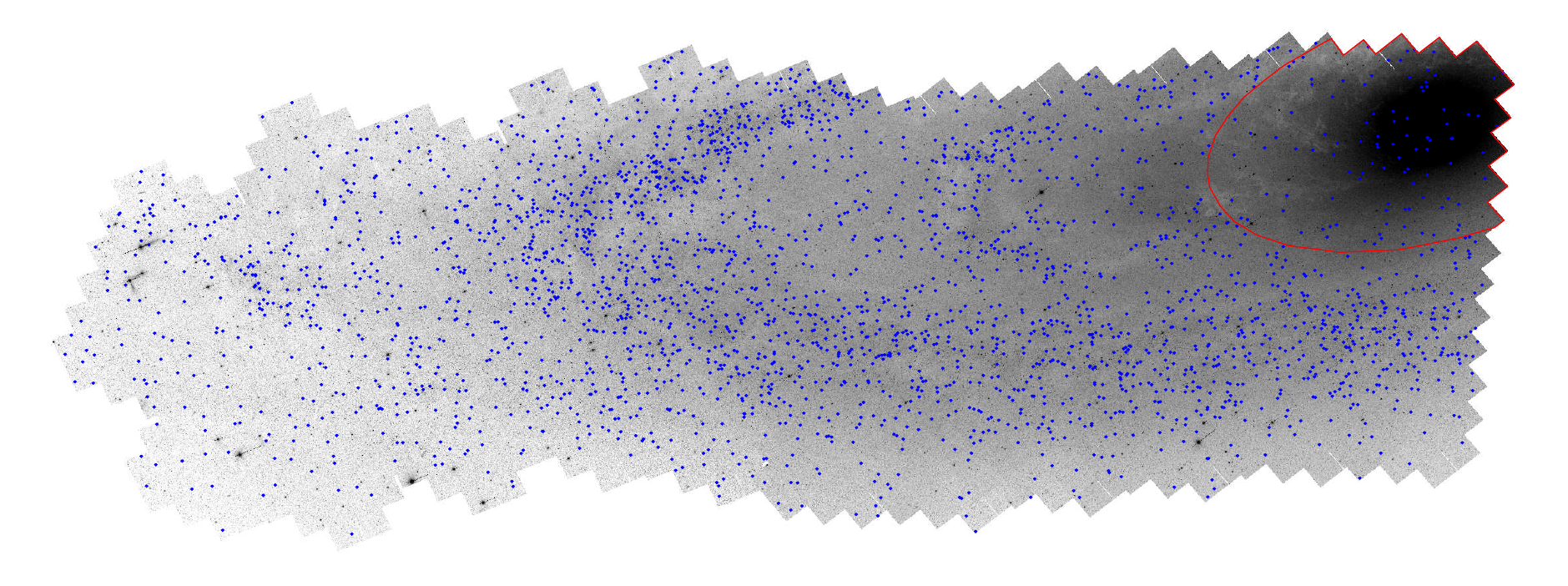}
\caption{Spatial distribution of AP cluster catalog overlaid on the PHAT survey-wide F475W image.  The red ellipse denotes the bulge region within which the catalog completeness and object recovery vary significantly from the rest of the survey.}
\label{figclst}
\end{figure*}

We apply the catalog construction techniques and user weighting methodology presented in Section \ref{catintro} to define an AP cluster catalog, adopting final selection criteria of:
\begin{equation}
f_{\rm cluster, W} > 0.6416
\text{ AND }
f_{\rm galaxy} < 0.3.
\end{equation}
These criteria yield a sample of 2,714 clusters.  We add two additional sets of clusters to these initial selections.  First, we add 35 clusters to the sample that are located in the bulge-dominated region within $\sim$3 kpc of M31's center, as defined by an ellipse with a center of (10.684575, +41.268972), semi-major axis of 815 arcsec, semi-minor axis of 410 arcsec, position angle of 45 degrees, and bounded by the PHAT footprint.  These objects are primarily globular clusters that were identified and confirmed by previous surveys.  These objects suffer from systematically low \ClusterFracW\ scores due to their atypical appearance (compact and smooth with few individually resolved stars), high-surface brightness backgrounds, and suboptimal search image scalings.  We decided that the most straightforward solution to correct for these missed objects was to include all previously confirmed clusters (high-quality Year 1 or RBC flag of 1) that lie within the defined region and evaluate all candidate or possible objects.  We confirmed by-eye that each of the previously confirmed objects has an appearance consistent with that of a cluster, and confirmed two additional candidate objects.  Second, we include 4 additional expert cluster identifications from the B03 tooth images that were not included in the AP search due to delayed data availability. 

The final AP cluster catalog includes 2,753 objects.  Figure \ref{figclst} shows the positions of the clusters within the PHAT survey footprint.  Andromeda's $\sim$10 kpc star-forming ring is a prominent feature visible in the cluster's spatial distribution.  We assign identifiers in descending order of their maximum per-image, uniformly-weighted \ViewFrac\ score.  Positions, aperture sizes, and other relevant catalog metadata are presented in Table \ref{ccat}.

All AP candidates with \ViewFrac\ $\ge$ 0.1 that are not included in the cluster catalog (or galaxy catalog; see Appendix \ref{bckgal}) are listed in an ancillary table in Appendix \ref{altcat}.  We include information on these additional candidates to allow other workers the opportunity to make different choices concerning catalog selection.

\subsection{Comparison to Previous Cluster Catalogs} \label{catcompare}

We cross-match our AP cluster catalog with a selection of previously published catalogs: the Year 1 catalog, the RBC, \citet{Caldwell09}, and the HKC.  We include alternate identifiers for previously classified objects in Table \ref{ccat} and summarize the high degree of consistency between the AP catalog and previous results below.

By design, the AP catalog bears a strong resemblance to the Year 1 catalog.  When we consider the portion of the AP catalog that lies within the Year 1 imaging footprint (including B01, differing slightly from the Section \ref{catintro} analysis), we count 688 clusters, which is a 14.5\% increase over the 601 object Year 1 catalog.  The agreement between the two samples is good: the AP catalog includes 88.5\% (532/601) of the good Year 1 clusters, and 91\% of the AP cluster catalog were previously classified as high-quality or possible Year 1 objects.  The AP catalog includes 39 Year 1 catalog rejections and 22 objects not classified in the Year 1 search.  While the majority of object-by-object classification differences are caused by clusters with \ClusterFracW\ scores that lie near the catalog cutoff, we discuss a number of meaningful systematic differences between the two catalogs in Section \ref{yr1diff}.

Comparison of the AP catalog to the RBC and the \citet{Caldwell09} catalog provides an opportunity to cross-reference with commonly cited sources, linking our present work to a wealth of ancillary information about these clusters, including a great deal of follow-up spectroscopy.  These ground-based catalogs do not reach the faint objects accessible to the PHAT imaging, therefore the following comparison mostly consists of verifying or discarding previously unconfirmed candidates that lie at the middle or bright end of the AP sample.

Cross-matching the AP catalog with the RBC, we find that 260 previously confirmed, candidate, or controversial clusters (RBC flag = 1, 2, or 3) match to AP clusters, while 42 AP classifications conflict with those from the RBC (40 AP clusters are not RBC clusters, 2 RBC clusters are not AP clusters), and 18 additional RBC candidate or controversial classifications were rejected.  PHAT's high spatial resolution imaging is often used as a definitive tool for classifying objects, so we defer to AP classifications for these conflicting cases.  We also find good agreement between the AP and the \citet{Caldwell09} catalogs.  Only 18 conflicts arise from the Caldwell catalog (8 AP clusters are not Caldwell clusters, 10 Caldwell clusters are not AP clusters), while 232 cluster classifications are common to both the Caldwell and the AP catalogs.

Finally, we compare the AP catalog with the HKC catalog compilation.  These clusters represent the low-mass additions to previous ground-based catalogs provided by early targeted HST observations, and therefore include many objects that lie at or near completeness limits.  As such, a direct comparison shows 156 previously identified clusters confirmed by our AP classifications, while 57 are not confirmed.  This 73\% yield is nearly identical to the 72\% yield we found for the Year 1 catalog during a similar comparison exercise.  A vast majority of HKC objects that were not confirmed by the AP catalog are borderline, marginal candidates where there is a subjective difference in opinion between the HKC authors and the consensus judgement of AP volunteers; rejected objects are distributed uniformly in \ClusterFracW, such that half of these rejected objects have \ClusterFracW\ $>$ 0.3.

Overall, the comparison between the AP catalog and previous non-PHAT M31 cluster catalogs shows good agreement with few conflicting classifications.  A total of 733 unique, previously cataloged objects (both cluster and non-cluster classifications) match to AP candidates; 468 were previous (confirmed) cluster classifications, of which 404 were confirmed by the AP catalog. Within the PHAT survey footprint, we have increased the sample of confirmed clusters by a factor of $\sim$6 (from 468 to 2753).
The HST-based AP catalog provides improvement in terms of catalog completeness and quality, and builds upon the firm foundation laid by these previous works.  Commentary on individual classification differences can be found Appendix \ref{knownc}.

\subsection{Integrated Photometry} \label{phot}

We perform integrated aperture photometry for each of the AP catalog entries.  Our photometry procedures are described in Paper I; we summarize the main ideas here, but refer the reader to that paper for additional details.  We use the mean center and median radius of an object's merged classifications to define the position and radius ($R_{\rm ap}$) of the photometric aperture.  The sky background is calculated within an annulus ten times the size of the photometric aperture, extending from 1.2 ${R_{\rm ap}}$ to $\sim$3.4 ${R_{\rm ap}}$.  Photometric uncertainties are dominated by uncertainties in the sky background determination; this source of uncertainty is often ignored in extragalactic cluster photometry.  Identical apertures (constant angular size) are employed across all six PHAT images.  Aperture magnitudes for significant detections (S/N $\ge$ 3 with respect to the variation in the sky background) are listed for each photometric passband in Table \ref{ccat}; 3$\sigma$ upper limits are provided for non-detections, and blank entries denote incomplete image coverage.

We obtain photometric \reff\ estimates by interpolating radial flux profiles.  These values are then used to derive aperture corrections, which help account for cluster light that falls outside of the photometric aperture.  We compare synthetic cluster input luminosities to measured magnitudes and find that this effect causes losses on the order of 0.1--0.3 mag.  Corrections assume a \citet{King62} profile with a concentration ($c = R_{\rm tidal}/R_{\rm core}$) of 7, scaled to match the cluster's photometrically determined F475W \reff, then extrapolated to radii beyond ${R_{\rm ap}}$ to obtain a magnitude correction, $m_{\rm ApCor}$.  Aperture corrections can be applied to raw aperture magnitudes to obtain total magnitude\footnote{$m_{\rm Total} = m_{\rm ap} + m_{\rm ApCor}$} estimates.  These estimates accurately recover the photometry of synthetic clusters with no bias at brighter magnitudes ($<$19) and $<$0.2 mag bias for fainter clusters (see Sec.~4.2 in Paper I).

We summarize the photometric measurements in Table \ref{photdat} where we tabulate the number of detections in each band, as well as the number of objects with detections spanning various combinations of photometric bands.

We found that accounting for the presence of image artifacts was critical to obtain accurate cluster photometry in the F275W, F336W, and F110W filter bandpasses.  Images in these three wavelengths proved problematic due to their small number of repeat observations and minimal spatial overlap between neighboring images, hindering typical artifact rejection techniques that require three or more exposures.  Interpolating over pipeline-rejected pixels in the F110W images was relatively straightforward, however detecting and rejecting UV cosmic ray image artifacts was more difficult.  We conservatively identify F275W and F336W artifacts by flagging bright, single-passband objects by comparing flux ratios of F275W, F336W, and F475W images.  This method allows us to reject hundreds of artifacts that tend to bias measurements to brighter magnitudes, however we caution that some uncorrected artifacts may continue to affect our UV photometry.

\begin{figure}
\centering
\includegraphics[scale=0.65]{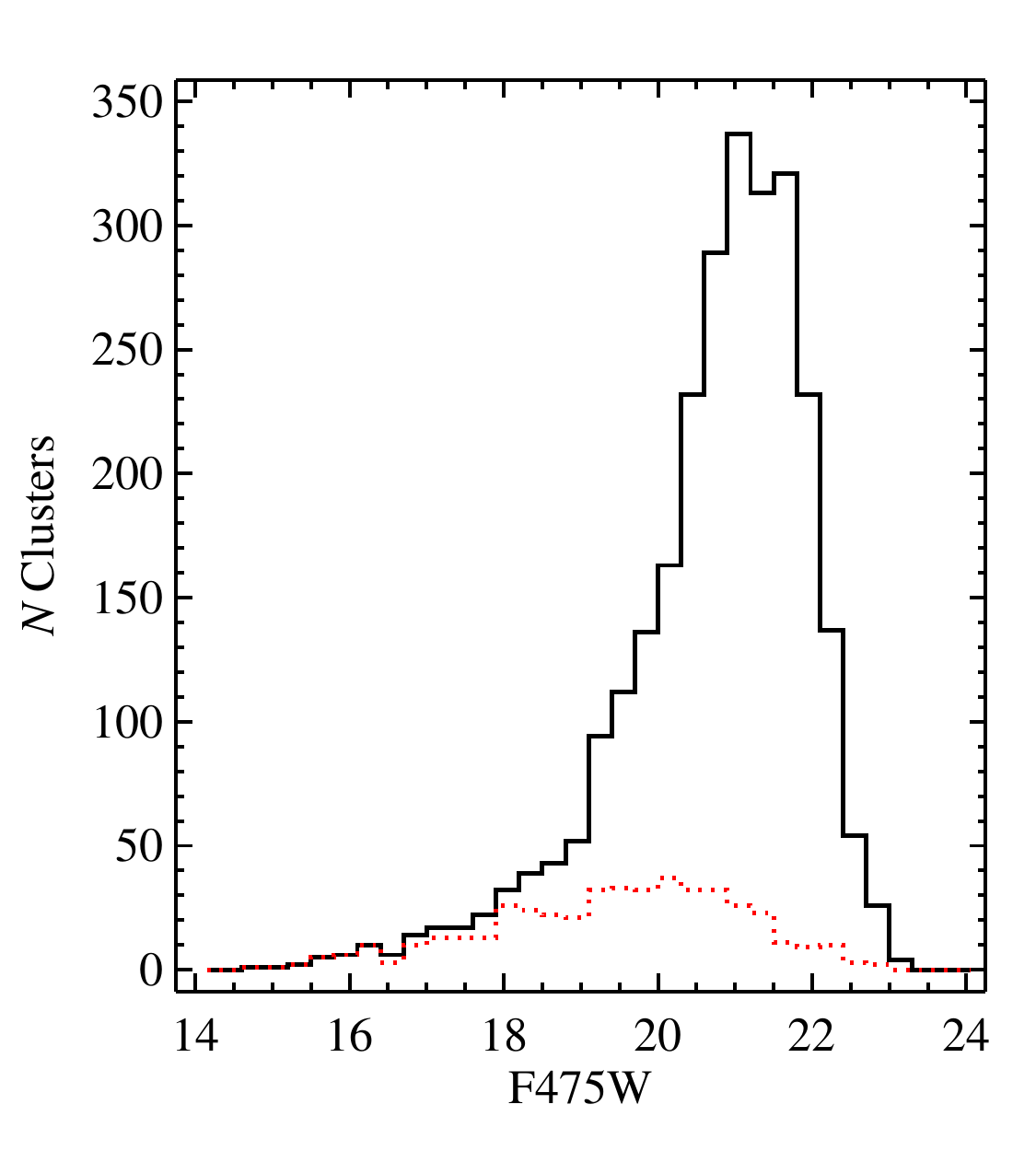}
\caption{Histogram of F475W integrated magnitudes for 2,717 AP clusters (out of 2,753 total).  The red dotted histogram represents the distribution of luminosities for 401 previously known clusters confirmed by the AP catalog (out of 404 total) that lie within the PHAT footprint, showing the vast improvement in cluster identification provided by the PHAT data.}
\label{fighistmag}
\end{figure}

\begin{figure}
\centering
\includegraphics[scale=0.65]{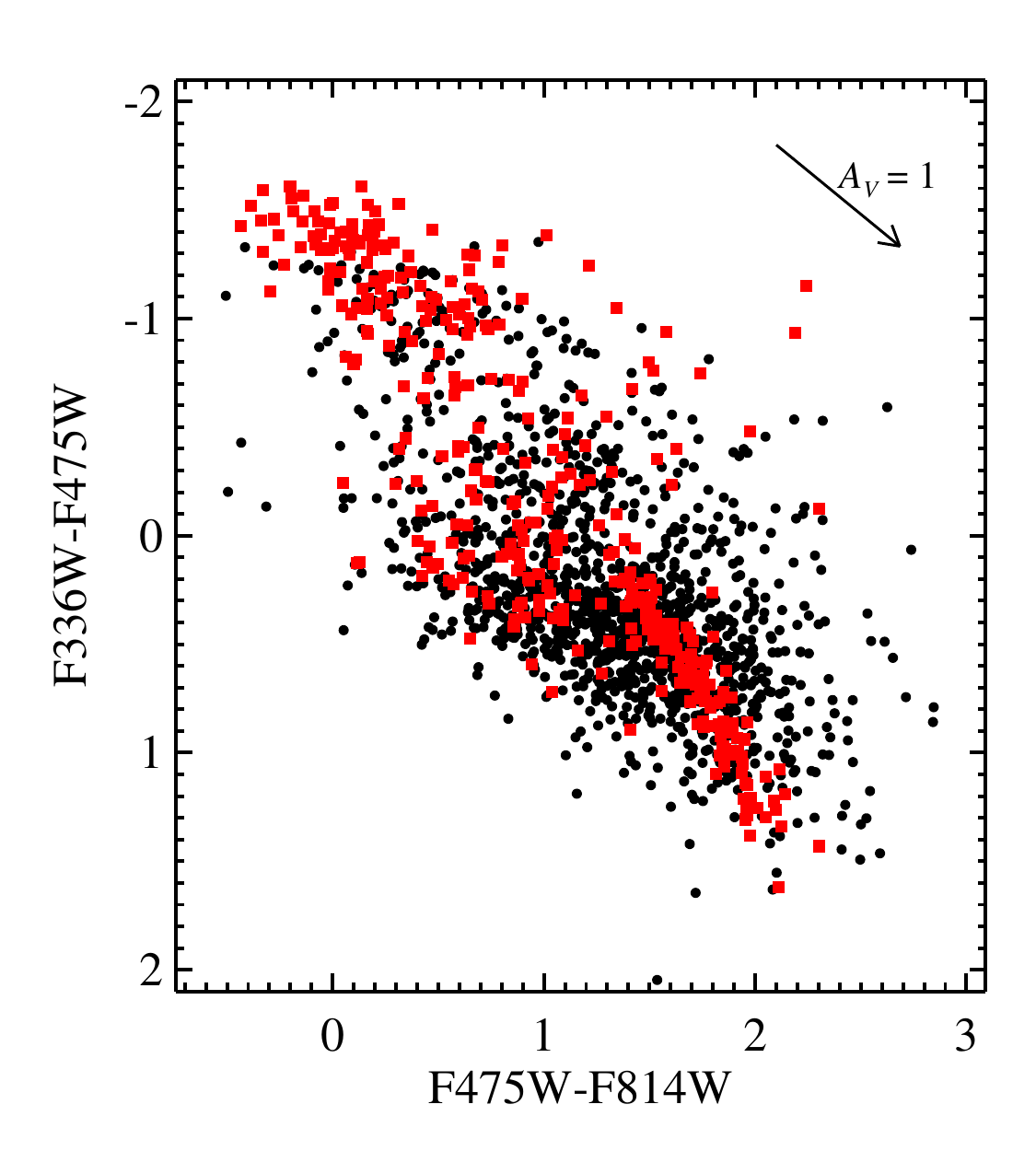}
\caption{Color-color diagram of 1,701 clusters with F336W, F475W, and F814W photometric detections.  The 378 clusters with F475W $<$ 19.5 are distinguished as red squares.  The color-color sequence of luminous globular clusters (see text in Sec.~\ref{phot}) is prominent in the sample of bright clusters.}
\label{figcolorcolor}
\end{figure}

\begin{figure}
\centering
\includegraphics[scale=0.7]{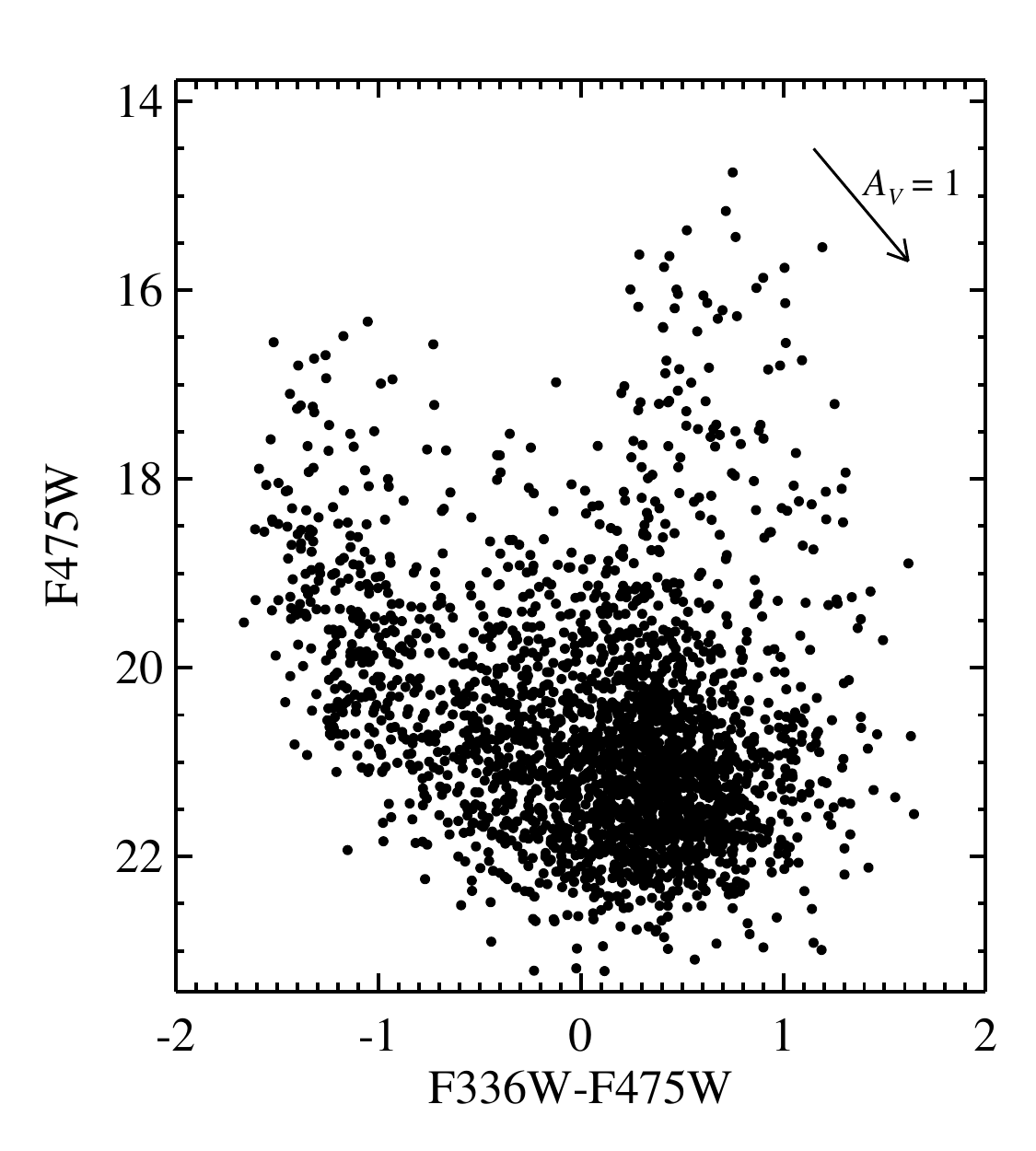}
\caption{Color-magnitude diagram of 2,464 clusters with F336W and F475W photometric detections.}
\label{figcolormag}
\end{figure}

In Figure \ref{fighistmag}, we compare the distributions of F475W magnitudes for previously known (and confirmed) clusters in the PHAT footprint and the new AP catalog.  The factor of $\sim$6 increase in the number of clusters shows the staggering improvement made possible by PHAT's high spatial resolution imaging.  The ability to differentiate between single bright stars and compact clusters allows us to include fainter, less massive clusters in the AP catalog.  Ground-based imaging limited previous efforts to clusters brighter than F475W $\sim$ 19.5, and while the HKC pushed that limit faint-ward, the amount of HST imaging available to those authors was significantly less than what is now available through PHAT.

Figures \ref{figcolorcolor} and \ref{figcolormag} show the color-color and color-magnitude distributions of AP clusters, providing a glimpse into the age composition of the catalog.  While the clusters span a wide range of colors that reflect a diversity in ages, a dominant portion of the catalog lies within the following color and magnitude range: 20$<$F475W$<$22, 0$<$F336W$-$F475W$<$1, and 1$<$F475W$-$F814W$<$2.  The specified region of color and magnitude parameter space points to a dominant population of $\sim$10$^3$ \solmass, $\sim$200--400 Myr old clusters that dominate the catalog by number, consistent with the age distribution found for the Year 1 sample \citep{Fouesneau14}.
This population dominates the cluster catalog because it represents a relatively large linear age range (leading to large number of clusters for a near constant formation history) where catalog completeness is still relatively high (50\% complete to $\sim$1,000 \solmass\ at 300 Myr).
We note that the large color dispersion shown in Figure \ref{figcolorcolor} agrees with predictions of stochastically-sampled cluster models \citep[see Fig. 4 in][]{Fouesneau14}.  In addition, the vertical sequence spanning 15$<$F475W$<$19 with a color range of 0.2$<$F336W$-$F475W$<$1.3 in Figure \ref{figcolormag} represents the old (10-14 Gyr), massive ($>10^5$ \solmass) globular clusters.  These massive, luminous systems also form a well-defined sequence of bright clusters in Figure \ref{figcolorcolor}, running from (F475W$-$F814W, F336W$-$F475W) coordinates of approximately (1.4,0.2) to (2.1,1.3), corresponding to a metallicity sequence running from -2.5$<$[Fe/H]$<$0.0 for these systems.

\begin{figure*}
\centering
\includegraphics[scale=0.40]{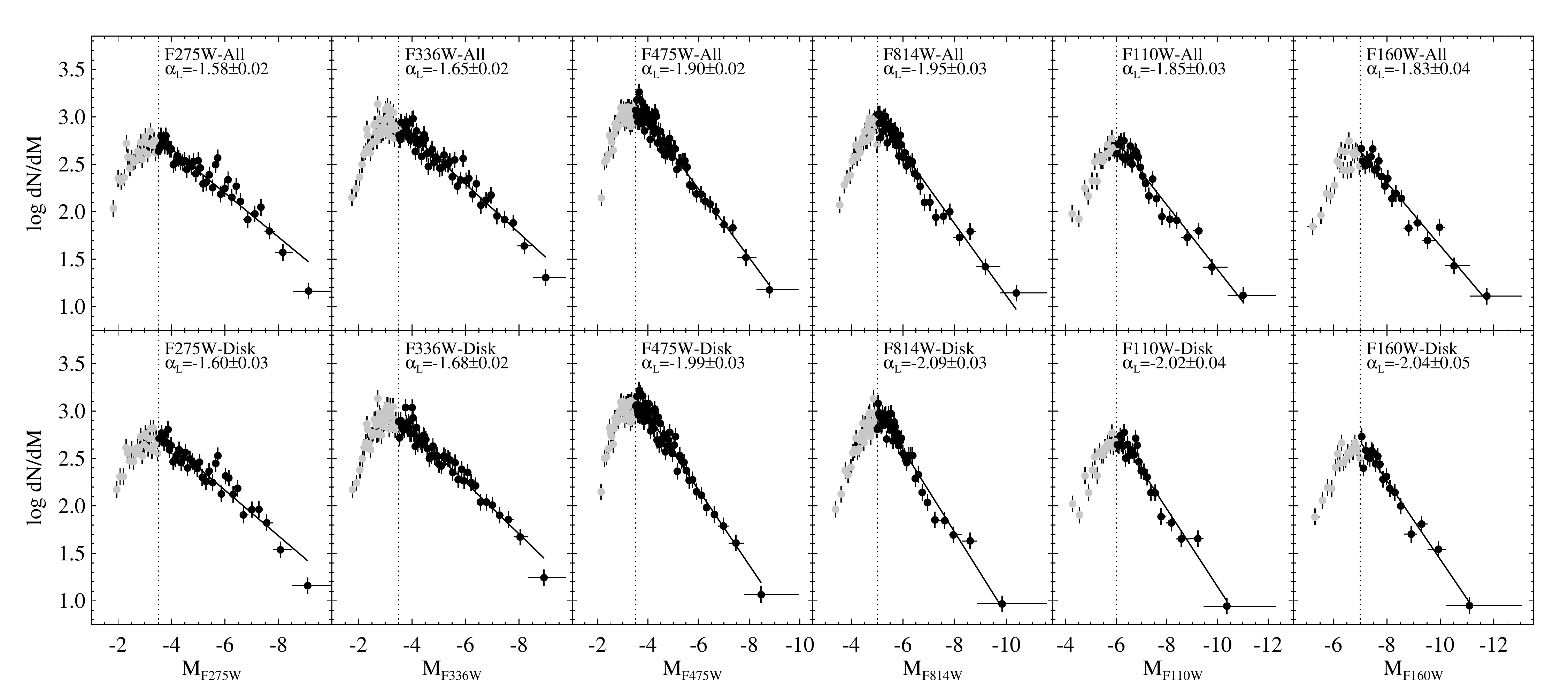}
\caption{Luminosity functions across six PHAT passbands.  Each plotted point represents an equal number of clusters ($N=25$) and linear fits are made to points above the adopted completeness limit (dotted line).  Top panels show results for the full AP cluster sample, while the bottom plots constrain the sample to those objects that lie within the disk, outside the inner bulge.}
\label{figlumfun}
\end{figure*}

We fit luminosity functions to the cluster photometry using a simple power law ($N \propto L^{-\alpha_{L}}$); we plot the results in Figure \ref{figlumfun}.  Notably, when we remove objects that lie within the previously defined bulge region (see Section \ref{catdata}), we find that luminosity functions steepen significantly.  As we argued in Paper I, old massive globular clusters dominate the bright end of the luminosity function; removing these objects, which reside primarily in the galaxy's bulge, allows us to examine the luminosity function behavior of younger ($\lesssim$3 Gyr) cluster populations.  The observed population-dependent variations in luminosity function indices affirm that factors such as the underlying cluster formation history, the intrinsic cluster mass function, and the stochastic conversion from mass to luminosity for less massive clusters all play a role in determining the overall distribution of cluster luminosities.  Untangling these various effects for the PHAT sample is possible through direct age and mass determinations of the individual clusters; we will perform this analysis as part of future work (Beerman et al., in prep.).

\section{Discussion} \label{discuss}

\subsection{Comparing the M31 Cluster Catalog to Extragalactic and Galactic Samples}

\begin{figure*}
\centering
\includegraphics[scale=0.75]{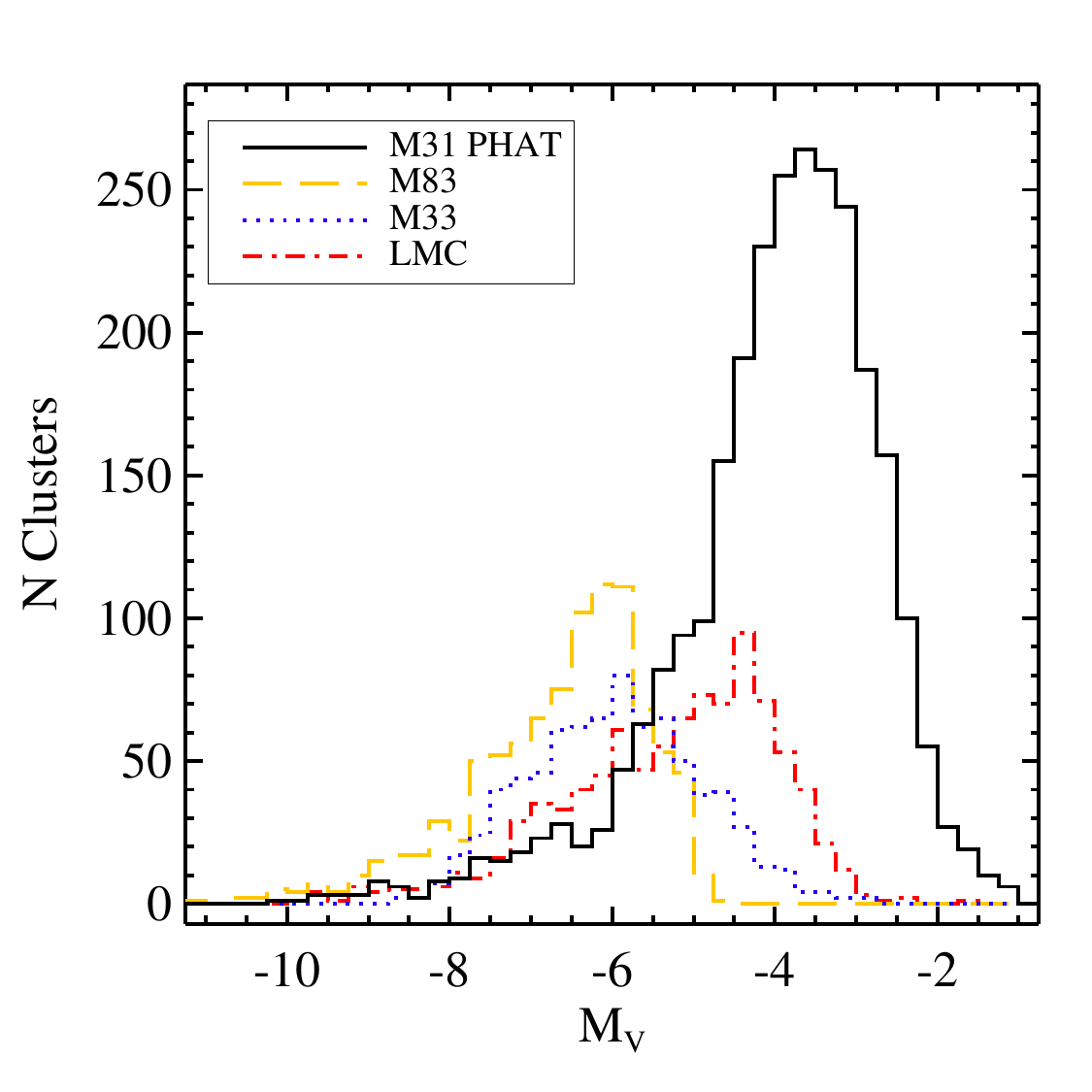}
\includegraphics[scale=0.75]{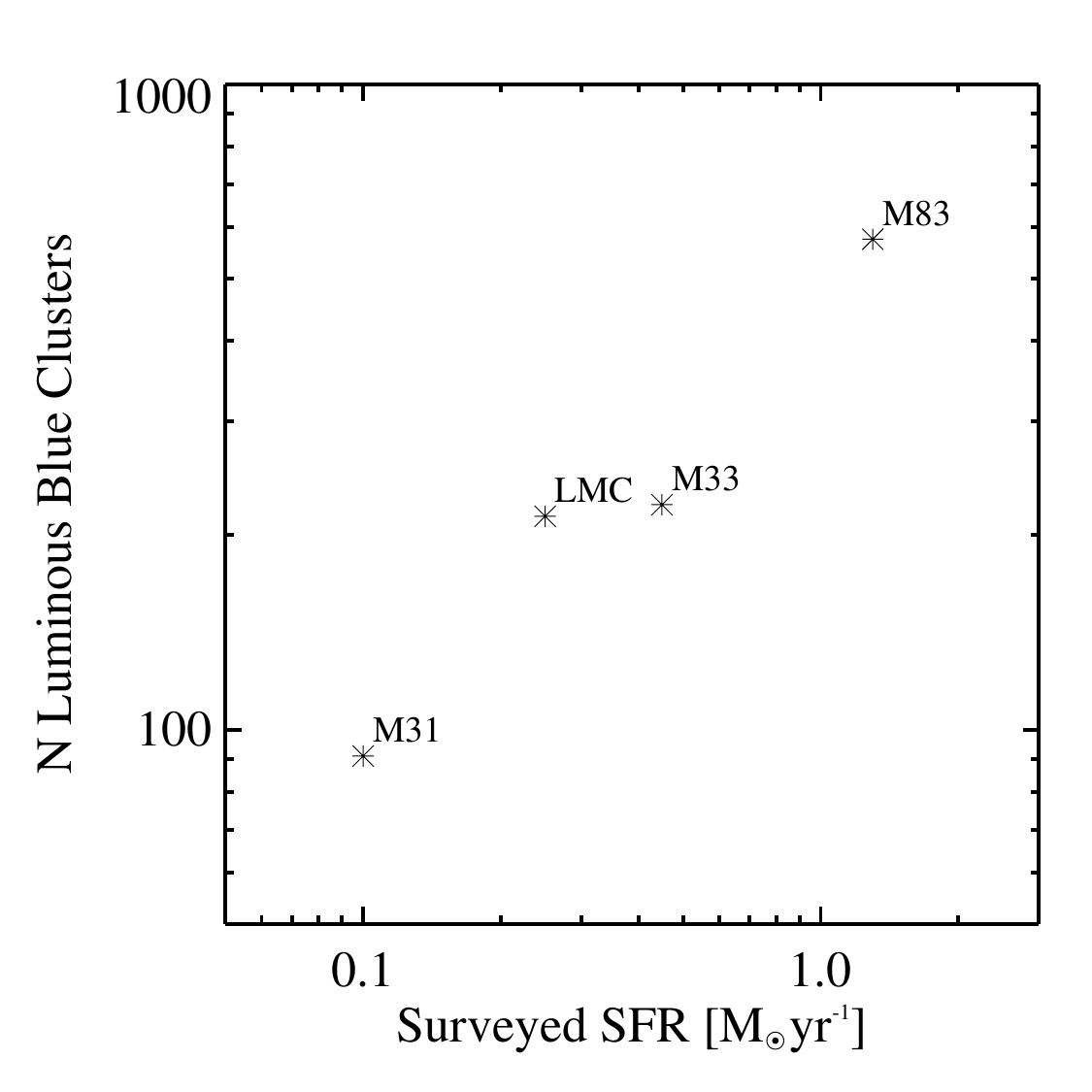}
\caption{Left: A comparison of luminosity functions for four extragalactic samples of star clusters: M31 (this work), M83 \citep{Bastian12}, M33 \citep{SanRoman10}, and the LMC \citep{Hunter03, Popescu12}. This plot shows the relative difference in detection limits and total number of clusters for each catalog.  Right: A comparison of the number of luminous blue clusters ($M_{V} < -6$, $B-V < 0.5$) in each galaxy sample as a function of SFR for the region that was surveyed in each galaxy.  This plot shows that the AP sample includes fewer luminous blue clusters due to the relatively low SFR found in the PHAT survey footprint.}
\label{figgalcompare}
\end{figure*}

To place the PHAT catalog of M31 star clusters into context, first we compare the luminosity distribution of our sample to those from three nearby star-forming galaxies: M83 \citep{Bastian12}, M33 \citep{SanRoman10}, and the LMC \citep{Hunter03, Popescu12}. We choose these three galaxies because they are well-known extragalactic cluster targets that have publicly-available cluster catalogs; we compare our sample to the much more heterogenous Milky Way catalog in the next subsection.  We compare the luminosity distributions of each sample in the left panel of Figure \ref{figgalcompare}, where we convert from PHAT's F475W to $V$-band apparent magnitudes using the following empirical relation:
\begin{equation}
m_{V} = m_{\rm F475W} - 0.363( m_{\rm F475W} - m_{\rm F814W}) - 0.111.
\end{equation}

Completeness limits for the three samples scale as a function of distance: M83 has the brightest completeness limit at $M_V \sim -6$, followed by M33's limit at $M_V \sim -5.5$, and the LMC's limit at $M_V \sim -4.5$.  The M31 detection limit of $M_V \sim -3.5$ leads to the inclusion of many more clusters, particularly those of moderate mass and intermediate ages: 10$^3$--10$^4$ \solmass\ between 100 Myr and 1-3 Gyr \citep{Fouesneau14}.  As a result, the PHAT sample contains $\sim$3 times more clusters than any of the other extragalactic samples compared here.

At bright magnitudes (${M_V} < -6$) where all four cluster samples are complete, we can compare the number of luminous blue ($B-V < 0.5$, or equivalently F475W$-$F814W $<$ 1.1) clusters in each sample.  This provides a first-order comparison of the young cluster populations captured by the catalogs of our set of comparison galaxies.  We show in the right panel of Figure \ref{figgalcompare} that the M83 catalog includes the largest number of blue clusters, followed in order by M33, the LMC, and M31.  Differences in the star formation rate (SFR) for the galaxy regions surveyed explain the differences observed in blue cluster populations.  The cluster sample from the starburst galaxy M83 corresponds to a SFR of 1.3 \solmassyr\ \citep[coverage fraction of 2/5 applied to galaxy-wide SFR of 3.3 \solmassyr;][]{Boissier05}, while M33's SFR is 0.45 \solmassyr\ \citep{Verley09} and the LMC's SFR is 0.25 \solmassyr\ \citep{Whitney08}.  Within the PHAT footprint, the current SFR is much lower at $\sim$0.1 \solmassyr\ \citep[coverage fraction of 1/3 applied to galaxy-wide SFR of 0.25 \solmassyr;][]{Ford13}.  Larger SFRs correlate with larger numbers of blue clusters; the relatively low number of luminous blue clusters in the AP catalog are a consequence of M31's relative quiescent SFR.

Next, we compare our PHAT cluster catalog to the sample of known Galactic clusters.  Without question, observations of Milky Way star clusters provide rich, detailed datasets for individual clusters and their member stars that cannot be matched in an extragalactic setting.  The ability to measure star-by-star proper motions, detect and resolve stars down to the hydrogen burning limit, and efficiently obtain detailed abundance information through spectroscopy of individual members are all major advantages of studying clusters in the Galaxy.  However, it is interesting to explore how Galactic cluster samples compare on galaxy-integrated scales.  Our Sun's position within the disk of the Milky Way, surrounded by obscuring gas and dust along the Galactic mid-plane, does not provide the optimal vantage point for observing the distribution of clusters throughout our galaxy.  In fact, we argue below that extragalactic samples provide a better assessment of overall cluster populations, due to the uniformity of selection and the ability to survey a wide range of galactic environments.

The recent catalog of Milky Way clusters by \citet{Kharchenko13} contains 2547 clusters\footnote{This total excludes associations, moving groups, and remnant cluster classifications from the catalog's 3006 overall entries.}, similar to the number of entries in the PHAT cluster catalog.  But while the sample sizes are comparable, the uniformity and selection function of the Milky Way clusters differ significantly from the AP clusters in M31.  The sample of Milky Way clusters is compiled from a heterogenous set of literature sources, including earlier work of \citep{Dias02}, leading to an ill-defined selection function and catalog completeness that is difficult to characterize.  Assuming a constant surface density of clusters, \citet{Kharchenko13} suggest that the sample is complete to a radius of $\sim$1.8~kpc around the sun thus covering an area of $\sim$10 kpc$^2$.  Not only is this area more than an order of magnitude smaller than the physical region covered by PHAT, but the surveyed region of Galaxy is limited to the Solar neighborhood.  Most of the area within 1.8~kpc of the Sun lies within a Galactic inter-arm region, limiting the amount of on-going star formation and range of environments one can study.

According to estimates compiled in a recent review by \citet{PZ10} \citep[based on the sample of][]{Dias02}, the young (excluding globular clusters) Milky Way cluster sample includes objects that range in mass from 25~M$_\odot$ to 5$\times$10$^4$~M$_\odot$.  Within a radius of 1.8~kpc, the complete cluster sample includes a mass range that varies over $<$3 orders of magnitude, up to 4000~M$_\odot$, the mass of the Orion Nebula Cluster.  The proximity of the Milky Way clusters allows for the inclusion of low mass objects that remain undetected in M31, however the accurate understanding of mass completeness and catalog selection for PHAT, along with the number and variety of clusters included, makes the AP catalog the best available resource for a wide range of cluster science studies: cluster dissolution, mass functions, cluster formation efficiency, and how cluster properties vary with environment.

\subsection{Catalog Differences: Year 1 \& AP} \label{yr1diff}

The cluster definition we use for the AP catalog, as described in Section \ref{clusterdef}, is more liberal than the one used in our previous Year 1 catalog.  In Paper I, we excluded three categories of candidate clusters that we do not explicitly reject from the AP catalog:
\begin{enumerate}
\item \textit{Loose Associations} --- Defined by their lack of centrally concentrated stars, these objects are likely to be gravitationally unbound due to their large spatial extents and low stellar densities, and were therefore rejected from inclusion in the Year 1 catalog.  The AP search yielded many high-significance candidates that were not identified during the Year 1 effort.
\item \textit{Emission Line Regions} --- Compact, high surface brightness HII regions show up prominently in F475W imaging ([OIII] and H$\beta$ emission lines lie within the F475W bandpass) and tends to enhance the visual appearance of associated clusters.  While line emission on its own does not provide explicit evidence for or against the presence of a cluster (non-cluster HII regions and line emitting clusters both exist), we find that cluster candidates associated with emission line flux are accepted more frequently into the AP catalog than by the expert-based Year 1 search. We document this tendency because it reveals a possible systematic affecting catalog completeness that is not captured by our synthetic cluster tests: low mass clusters that produce line emission may be systematically overrepresented in the AP catalog with respect to the completeness function derived in Section \ref{comp}.
\item \textit{Small Clusters} --- While we emphasized a liberal, inclusive approach to cluster identification in Paper I, small candidate clusters were often discarded, with a loosely-defined limit requiring 3-4 spatially correlated stars to trigger inclusion in the catalog.  For the AP search, no star count limit was ever discussed.
\end{enumerate}

These three categories of objects represent systematic differences between the Year 1 and AP catalogs.  Of these three, the loose associations represent the most conspicuous difference: the number of bright blue objects (F336W$-$F475W $<$ -0.5 and F475W $<$ 19.75) identified within the Year 1 footprint more than doubled, from 15 to 35 clusters, many of which appear extended and poorly concentrated.  In an effort to clearly identify these uncertain and controversial AP clusters, we flag objects that match the following criteria as possible associations: bright (F475W $<$ 19.75), blue (F336W$-$F475W $<$ -0.5), and spatially extended.  A cluster is characterized as spatially extended either through its light profile, according to its half-light radius, or its profile of resolved main sequence stars, according to the radius that contains 60\% of the cluster's main sequence stars ($R_{0.6 N{\rm (MS)}}$).  We adopt the following criteria for spatial extension: \reff\ $>$ 1.05 arcsec (4 pc), or $R_{0.6 N{\rm (MS)}} > 0.5 R_{\rm ap}$ for stars with F475W$-$F814W $<$ 1 and F475W $<$ 24.  The combined color, magnitude, and spatial extension criteria identify 64 association-like objects; flags identifying these objects are included in Table \ref{ccat}.  These extended candidates are the most likely examples of regions hosting spatially correlated star formation, but where the stars may not have ever been gravitationally bound to one another.  As such, one should carefully evaluate the possibility that these candidates may not be the young progenitors of the older clusters we identify in this catalog.

\section{Summary} \label{summary}

We presented our methodology for transforming crowd-sourced effort into cluster catalogs for the AP-based analysis of the PHAT survey data.  We show the validity of our crowd-sourced cluster identification methodology and show good consistency between our results and expert-derived by-eye catalogs.  In addition, we present a thorough analysis of the resulting completeness characteristics of our cluster catalog, an essential component to any study of galaxy-wide star cluster populations.  Our completeness tests demonstrate that our PHAT cluster catalog is mass-limited and 50\% complete to $\sim$500 \solmass\ up to an age of 100 Myr, at which point the catalog becomes luminosity-limited at F475W $\sim$21.5.

The final cluster catalog includes 2753 entries, spanning more than three orders of magnitude in F475W luminosity.  Our use of HST imaging provides access to systems spanning the range from massive globular clusters to low-mass ($<10^3$ \solmass) clusters in the disk, similar to Milky Way open clusters.  Analysis of this sample provides a unique and unmatched opportunity to obtain a comprehensive understanding of star cluster populations within a large spiral galaxy.  The AP catalog serves as the definitive base data product that will enable an array of stellar cluster studies within M31.

\acknowledgements
{We owe a great debt of gratitude to the $\sim$30,000 AP volunteers who made this work possible.  Their contributions are acknowledged individually at \url{http:// www.andromedaproject.org/\#!/authors}. In addition, we acknowledge the collective efforts of the Zooniverse team and members of the PHAT collaboration for their assistance throughout the project.  We thank the anonymous referee for a prompt and helpful report.  We also thank Phil Marshall for insightful discussions. Support for this work was provided by NASA through grant number HST-GO-12055 from the Space Telescope Science Institute, which is operated by AURA, Inc., under NASA contract NAS5-26555.  DRW is supported by NASA through Hubble Fellowship grant HST-HF-51331.01 awarded by the Space Telescope Science Institute.  D.A.G. kindly acknowledges financial support by the German Research Foundation through grant GO 1659/3-1.  We acknowledges the Institute of Astronomy and Astrophysics, Academia Sinica (ASIAA) and Taiwan's Ministry of Science and Technology (MOST) for support of the Citizen Science in Astronomy Workshop.  Some of the data presented in this paper were obtained from the Mikulski Archive for Space Telescopes (MAST).  This research made use of the TOPCAT \citep{TOPCAT05} application and NASAÕs Astrophysics Data System (ADS) bibliographic services.}

{\it Facilities:} \facility{HST (ACS, WFC3)}.

\appendix

\section{Catalog Construction Procedure} \label{buildcat}

\begin{figure*}
\centering
\includegraphics[scale=0.47]{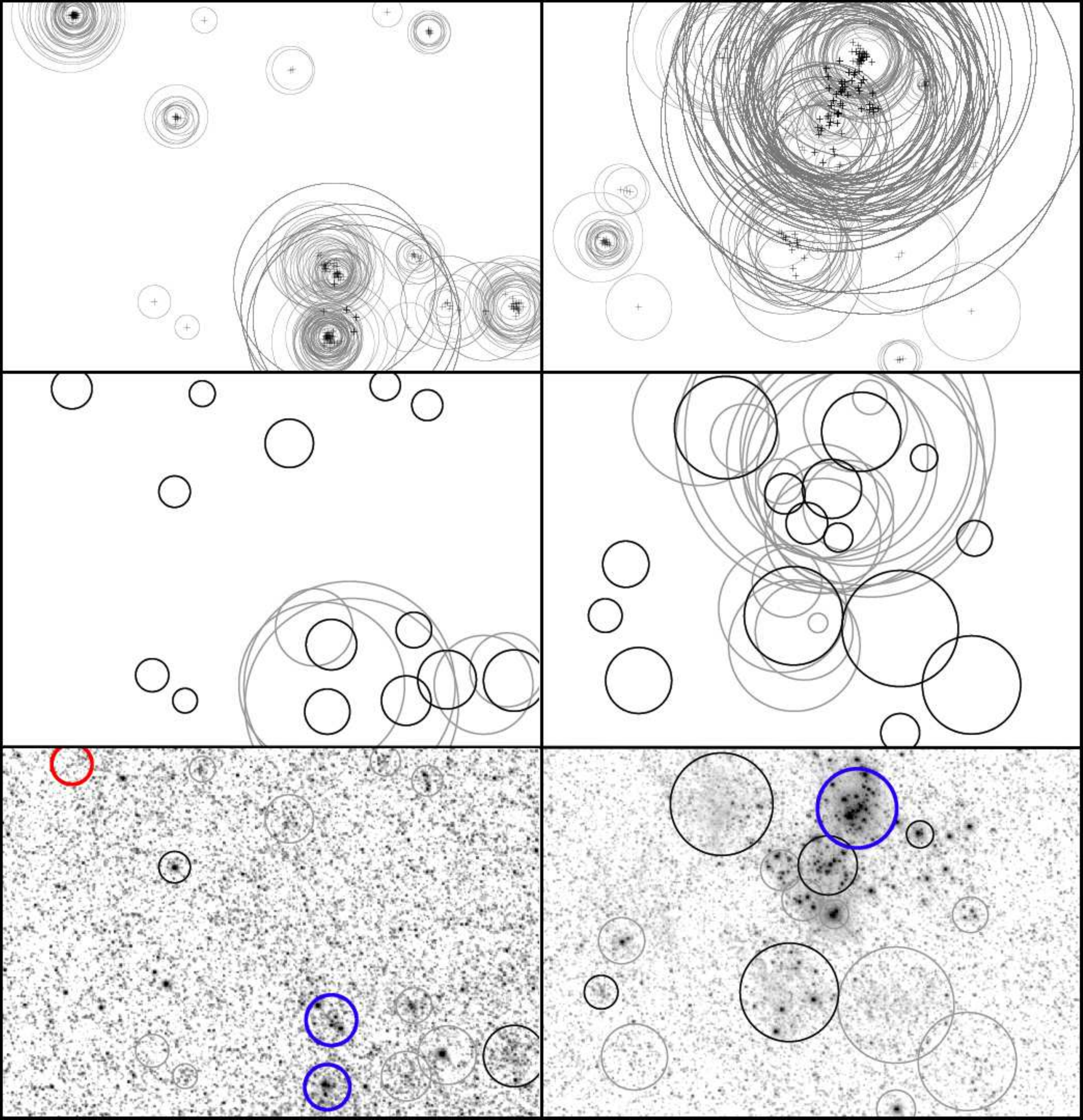}
\caption{Catalog construction examples, featuring identifications from B02-F11\_22 (left) and B06-F16\_22 (right).  Top panels: We plot all individual cluster and galaxy markings along with their centers.  Middle panels: We plot all initial object candidates that result from the grouping of center positions. Gray circles represent candidates that were pruned, while black circles represent candidates that go on to become final catalog entries.  Bottom panels: Each final candidate is shown, color coded by its final status: clusters (thick blue), galaxies (thick red), ancillary candidates with \ViewFrac\ $\ge$ 0.1 (thin black), and low significance candidates with \ViewFrac\ $<$ 0.1 (thin gray).}
\label{buildex}
\end{figure*}

AP catalog construction occurs in two phases: merging identifications on an image-by-image basis, followed by the merging of per-image catalogs into a single survey-wide catalog.  Throughout this description, we use the terms ``click'' and ``identification'' interchangeably to represent image markers placed by AP volunteers.  We begin by describing the first phase, which consists of three steps:

\begin{enumerate}
\item \textit{Create Candidate List} --- From the set of all cluster and galaxy identifications recorded for a given image, we construct a list of initial candidate objects by grouping center positions using a matching radius of 20 pixels (equivalent to 1 arcsec or 3.81 pc).  Our choice of matching radius was tuned such that clicks representing the same object were merged together, but distinct neighboring objects were not merged.  We observed that the positioning of marker centers are quite precise; the distribution of user-determined centers for well-defined image features can be described as a 2D Gaussian with $\sigma$=2 pixels (equivalent to 0.1 arcsec or 0.4 pc).  We iterate through the list of identifications, sorted from smallest to largest radius under the assumption that small radii identifications will have the most accurate center positions.  Each resulting candidate consists of a central position and radius, represented by the mean X and Y image coordinates and the median radius of each set of merged clicks.

\item \textit{Prune Candidates} --- Next we prune duplicate objects from the initial set of candidate objects.  Here, a duplicate object is a candidate that corresponds to the same image feature as another in the initial list, but the component clicks were not merged during the previous step.  This process begins by iterating through the initial candidate list in order of decreasing \ViewFrac.  For each iteration, we define the candidate in question as the primary object, and search for secondary objects, which are any other initial candidates whose circular boundary encloses the primary's center.  If we identify any secondary objects with a \ViewFrac\ less than that of the primary, the secondary is dropped from the candidate list.  If a secondary candidate has a higher \ViewFrac\ than the primary, the primary is dropped from the candidate list.  Once we've iterated through all initial candidates, the result of this pruning procedure is a list of spatially-unique candidates.

\item \textit{Re-associate Identifications with Final Candidates} --- To calculate final hit-rate statistics for each surviving candidate, we identify all original identifications where the candidate aperture encloses the identification center and vice versa and use these clicks to calculate \ViewFrac, \ClusterFrac, and \GalFrac values.  However, candidates retain their previous center and radius values.  Finally, we remove any candidates with only one associated identification (i.e., single click candidates), while remaining multi-click candidates go on to join the final per-image catalog.
\end{enumerate}

We present two image examples that show how our catalog construction algorithm works.  The top row of Figure \ref{buildex} shows all object identifications and their associated centers for each image.  The second row shows the full list of merged candidates that result from the first step described above, where those that survive the pruning process are highlighted in bold.  Finally, we show the final list of candidates that survive the candidate pruning and subsequent single-click cut overlaid on top of the field's single-band F475W image, where the most significant detections (\ViewFrac\ is $\ge$0.1) are shown in red.  The left column of Figure \ref{buildex} shows the B02-F11\_22 sub-image, a field that consists primarily of well-separated, well-defined candidates.  The right column shows B06-F16\_22, which highlights a challenging case with many non-unique, overlapping feature identifications.

The B06-F16\_22 image example presents a particularly informative example of our cataloging algorithm in action. The transition from the raw identification data in the top panel to the initial candidate list in the middle panel shows that our methodology for merging clicks (using a small 20 pixel matching radius) is quite conservative, insuring that nearby objects are not incorrectly combined.  Next, this initial candidate list is pruned to remove true duplications, cutting the first set of candidates down to those plotted in black in the middle panel.  This operation takes the significance of each candidate into account (according to \ViewFrac\ scores, reflecting total numbers of clicks), and yields a final list of objects that are spatially unique.  Identifications associated with the dropped duplicate candidates are not discarded, as most are re-associated during the final step of per-image processing.  Finally, the bottom panel shows the output of catalog processing, showing reasonable results even for this complex set of inputs.  While the low and moderate significance identifications (gray and black circles, respectively) are not included in the AP catalog published in Table \ref{ccat}, these objects are all recoverable due to their inclusion in the publicly available ancillary catalog presented in Appendix \ref{altcat}.

The primary AP base data product is produced in the second phase of the construction process: merging per-image catalogs into a final survey-wide catalog.  We perform this merge in a two-step process:

\begin{enumerate}
\item \textit{Match Per-Image Candidates} --- We compile a list of all sub-image catalog entries, and iterate through each entry in order from highest \ViewFrac\ to lowest.  During each iteration, we define the candidate in question as the primary object, and search for secondary objects, which are any other candidates whose circular boundary encloses the primary's center and vice versa.   If we identify any secondary objects, these matches are immediately removed from the list.  When complete, the resulting list of surviving objects represents our final list of spatially-unique catalog entries.

\item \textit{Merge Candidate Properties} --- To determine the properties of each final catalog object, consider each entry and its set of associated secondary entries.  From this set of per-image objects, identify those that lie completely within the bounds of their host sub-image (whose radius is less than the distance to the closest image edge) and merge their positions (in RA/Dec coordinates) and radii using the mean of their individual values, and assign final \ViewFrac, \ClusterFrac, and \GalFrac\ values and using a mean weighted by the number of total sub-image views.  Excluding objects that do not fall completely within their host image allows us to limit the influence of edge effects and biases on the final cataloged properties.  If none of the merged per-image entries pass this edge criteria, we adopt the properties of the entry that lies furthest from an image edge.
\end{enumerate}

\section{PHAT Background Galaxy Catalog} \label{bckgal}

\begin{figure}
\centering
\includegraphics[scale=0.55]{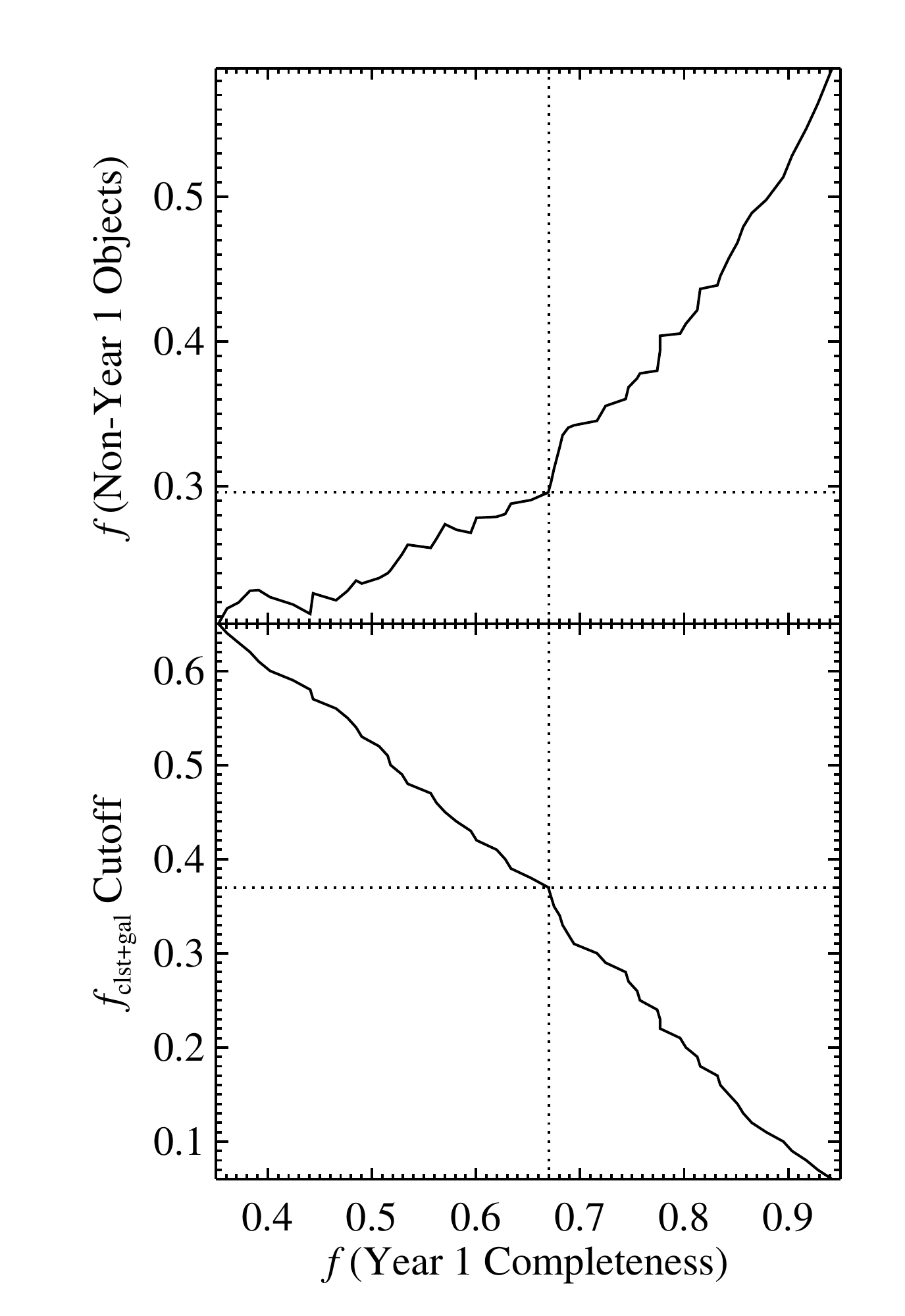}
\caption{Top: Completeness versus contamination curve for galaxy sample.  Bottom: Completeness versus \ViewFrac\ cutoff values.}
\label{figgalcomp}
\end{figure}

To define an AP galaxy sample, we base our selection on a combination of \ViewFrac\ and \GalFrac\ criteria.  Utilizing the \ViewFrac\ metric allows for better recovery of moderate and high \ViewFrac\ objects with \GalFrac\ scores that lie on the tail (0.3--0.8) of the distribution.  Adopting an \GalFrac\ threshold of 0.3, as discussed in Section \ref{catintro}, we construct a completeness curve for the galaxy sample to choose an appropriate \ViewFrac\ cutoff.  Similar to our cluster analysis, we compare the AP sample to the Year 1 galaxy sample.  The Year 1 galaxy sample was not a focused effort to identify all possible galaxies, therefore we do not categorize AP identifications that do not match Year 1 galaxies as contaminants, but study the behavior of the relative completeness fraction of these objects in a similar way.  We do not pursue the application of user weights for these results, but derive a single unweighted curve presented in Figure \ref{figgalcomp}.  

\begin{figure*}
\centering
\includegraphics[scale=1.0]{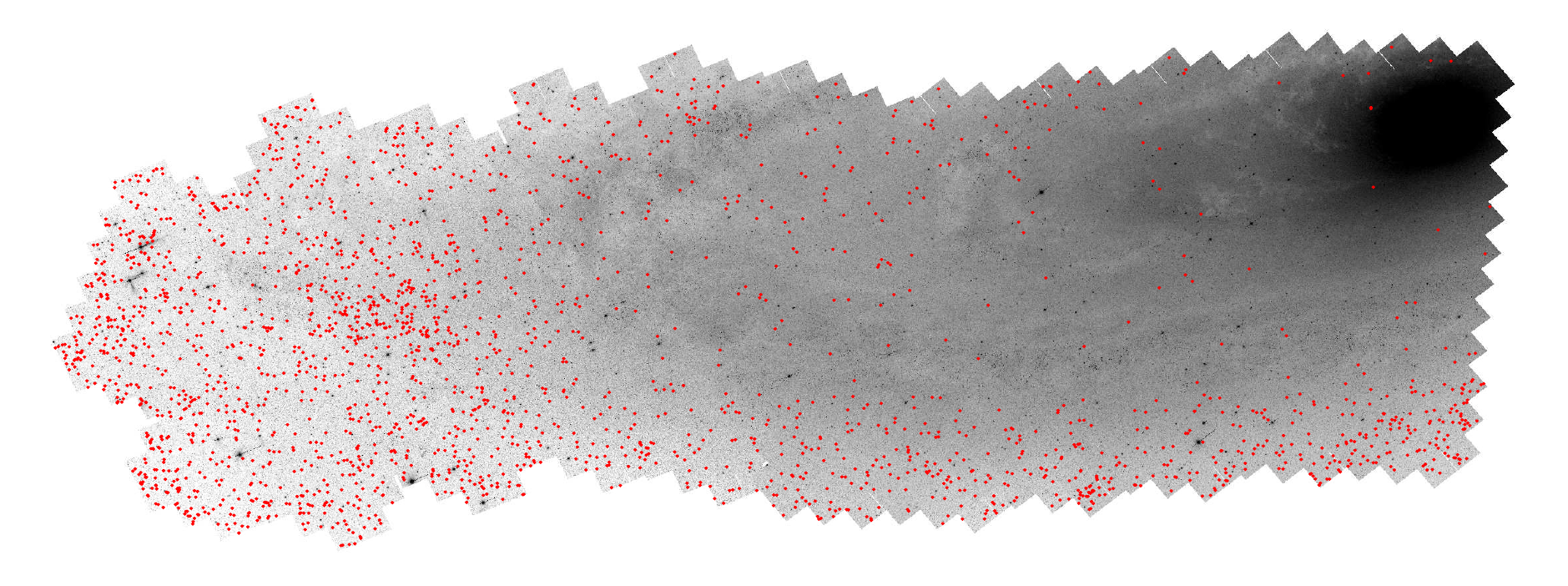}
\caption{Spatial distribution of AP background galaxy catalog overlaid on the PHAT survey-wide F475W image.}
\label{figgal}
\end{figure*}

We observe a transition in the behavior of the completeness curve at a Year 1 completeness of $\sim$0.67.  The slope of the completeness curve becomes steeper; quantitatively, this transition represents the point of diminishing returns, where more non-Year 1 objects are being added to the sample than previously identified Year 1 objects.  We choose this transition point as a suitable limit for catalog inclusion. Therefore, we define the AP galaxy sample using the following selection criteria:
\begin{equation}
f_{\rm clst+gal} > 0.37
\text{ AND }
f_{\rm galaxy} \ge 0.3
\end{equation}

These criteria select a sample of 2,270 background galaxies.  The catalog is presented in Table \ref{galcat} and their spatial distribution is shown in Figure \ref{figgal}.

\section{Ancillary Catalog Data} \label{altcat}

\textbf{Additional Candidate Catalog}: In addition to the AP clusters presented in Table \ref{ccat} and background galaxies presented in Table \ref{galcat}, we present Table \ref{altdat} containing 8775 other candidate object identifications with \ViewFrac\ $\ge$ 0.1 that were not included in either of the other data tables.

\textbf{Synthetic Cluster Results}: We present cluster-by-cluster synthetic recovery results in Table \ref{syndat} to allow for custom completeness analyses.  The table includes input cluster parameter information (i.e., age, mass, \reff, position, etc.) as well as AP catalog metadata.

\section{Commentary on Existing Cluster Catalogs} \label{knownc}

To supplement the broad comparison to previously published cluster catalogs that we presented in Section \ref{catcompare}, here we provide commentary on conflicting M31 object classifications.  We summarize these differences numerically in Table \ref{oldcat-summary} and present classification updates in Table \ref{oldcat-update}.

\section{Archival Image Search and Catalogs} \label{archivecat}

As part of the second round of AP data collection, we included additional, non-PHAT ACS images obtained from the HST archive.  These images were obtained by a program (PID: 10273, PI: Crotts) that observed four contiguous stripes within M31, each composed of 8 side-by-side ACS fields.  Please see Figure \ref{fig_footprint} for the locations of the strips with respect to the PHAT survey footprint.  This program utilized a F555W, F814W filter combination and exposure times that are shorter than those of PHAT: F555W varying between 81 and 413 sec, F814W varying between 457 and 502 sec.  We divided each of the 32 ACS images into 54 sub-images, yielding a total of 1,728 search images.

Following the same catalog construction procedures (see Section \ref{catintro}) and selection criteria (see Section \ref{catdata}) used for the PHAT classification data, we construct catalogs of clusters and background galaxies.  We present the cluster sample in Table \ref{archccat}, the background galaxy sample in Table \ref{archgalcat}, and compile an ancillary sample of all other identifications with \ViewFrac\ $\ge$ 0.1 in Table \ref{archaltdat}.


\bibliographystyle{apj}
\bibliography{clusterlit_apcat.bib}

\begin{thebibliography}{50}
\expandafter\ifx\csname natexlab\endcsname\relax\def\natexlab#1{#1}\fi

\bibitem[{{Allison} {et~al.}(2010){Allison}, {Goodwin}, {Parker}, {Portegies
  Zwart}, \& {de Grijs}}]{Allison10}
{Allison}, R.~J., {Goodwin}, S.~P., {Parker}, R.~J., {Portegies Zwart}, S.~F.,
  \& {de Grijs}, R. 2010, \mnras, 407, 1098

\bibitem[{{Bastian} {et~al.}(2012){Bastian}, {Adamo}, {Gieles}, {Silva-Villa},
  {Lamers}, {Larsen}, {Smith}, {Konstantopoulos}, \& {Zackrisson}}]{Bastian12}
{Bastian}, N., {et~al.} 2012, \mnras, 419, 2606

\bibitem[{{Beerman} {et~al.}(2012){Beerman}, {Johnson}, {Fouesneau},
  {Dalcanton}, {Weisz}, {Seth}, {Williams}, {Bell}, {Bianchi}, {Caldwell},
  {Dolphin}, {Gouliermis}, {Kalirai}, {Larsen}, {Melbourne}, {Rix}, \&
  {Skillman}}]{Beerman12}
{Beerman}, L.~C., {et~al.} 2012, \apj, 760, 104

\bibitem[{{Boissier} {et~al.}(2005){Boissier}, {Gil de Paz}, {Madore},
  {Boselli}, {Buat}, {Burgarella}, {Friedman}, {Barlow}, {Bianchi}, {Byun},
  {Donas}, {Forster}, {Heckman}, {Jelinsky}, {Lee}, {Malina}, {Martin},
  {Milliard}, {Morrissey}, {Neff}, {Rich}, {Schiminovich}, {Siegmund}, {Small},
  {Szalay}, {Welsh}, \& {Wyder}}]{Boissier05}
{Boissier}, S., {et~al.} 2005, \apjl, 619, L83

\bibitem[{{Boutloukos} \& {Lamers}(2003)}]{Boutloukos03}
{Boutloukos}, S.~G., \& {Lamers}, H.~J.~G.~L.~M. 2003, \mnras, 338, 717

\bibitem[{{Bressert} {et~al.}(2010){Bressert}, {Bastian}, {Gutermuth},
  {Megeath}, {Allen}, {Evans}, {Rebull}, {Hatchell}, {Johnstone}, {Bourke},
  {Cieza}, {Harvey}, {Merin}, {Ray}, \& {Tothill}}]{Bressert10}
{Bressert}, E., {et~al.} 2010, \mnras, 409, L54

\bibitem[{{Caldwell} {et~al.}(2009){Caldwell}, {Harding}, {Morrison}, {Rose},
  {Schiavon}, \& {Kriessler}}]{Caldwell09}
{Caldwell}, N., {Harding}, P., {Morrison}, H., {Rose}, J.~A., {Schiavon}, R.,
  \& {Kriessler}, J. 2009, \aj, 137, 94

\bibitem[{{Chandar} {et~al.}(2010{\natexlab{a}}){Chandar}, {Fall}, \&
  {Whitmore}}]{Chandar10}
{Chandar}, R., {Fall}, S.~M., \& {Whitmore}, B.~C. 2010{\natexlab{a}}, \apj,
  711, 1263

\bibitem[{{Chandar} {et~al.}(2010{\natexlab{b}}){Chandar}, {Whitmore}, {Kim},
  {Kaleida}, {Mutchler}, {Calzetti}, {Saha}, {O'Connell}, {Balick}, {Bond},
  {Carollo}, {Disney}, {Dopita}, {Frogel}, {Hall}, {Holtzman}, {Kimble},
  {McCarthy}, {Paresce}, {Silk}, {Trauger}, {Walker}, {Windhorst}, \&
  {Young}}]{Chandar10-M83}
{Chandar}, R., {et~al.} 2010{\natexlab{b}}, \apj, 719, 966

\bibitem[{{Dalcanton} {et~al.}(2012){Dalcanton}, {Williams}, {Lang}, {Lauer},
  {Kalirai}, {Seth}, {Dolphin}, {Rosenfield}, {Weisz}, {Bell}, {Bianchi},
  {Boyer}, {Caldwell}, {Dong}, {Dorman}, {Gilbert}, {Girardi}, {Gogarten},
  {Gordon}, {Guhathakurta}, {Hodge}, {Holtzman}, {Johnson}, {Larsen}, {Lewis},
  {Melbourne}, {Olsen}, {Rix}, {Rosema}, {Saha}, {Sarajedini}, {Skillman}, \&
  {Stanek}}]{Dalcanton12}
{Dalcanton}, J.~J., {et~al.} 2012, \apjs, 200, 18

\bibitem[{{Dias} {et~al.}(2002){Dias}, {Alessi}, {Moitinho}, \&
  {L{\'e}pine}}]{Dias02}
{Dias}, W.~S., {Alessi}, B.~S., {Moitinho}, A., \& {L{\'e}pine}, J.~R.~D. 2002,
  \aap, 389, 871

\bibitem[{{Dolphin}(2000)}]{Dolphin00}
{Dolphin}, A.~E. 2000, \pasp, 112, 1383

\bibitem[{{Fall} {et~al.}(2009){Fall}, {Chandar}, \& {Whitmore}}]{Fall09}
{Fall}, S.~M., {Chandar}, R., \& {Whitmore}, B.~C. 2009, \apj, 704, 453

\bibitem[{{Ford} {et~al.}(2013){Ford}, {Gear}, {Smith}, {Eales}, {Baes},
  {Bendo}, {Boquien}, {Boselli}, {Cooray}, {De Looze}, {Fritz}, {Gentile},
  {Gomez}, {Gordon}, {Kirk}, {Lebouteiller}, {O'Halloran}, {Spinoglio},
  {Verstappen}, \& {Wilson}}]{Ford13}
{Ford}, G.~P., {et~al.} 2013, \apj, 769, 55

\bibitem[{{Fouesneau} \& {Lan{\c c}on}(2010)}]{Fouesneau10}
{Fouesneau}, M., \& {Lan{\c c}on}, A. 2010, \aap, 521, A22+

\bibitem[{{Fouesneau} {et~al.}(2014){Fouesneau}, {Johnson}, {Weisz},
  {Dalcanton}, {Bell}, {Bianchi}, {Caldwell}, {Gouliermis}, {Guhathakurta},
  {Kalirai}, {Larsen}, {Rix}, {Seth}, {Skillman}, \& {Williams}}]{Fouesneau14}
{Fouesneau}, M., {et~al.} 2014, \apj, 786, 117

\bibitem[{{Galleti} {et~al.}(2004){Galleti}, {Federici}, {Bellazzini}, {Fusi
  Pecci}, \& {Macrina}}]{Galleti04}
{Galleti}, S., {Federici}, L., {Bellazzini}, M., {Fusi Pecci}, F., \&
  {Macrina}, S. 2004, \aap, 416, 917

\bibitem[{{Gieles} {et~al.}(2012){Gieles}, {Moeckel}, \& {Clarke}}]{Gieles12}
{Gieles}, M., {Moeckel}, N., \& {Clarke}, C.~J. 2012, \mnras, 426, L11

\bibitem[{{Gieles} \& {Portegies Zwart}(2011)}]{Gieles11}
{Gieles}, M., \& {Portegies Zwart}, S.~F. 2011, \mnras, 410, L6

\bibitem[{{Girardi} {et~al.}(2010){Girardi}, {Williams}, {Gilbert},
  {Rosenfield}, {Dalcanton}, {Marigo}, {Boyer}, {Dolphin}, {Weisz},
  {Melbourne}, {Olsen}, {Seth}, \& {Skillman}}]{Girardi10}
{Girardi}, L., {et~al.} 2010, \apj, 724, 1030

\bibitem[{{Hodge} {et~al.}(2010){Hodge}, {Krienke}, {Bianchi}, {Massey}, \&
  {Olsen}}]{Hodge10}
{Hodge}, P., {Krienke}, O.~K., {Bianchi}, L., {Massey}, P., \& {Olsen}, K.
  2010, \pasp, 122, 745

\bibitem[{{Hodge} {et~al.}(2009){Hodge}, {Krienke}, {Bellazzini}, {Perina},
  {Barmby}, {Cohen}, {Puzia}, \& {Strader}}]{Hodge09}
{Hodge}, P.~W., {Krienke}, O.~K., {Bellazzini}, M., {Perina}, S., {Barmby}, P.,
  {Cohen}, J.~G., {Puzia}, T.~H., \& {Strader}, J. 2009, \aj, 138, 770

\bibitem[{{Hunter} {et~al.}(2003){Hunter}, {Elmegreen}, {Dupuy}, \&
  {Mortonson}}]{Hunter03}
{Hunter}, D.~A., {Elmegreen}, B.~G., {Dupuy}, T.~J., \& {Mortonson}, M. 2003,
  \aj, 126, 1836

\bibitem[{{Huxor} {et~al.}(2014){Huxor}, {Mackey}, {Ferguson}, {Irwin},
  {Martin}, {Tanvir}, {Veljanoski}, {McConnachie}, {Fishlock}, {Ibata}, \&
  {Lewis}}]{Huxor14}
{Huxor}, A.~P., {et~al.} 2014, \mnras, 442, 2165

\bibitem[{{Johnson} {et~al.}(2012){Johnson}, {Seth}, {Dalcanton}, {Caldwell},
  {Fouesneau}, {Gouliermis}, {Hodge}, {Larsen}, {Olsen}, {San Roman},
  {Sarajedini}, {Weisz}, {Williams}, {Beerman}, {Bianchi}, {Dolphin},
  {Girardi}, {Guhathakurta}, {Kalirai}, {Lang}, {Monachesi}, {Nanda}, {Rix}, \&
  {Skillman}}]{Johnson12}
{Johnson}, L.~C., {et~al.} 2012, \apj, 752, 95

\bibitem[{{Kharchenko} {et~al.}(2013){Kharchenko}, {Piskunov}, {Schilbach},
  {R{\"o}ser}, \& {Scholz}}]{Kharchenko13}
{Kharchenko}, N.~V., {Piskunov}, A.~E., {Schilbach}, E., {R{\"o}ser}, S., \&
  {Scholz}, R.-D. 2013, \aap, 558, A53

\bibitem[{{King}(1962)}]{King62}
{King}, I. 1962, \aj, 67, 471

\bibitem[{{Krienke} \& {Hodge}(2007)}]{Krienke07}
{Krienke}, O.~K., \& {Hodge}, P.~W. 2007, \pasp, 119, 7

\bibitem[{{Krienke} \& {Hodge}(2008)}]{Krienke08}
---. 2008, \pasp, 120, 1

\bibitem[{{Krienke} \& {Hodge}(2013)}]{Krienke13}
---. 2013, \pasp, 125, 636

\bibitem[{{Kroupa}(2001)}]{Kroupa01}
{Kroupa}, P. 2001, \mnras, 322, 231

\bibitem[{{Kruijssen}(2012)}]{Kruijssen12}
{Kruijssen}, J.~M.~D. 2012, \mnras, 426, 3008

\bibitem[{{Lada} \& {Lada}(2003)}]{Lada03}
{Lada}, C.~J., \& {Lada}, E.~A. 2003, \araa, 41, 57

\bibitem[{{Lintott} {et~al.}(2008){Lintott}, {Schawinski}, {Slosar}, {Land},
  {Bamford}, {Thomas}, {Raddick}, {Nichol}, {Szalay}, {Andreescu}, {Murray}, \&
  {Vandenberg}}]{Lintott08}
{Lintott}, C.~J., {et~al.} 2008, \mnras, 389, 1179

\bibitem[{{Marigo} {et~al.}(2008){Marigo}, {Girardi}, {Bressan}, {Groenewegen},
  {Silva}, \& {Granato}}]{Marigo08}
{Marigo}, P., {Girardi}, L., {Bressan}, A., {Groenewegen}, M.~A.~T., {Silva},
  L., \& {Granato}, G.~L. 2008, \aap, 482, 883

\bibitem[{{McConnachie} {et~al.}(2005){McConnachie}, {Irwin}, {Ferguson},
  {Ibata}, {Lewis}, \& {Tanvir}}]{McConnachie05}
{McConnachie}, A.~W., {Irwin}, M.~J., {Ferguson}, A.~M.~N., {Ibata}, R.~A.,
  {Lewis}, G.~F., \& {Tanvir}, N. 2005, \mnras, 356, 979

\bibitem[{{Popescu} {et~al.}(2012){Popescu}, {Hanson}, \&
  {Elmegreen}}]{Popescu12}
{Popescu}, B., {Hanson}, M.~M., \& {Elmegreen}, B.~G. 2012, \apj, 751, 122

\bibitem[{{Portegies Zwart} {et~al.}(2010){Portegies Zwart}, {McMillan}, \&
  {Gieles}}]{PZ10}
{Portegies Zwart}, S.~F., {McMillan}, S.~L.~W., \& {Gieles}, M. 2010, \araa,
  48, 431

\bibitem[{{San Roman} {et~al.}(2010){San Roman}, {Sarajedini}, \&
  {Aparicio}}]{SanRoman10}
{San Roman}, I., {Sarajedini}, A., \& {Aparicio}, A. 2010, \apj, 720, 1674

\bibitem[{{Schlafly} \& {Finkbeiner}(2011)}]{Schlafly11}
{Schlafly}, E.~F., \& {Finkbeiner}, D.~P. 2011, \apj, 737, 103

\bibitem[{{Schwamb} {et~al.}(2012){Schwamb}, {Lintott}, {Fischer}, {Giguere},
  {Lynn}, {Smith}, {Brewer}, {Parrish}, {Schawinski}, \& {Simpson}}]{Schwamb12}
{Schwamb}, M.~E., {et~al.} 2012, \apj, 754, 129

\bibitem[{{Silva-Villa} \& {Larsen}(2011)}]{Silva-Villa11}
{Silva-Villa}, E., \& {Larsen}, S.~S. 2011, \aap, 529, A25+

\bibitem[{{Simpson} {et~al.}(2012){Simpson}, {Povich}, {Kendrew}, {Lintott},
  {Bressert}, {Arvidsson}, {Cyganowski}, {Maddison}, {Schawinski}, {Sherman},
  {Smith}, \& {Wolf-Chase}}]{Simpson12}
{Simpson}, R.~J., {et~al.} 2012, \mnras, 424, 2442

\bibitem[{{Taylor}(2005)}]{TOPCAT05}
{Taylor}, M.~B. 2005, in Astronomical Society of the Pacific Conference Series,
  Vol. 347, Astronomical Data Analysis Software and Systems XIV, ed.
  P.~{Shopbell}, M.~{Britton}, \& R.~{Ebert}, 29

\bibitem[{{Verley} {et~al.}(2009){Verley}, {Corbelli}, {Giovanardi}, \&
  {Hunt}}]{Verley09}
{Verley}, S., {Corbelli}, E., {Giovanardi}, C., \& {Hunt}, L.~K. 2009, \aap,
  493, 453

\bibitem[{{Whitmore} {et~al.}(2014){Whitmore}, {Chandar}, {Bowers}, {Larsen},
  {Lindsay}, {Ansari}, \& {Evans}}]{Whitmore14}
{Whitmore}, B.~C., {Chandar}, R., {Bowers}, A.~S., {Larsen}, S., {Lindsay}, K.,
  {Ansari}, A., \& {Evans}, J. 2014, \aj, 147, 78

\bibitem[{{Whitney} {et~al.}(2008){Whitney}, {Sewilo}, {Indebetouw},
  {Robitaille}, {Meixner}, {Gordon}, {Meade}, {Babler}, {Harris}, {Hora},
  {Bracker}, {Povich}, {Churchwell}, {Engelbracht}, {For}, {Block}, {Misselt},
  {Vijh}, {Leitherer}, {Kawamura}, {Blum}, {Cohen}, {Fukui}, {Mizuno},
  {Mizuno}, {Srinivasan}, {Tielens}, {Volk}, {Bernard}, {Boulanger}, {Frogel},
  {Gallagher}, {Gorjian}, {Kelly}, {Latter}, {Madden}, {Kemper}, {Mould},
  {Nota}, {Oey}, {Olsen}, {Onishi}, {Paladini}, {Panagia}, {Perez-Gonzalez},
  {Reach}, {Shibai}, {Sato}, {Smith}, {Staveley-Smith}, {Ueta}, {Van Dyk},
  {Werner}, {Wolff}, \& {Zaritsky}}]{Whitney08}
{Whitney}, B.~A., {et~al.} 2008, \aj, 136, 18

\bibitem[{{Willett} {et~al.}(2013){Willett}, {Lintott}, {Bamford}, {Masters},
  {Simmons}, {Casteels}, {Edmondson}, {Fortson}, {Kaviraj}, {Keel}, {Melvin},
  {Nichol}, {Raddick}, {Schawinski}, {Simpson}, {Skibba}, {Smith}, \&
  {Thomas}}]{Willett13}
{Willett}, K.~W., {et~al.} 2013, \mnras, 435, 2835

\bibitem[{{Williams} \& {Hodge}(2001)}]{Williams01}
{Williams}, B.~F., \& {Hodge}, P.~W. 2001, \apj, 559, 851

\bibitem[{{Williams} {et~al.}(2014){Williams}, {Lang}, {Dalcanton}, {Dolphin},
  \& {Weisz}}]{Williams14}
{Williams}, B.~F., {Lang}, D., {Dalcanton}, J.~J., {Dolphin}, A., \& {Weisz},
  D.~R. 2014, \apjs

\end{thebibliography}

\newpage


\begin{deluxetable*}{lcccccccc}
\centering
\tablecaption{User Weighting Parameters\label{compconttable}}
\tablehead{
\colhead{Name} & \multicolumn{2}{c}{Detection} & \multicolumn{2}{c}{Non-detection} & \colhead{$d_{optimal}$} & \colhead{Optimal} & \colhead{Optimal}  & \colhead{\ClusterFrac} \\
\colhead{} & \colhead{$m_{\rm{logistic}}$} & \colhead{$b_{\rm{logistic}}$} & \colhead{$m_{\rm{logistic}}$} & \colhead{$b_{\rm{logistic}}$} & \colhead{} & \colhead{Completeness} & \colhead{Contamination} & \colhead{Cutoff}
}
\tablewidth{0pt}
\startdata
Uniform Weights & \nodata & \nodata & \nodata & \nodata & 0.1809 & 0.8528 & 0.1052 & 0.5114 \\
Uniform Weights (Match Comp) & \nodata & \nodata & \nodata & \nodata & 0.1928 & 0.8811 & 0.1518 & 0.4512 \\
Uniform Weights (Match Cont) & \nodata & \nodata & \nodata & \nodata & 0.1828 & 0.8453 & 0.0974 & 0.5214 \\
Best Weights & 16.0 & 0.6 & 39.0 & 1.1 & 0.1543 & 0.8811 & 0.0984 & 0.6416
\enddata
\end{deluxetable*}


\begin{deluxetable*}{cccccccccccc}
\tabletypesize{\tiny}
\setlength{\tabcolsep}{0.05in}
\tablecaption{AP Cluster Catalog \label{ccat}}
\tablewidth{0pt}

\tablehead{
\colhead{AP ID} & \colhead{RA (J2000)} & \colhead{Dec (J2000)} & \colhead{$R_{\rm ap}$ ($''$)} & \colhead{$R_{\rm eff}$ ($''$)} &  \colhead{$m_{\rm ApCor}$\tablenotemark{a}} & \colhead{F275W$_{\rm ap}$}  & \colhead{$\sigma$} & \colhead{F336W$_{\rm ap}$}  & \colhead{$\sigma$} & \colhead{F475W$_{\rm ap}$} & \colhead{$\sigma$} \\
\colhead{\ViewFrac} & \colhead{\GalFrac} & \colhead{\ClusterFracW} & \colhead{PC ID}& \colhead{Alt ID} & \colhead{Flags} & \colhead{F814W$_{\rm ap}$}  & \colhead{$\sigma$} & \colhead{F110W$_{\rm ap}$}  & \colhead{$\sigma$} & \colhead{F160W$_{\rm ap}$} & \colhead{$\sigma$}
}

\startdata
    1 & 11.435516 & 41.698562 &  2.19 &  0.60 & -0.01 & 20.12 & 0.04 & 19.16 & 0.01 & 18.77 & 0.01 \\
1.0000 & 0.0000 & 1.0000 & \nodata &    B374-G306         & \nodata & 17.69 & 0.08 & 17.19 & 0.15 & 16.59 & 0.20 \\ \hline
    2 & 11.366514 & 41.701013 &  1.86 &  0.61 & -0.04 & 20.91 & 0.10 & 20.25 & 0.02 & 20.01 & 0.10 \\
0.9717 & 0.0083 & 0.9894 & \nodata &    M085              & \nodata & 19.05 & 0.21 & $>$18.06 & \nodata & $>$17.06 & \nodata \\ \hline
    3 & 11.471290 & 42.049246 &  1.95 &  0.88 & -0.14 & 21.27 & 0.14 & 20.81 & 0.03 & 20.67 & 0.08 \\
1.0000 & 0.0000 & 1.0000 & \nodata &    \nodata           & \nodata & 20.07 & 0.31 & $>$18.97 & \nodata & $>$18.07 & \nodata \\ \hline
    4 & 11.474664 & 42.038351 &  2.87 &  0.98 & -0.05 & 20.10 & 0.04 & 19.10 & 0.02 & 18.75 & 0.03 \\
1.0000 & 0.0227 & 0.9909 & \nodata &    B483-D085         & \nodata & 17.78 & 0.08 & 17.29 & 0.16 & 16.66 & 0.21 \\ \hline
    5 & 10.991636 & 41.359328 &  1.46 &  0.40 & -0.01 & 20.88 & 0.04 & 20.29 & 0.03 & 20.09 & 0.06 \\
1.0000 & 0.0000 & 1.0000 & \nodata &    M005              & \nodata & 19.36 & 0.10 & 18.73 & 0.21 & 17.72 & 0.25
\enddata

\tablecomments{Table \ref{ccat} is published in its entirety in the electronic edition of the {\it Astrophysical Journal}.  A portion is shown here for guidance regarding its form and content.  Three-sigma upper limits are denoted by a ``$>$'' symbol.  PC ID and Alt ID refer to cluster identifiers from the Year 1 catalog and other literature sources, respectively.  Flags: E = Extended Object (see Section \ref{yr1diff}); B = Bulge or B03 object manually added (see Section \ref{catdata}).}
\tablenotetext{a}{Aperture Corrections are provided such that $m_{\rm Total} = m_{\rm ap} + m_{\rm ApCor}$.}

\end{deluxetable*}


\begin{deluxetable*}{lc}
\tablecaption{Passband Photometric Quality Comparison for Cluster Sample\label{photdat}}
\tablehead{\colhead{Passband} & \colhead{$N$(Detections)}}
\tablewidth{0pt}
\startdata
            F275W & 1733 (62.9\%) \\
            F336W & 2481 (90.1\%) \\
            F475W & 2717 (98.7\%) \\
            F814W & 1871 (68.0\%) \\
            F110W & 1209 (43.9\%) \\
            F160W & 1035 (37.6\%) \\
      F336W+F475W & 2464 (89.5\%) \\
      F475W+F814W & 1867 (67.8\%) \\
F336W+F475W+F814W & 1701 (61.8\%)
\enddata
\end{deluxetable*}


\begin{deluxetable*}{lccccccc}
\tablecolumns{8}
\tablecaption{AP Background Galaxy Catalog \label{galcat}}
\tablewidth{0pt}

\tablehead{
\colhead{AP ID} & \colhead{RA (J2000)} & \colhead{Dec (J2000)} & \colhead{$R_{\rm ap}$ ($''$)} & \colhead{\ViewFrac} & \colhead{\GalFrac} & \colhead{F814W$_{\rm ap}$}  & \colhead{$\sigma$}
}

\startdata
    8 & 11.447226 & 42.268672 &  2.75 & 1.0000 & 0.9884 & 18.34 & 0.05 \\
   10 & 11.922144 & 42.102526 &  3.53 & 1.0000 & 1.0000 & 18.53 & 0.11 \\
   20 & 11.911096 & 42.076717 &  2.00 & 0.9902 & 0.9604 & 20.07 & 0.29 \\
   21 & 11.585498 & 41.726941 &  3.82 & 0.9901 & 0.9900 & 16.04 & 0.02 \\
   22 & 11.270065 & 41.312829 &  2.38 & 0.9901 & 0.9500 & 18.67 & 0.11
\enddata

\tablecomments{Table \ref{galcat} is published in its entirety in the electronic edition of the {\it Astrophysical Journal}.  A portion is shown here for guidance regarding its form and content.}
\end{deluxetable*}


\begin{deluxetable*}{cccccccccccc}
\tabletypesize{\tiny}
\setlength{\tabcolsep}{0.05in}
\tablecaption{AP Ancillary Catalog \label{altdat}}
\tablewidth{0pt}

\tablehead{
\colhead{AP ID} & \colhead{RA (J2000)} & \colhead{Dec (J2000)} & \colhead{$R_{\rm ap}$ ($''$)} & \colhead{$R_{\rm eff}$ ($''$)} &  \colhead{$m_{\rm ApCor}$\tablenotemark{a}} & \colhead{F275W$_{\rm ap}$}  & \colhead{$\sigma$} & \colhead{F336W$_{\rm ap}$}  & \colhead{$\sigma$} & \colhead{F475W$_{\rm ap}$} & \colhead{$\sigma$} \\
\colhead{\ViewFrac} & \colhead{\GalFrac} & \colhead{\ClusterFracW} & \colhead{PC ID}& \colhead{Alt ID} & \colhead{Flags} & \colhead{F814W$_{\rm ap}$}  & \colhead{$\sigma$} & \colhead{F110W$_{\rm ap}$}  & \colhead{$\sigma$} & \colhead{F160W$_{\rm ap}$} & \colhead{$\sigma$}
}

\startdata
 1706 & 11.393493 & 41.774981 &  1.25 &  0.36 & -0.02 & $>$21.62 & \nodata & 21.23 & 0.24 & 21.21 & 0.20 \\
0.3415 & 0.0000 & 0.5252 & \nodata &    \nodata           & \nodata & $>$20.08 & \nodata & $>$19.39 & \nodata & $>$18.73 & \nodata \\ \hline
 2073 & 11.701786 & 41.963523 &  1.09 &  0.46 & -0.11 & 23.85 & 0.39 & 23.21 & 0.07 & 22.57 & 0.16 \\
0.7738 & 0.1538 & 0.6389 & \nodata &    \nodata           & \nodata & 20.95 & 0.33 & 20.09 & 0.34 & 19.16 & 0.31 \\ \hline
 2149 & 11.133017 & 41.395088 &  2.51 &  1.69 & -0.38 & 16.76 & 0.04 & 17.28 & 0.15 & 18.80 & 0.04 \\
0.5294 & 0.0000 & 0.4364 & \nodata &    \nodata           & \nodata & $>$19.62 & \nodata & $>$18.12 & \nodata & $>$16.93 & \nodata \\ \hline
 2486 & 11.554857 & 41.873578 &  1.81 &  0.23 & -0.00 & 19.33 & 0.11 & 19.23 & 0.11 & $>$20.12 & \nodata \\
0.3372 & 0.0345 & 0.3793 & \nodata &    \nodata           & \nodata & $>$19.57 & \nodata & $>$18.62 & \nodata & $>$17.77 & \nodata \\ \hline
 2532 & 10.915584 & 41.487991 &  2.08 &  1.08 & -0.21 & 21.71 & 0.34 & 20.63 & 0.17 & 20.06 & 0.16 \\
0.7263 & 0.2754 & 0.5502 & \nodata &    SK070A            & \nodata & 18.14 & 0.09 & 17.46 & 0.11 & 16.78 & 0.22
\enddata

\tablecomments{Table \ref{altdat} is published in its entirety in the electronic edition of the {\it Astrophysical Journal}.  A portion is shown here for guidance regarding its form and content.  Three-sigma upper limits are denoted by a ``$>$'' symbol.  PC ID and Alt ID refer to cluster identifiers from the Year 1 catalog and other literature sources, respectively.}
\tablenotetext{a}{Aperture Corrections are provided such that $m_{\rm Total} = m_{\rm ap} + m_{\rm ApCor}$.}

\end{deluxetable*}


\begin{deluxetable*}{ccccccccc}
\tabletypesize{\tiny}
\setlength{\tabcolsep}{0.05in}
\tablecaption{Synthetic Cluster Results \label{syndat}}
\tablewidth{0pt}

\tablehead{
\colhead{ID} & \colhead{log(Mass/\solmass)} & \colhead{log(Age/yr)} & \colhead{$Z$} & \colhead{$A_{V}$} & \colhead{$R_{\rm eff,in}$ ($''$)} & \colhead{F475W$_{\rm in}$} & \colhead{F814W$_{\rm in}$} & \colhead{F475W$_{-3{\rm ,in}}$} \\
\colhead{RA (J2000)} & \colhead{Dec (J2000)} & \colhead{$R_{\rm ap}$ ($''$)} & \colhead{\ViewFrac} & \colhead{\GalFrac} & \colhead{\ClusterFracW} & \colhead{$N$(MS)} & \colhead{$N$(RGB)} & \colhead{Detected} 
}

\startdata
    1 &  3.17 &  7.30 & 0.019 & 1.612 & 0.319 & 20.75 & 20.06 & 21.07 \\
11.012636 & 41.181335 &  1.39 & 0.9333 & 0.0000 & 0.9997 &   169 &   374 & T \\ \hline
    2 &  4.51 &  9.60 & 0.019 & 0.253 & 1.397 & 20.19 & 18.34 & 20.21 \\
11.003787 & 41.184849 &  1.99 & 0.7326 & 0.0159 & 0.9386 &   192 &   418 & T \\ \hline
    3 &  2.17 &  8.10 & 0.019 & 0.230 & 0.467 & 21.18 & 21.18 & 22.36 \\
10.985614 & 41.192447 &  1.42 & 0.5222 & 0.0000 & 0.6968 &   145 &   468 & T \\ \hline
    4 &  3.92 & 10.05 & 0.0001 & 1.366 & 0.343 & 22.92 & 21.02 & 23.22 \\
11.004518 & 41.190121 &  1.16 & 0.1786 & 0.0000 & 0.1575 &   200 &   398 & F \\ \hline
    5 &  4.47 & 10.05 & 0.004 & 0.370 & 0.683 & 20.72 & 18.82 & 20.87 \\
10.990636 & 41.195372 &  1.47 & 0.7126 & 0.0806 & 0.9198 &   199 &   497 & T
\enddata

\tablecomments{Table \ref{syndat} is published in its entirety in the electronic edition of the {\it Astrophysical Journal}.  A portion is shown here for guidance regarding its form and content.  $N$(MS) and $N$(RGB) values are evaluated per search image, as defined in Section \ref{comp}.}

\end{deluxetable*}


\begin{deluxetable*}{l|lcc}
\tabletypesize{\footnotesize}
\tablecaption{Summary of Existing Cluster Catalog Classifications and Revisions \label{oldcat-summary}}
\tablewidth{0pt}
\tablecolumns{4}

\tablehead{
\colhead{Catalog} & \colhead{Clusters} & \colhead{Candidates\tablenotemark{a}} & \colhead{Non-Cluster\tablenotemark{b}} \\
\colhead{} & \multicolumn{3}{c}{\# Accepted as AP Cluster (\# Rejected as Not AP Cluster)}
}

\startdata
Year 1         & 532 (69) & 95 (142) & 39 \\
RBC            & 232 (2) & 28 (18) & 40 \\
Caldwell     & 232 (10) & \nodata & 8 \\
HKC            & 156 (57) & \nodata & \nodata
\enddata

\tablenotetext{a}{The ``candidate'' classification refers to PHAT Year 1 possible cluster table and RBC flags 2 and 3.}
\tablenotetext{b}{The ``non-cluster'' classification refers to galaxy, star, HII region, and other non-cluster classifications.}

\end{deluxetable*}


\begin{deluxetable*}{lcccl}
\tablecaption{RBC Flag Revisions and Commentary \label{oldcat-update}}
\tablewidth{0pt}

\tablehead{
\colhead{AP ID} & \colhead{RBC Name} & \colhead{New Flag} & \colhead{Old Flag} & \colhead{Comments}
}

\startdata
   55 &       SK102A    &  1 &  6 &         \nodata \\
   68 &       SK213B    &  1 &  2 &         \nodata \\
   77 &       SK182B    &  1 &  6 &         \nodata \\
   89 &       M065      &  1 &  2 &         \nodata \\
   91 &       M004      &  1 &  6 &         \nodata
\enddata

\tablecomments{Table \ref{oldcat-update} is published in its entirety in the electronic edition of the {\it Astrophysical Journal}.  A portion is shown here for guidance regarding its form and content.}
\end{deluxetable*}

\clearpage

\begin{deluxetable*}{lccccccc}
\tabletypesize{\tiny}
\setlength{\tabcolsep}{0.05in}
\tablecaption{Archival AP Cluster Catalog \label{archccat}}
\tablewidth{0pt}

\tablehead{
\colhead{AAP ID} & \colhead{RA (J2000)} & \colhead{Dec (J2000)} & \colhead{$R_{ap}$ ($''$)} & \colhead{\ViewFrac} & \colhead{\GalFrac} & \colhead{\ClusterFracW} & \colhead{Alt ID}
}

\startdata
    1 & 10.522610 & 41.435868 &  2.42 & 0.9903 & 0.0294 & 0.9730 & B069-G132 \\
    2 & 10.541597 & 40.907603 &  2.82 & 0.9804 & 0.0000 & 0.9905 &    \nodata \\
    8 & 10.509294 & 40.896004 &  1.97 & 0.9800 & 0.0408 & 0.9908 & SK041A \\
    9 & 10.753904 & 41.656852 &  1.42 & 0.9126 & 0.0000 & 0.9915 &    \nodata \\
   11 & 10.521275 & 40.885136 &  1.66 & 0.9712 & 0.0594 & 0.9828 &    \nodata
\enddata

\tablecomments{Table \ref{archccat} is published in its entirety in the electronic edition of the {\it Astrophysical Journal}.  A portion is shown here for guidance regarding its form and content.}

\end{deluxetable*}


\begin{deluxetable*}{lccccc}
\tablecolumns{6}
\tablecaption{Archival AP Background Galaxy Catalog \label{archgalcat}}
\tablewidth{0pt}

\tablehead{
\colhead{AAP ID} & \colhead{RA (J2000)} & \colhead{Dec (J2000)} & \colhead{$R_{\rm ap}$ ($''$)} & \colhead{\ViewFrac} & \colhead{\GalFrac}
}

\startdata
    3 & 10.846248 & 41.040394 &  1.99 & 0.9804 & 0.8900 \\
    4 & 10.536274 & 41.444508 &  5.10 & 0.9802 & 0.9596 \\
    5 & 10.595225 & 40.953062 &  2.44 & 0.9802 & 0.9192 \\
    6 & 10.465582 & 41.411634 &  4.50 & 0.9802 & 0.9495 \\
    7 & 10.463831 & 41.416378 &  4.79 & 0.9802 & 0.9697
\enddata

\tablecomments{Table \ref{archgalcat} is published in its entirety in the electronic edition of the {\it Astrophysical Journal}.  A portion is shown here for guidance regarding its form and content.}
\end{deluxetable*}


\begin{deluxetable*}{lccccccc}
\tabletypesize{\tiny}
\setlength{\tabcolsep}{0.05in}
\tablecaption{Archival AP Ancillary Catalog \label{archaltdat}}
\tablewidth{0pt}

\tablehead{
\colhead{AAP ID} & \colhead{RA (J2000)} & \colhead{Dec (J2000)} & \colhead{$R_{\rm ap}$ ($''$)} & \colhead{\ViewFrac} & \colhead{\GalFrac} & \colhead{\ClusterFracW} & \colhead{Alt ID}
}

\startdata
  238 & 10.491200 & 41.439039 &  1.01 & 0.7788 & 0.2716 & 0.6317 & KHM31-357 \\
  334 & 10.750010 & 41.001666 &  1.35 & 0.6923 & 0.1389 & 0.5875 & KHM31-453 \\
  389 & 10.651615 & 41.552907 &  1.36 & 0.6400 & 0.2188 & 0.5784 &    \nodata \\
  399 & 10.546207 & 41.509361 &  1.41 & 0.6341 & 0.0385 & 0.6197 & KHM31-367 \\
  400 & 10.950410 & 41.192429 &  1.36 & 0.6337 & 0.2031 & 0.5965 &    \nodata
\enddata

\tablecomments{Table \ref{archaltdat} is published in its entirety in the electronic edition of the {\it Astrophysical Journal}.  A portion is shown here for guidance regarding its form and content.}

\end{deluxetable*}


\end{document}